\documentclass[journal,10pt]{IEEEtran}

\ifCLASSINFOpdf
\else
   \usepackage[dvips]{graphicx}
\fi
\usepackage{url}
\usepackage{algorithmicx}
\usepackage{algpseudocode}
\usepackage{algorithm}
\algdef{SE}[DOWHILE]{Do}{doWhile}{\algorithmicdo}[1]{\algorithmicwhile\ #1}%
\usepackage{amsmath}
\usepackage{dsfont, cuted}
\usepackage{amsfonts}
\usepackage{amsthm}
\usepackage[utf8]{inputenc}
\usepackage{breqn}
\usepackage{subfigure}
\usepackage{multirow}
\usepackage{multicol}
\usepackage{booktabs}
\usepackage{fullwidth}
\usepackage[noadjust]{cite}
\usepackage{setspace}
\usepackage{flushend}
\usepackage{hyperref}
\usepackage{bmpsize}
\usepackage{xcolor}
\usepackage{graphicx}

\usepackage{multirow,tabularx}
\newcolumntype{C}{>{\centering\arraybackslash}X}

\begin{document}

\title{Modeling and SAR Imaging of the Sea Surface: \\a Review of the State-of-the-Art with Simulations}

\author{Igor~G.~Rizaev,~Oktay~Karakuş,~S.~John~Hogan,
Alin~Achim
        \thanks{This work was supported by the Engineering and Physical Sciences Research Council (EPSRC) under grant EP/R009260/1 (AssenSAR).}
        \thanks{Igor G. Rizaev, and Alin Achim are with the Visual Information Laboratory, University of Bristol, Bristol BS1 5DD, U.K. (e-mail: i.g.rizaev@bristol.ac.uk; alin.achim@bristol.ac.uk)}
        \thanks{Oktay Karakuş is with the School of Computer Science and Informatics, Cardiff University, Cardiff, U.K. (e-mail: karakuso@cardiff.ac.uk)}
        \thanks{S. John Hogan is with the Department of Engineering Mathematics, University of Bristol, Bristol BS1 5DD, U.K. (e-mail: s.j.hogan@bristol.ac.uk)}
}
\markboth{}
{Shell \MakeLowercase{\textit{et al.}}: Bare Demo of IEEEtran.cls for IEEE Journals}

\maketitle

\begin{abstract}
Among other remote sensing technologies, synthetic aperture radar (SAR) has become firmly established in the practice of oceanographic research. Despite solid experience in this field, comprehensive knowledge and interpretation of ocean/sea and vessel wave signatures on radar images are still very challenging. This is not only due to the complex mechanisms involved in the SAR imaging of moving waves: Many technical parameters and scanning conditions vary for different SAR platforms, which also imposes some restrictions on the cross-analysis of their respective images. Numerical simulation of SAR images, on the other hand, allows the analysis of many radar imaging parameters including environmental, ship, or platform related. In this paper, we present a universal simulation framework for SAR imagery of the sea surface, which includes the superposition of sea-ship waves. This paper is the first attempt to cover exhaustively all SAR imaging effects for the sea waves and ship Kelvin wakes scene. The study is based on well proven concepts: the linear theory of sea surface modeling, Michell thin-ship theory for Kelvin wake modeling, and ocean SAR imaging theory. We demonstrate the role of two main factors that affect imaging of both types of waves: (i) SAR parameters and (ii) Hydrodynamic related parameters such as wind state and Froude number. The SAR parameters include frequency (X, C, and L-band), signal polarization (VV, HH), mean incidence angle, image resolution (2.5, 5 and 10 m), variation by scanning platform (airborne or spaceborne) of the range-to-velocity ($R/V$) ratio, and velocity bunching  with associated shifting, smearing and azimuthal cutoff effects. We perform modeling for five wave frequency spectra and four ship models. We also compare spectra in two aspects: with Cox and Munk’s probability density function (PDF), and with a novel proposed evaluation of ship wake detectability. The simulation results agree well with SAR imaging theory and the example of a real SAR image. The study gives a fuller understanding of radar imaging mechanisms for sea waves and ship wakes.
\end{abstract}

\IEEEpeerreviewmaketitle

\section{Introduction}\label{sec:introduction}
\IEEEPARstart{R}{emote} sensing for oceanography and oceanic engineering has seen many studies focusing on Synthetic Aperture Radar (SAR) imaging technologies in recent decades. SAR images provide useful information on ocean state, including wind speed and direction, gravity waves, swells, currents, sea-ice structures, and meteorological and environmental conditions in bodies of water. SAR imaging is used extensively to observe vessels and their wave signatures, in order to monitor port traffic and so on. Although SAR has been used in ocean research for over fifty years, the understanding and application of ocean radar scenes are still being actively developed, and there remain unresolved issues in the interpretation of SAR images. The physical principle of SAR imaging of moving waves is complicated, involving multiple factors, which are difficult to disambiguate. This problem can generally be approached from two perspectives. The first approach involves consideration of the SAR configuration geometry such as the direction of scanning (left or right-looking antenna), choice of platform (airborne or satellite-based), in conjunction with the range-to-velocity ($R/V$) ratio and scanning parameters including image resolution, incidence angle, and electromagnetic properties (i.e. frequency and polarization of the signal). The second approach involves a combination of linear and nonlinear radar imaging mechanisms such as real aperture radar (RAR), including tilt and hydrodynamic modulations (predominately in scanning or range direction), and specific SAR imaging process. The latter mainly relate to the azimuth direction and arise from variation of the radial velocities of the sea surface scatterers. These are commonly referred to in the literature as facets.  This mechanism is usually approximated using time-dependent or velocity bunching models \cite{zurk1996comparison, bruning1990monte, bruning1994estimation}, although other imaging models \cite{kasilingam1990models} have also been proposed in the past. The contribution of the hydrodynamic component of moving waves to the SAR image formation must also be considered, with larger orbital velocities of water particles corresponding to greater image degradation. The existence of ship wakes signatures in SAR images is determined by vessel parameters and ambient sea state, as well as SAR scanning parameters. Higher wind velocity~\cite{pichel2004ship, zilman2014detectability, panico2017sar, tings2019extension} or high sea wave amplitude reduces the visibility of transverse and divergent waves, leaving only the central turbulent wake, and then the external boundary of the wake, which is formed by cusp waves. All the parameters and processes described above are usually closely related and in combination can either distort or enhance SAR images of wakes. This creates difficulties in the analysis and interpretation of real radar images. One of the main limitations of SAR systems, therefore, is their sensitivity to the motion of targets, which results in a distorted phase history of the backscattered signal.

There is now a wide range of spaceborne SAR missions, such as TerraSARX, COSMO-SkyMed, NovaSAR-1, ICEYE-1, ALOS-2 and Sentinel-1, which are capable of producing very high spatial resolutions. However, not all SAR missions can provide the best wave imaging results, the reasons for which are complex and go beyond the single factor of "image resolution". For example, higher satellite orbital altitude increases $R/V$ ratio, as is the case for example with Sentinel-1, and this leads to additional image degradation and azimuthal cut-off of gravity waves with small wavelengths, which includes vessel wakes. Although in recent years there has been an increasing number of calls to make data freely available for research purposes, it remains the case that most sources of real SAR data are proprietary, and access to them is, therefore, limited. This incentivizes the simulation of SAR images. The use of simulators also advances the state-of-the-art by developing new methods for processing real SAR images, and simulation is useful for the planning of new SAR missions. The central factor in forming radar images of the sea is an approximation of the backscattering power from the water surface. The methodological basis for simulating reflections of a radar signal from the sea surface is derived from the fundamental theoretical framework of electromagnetic (EM) wave scattering from rough surfaces, including: Geometrical Optics (GO) \cite{kodis1966note, barrick1968rough}, Phase Perturbation Model (PPM) \cite{broschat1993phase, chen1995comparison}, Integral Equation Model (IEM) \cite{fung1992backscattering}, Full Wave Model (FWM) \cite{bahar1991scattering}, Facet Approach (FA) \cite{clarizia2011simulation}, Small Slope Approximation (SSA), for first order SSA-1 \cite{voronovich1994small} and second order SSA-2 \cite{voronovich2001theoretical}, and many others \cite{elfouhaily2004critical}.
It is first necessary to note that the radar scattering mechanisms can be divided into Bragg scattering and non-Bragg scattering, associated with specular (or quasi-specular) reflection, and breaking waves \cite{kudryavtsev2003semiempirical}. In the literature, for both scattering mechanisms a good amount of theoretical work has already been conducted. For the specular reflection region, with incidence angles about $0^{\circ}\sim20^{\circ}$, the Kirchoff Approximation (KA) \cite{brekhovskikh1952wave}, also known as the Physical Optics (PO) method, has been developed. The contribution of breaking waves in a normalized radar cross-section (NRCS) has been investigated in numerous works \cite{phillips1988radar, kudryavtsev2003semiempirical, hwang2015surface, li2017facet, zhang2019volume}. At moderate incidence angles ($20^{\circ}\sim70^{\circ}$ for VV, and $20^{\circ}\sim60^{\circ}$ for HH polarization), the dominant mechanism is Bragg scattering. Here, the Small Perturbation Model (SPM) \cite{rice1951reflection} was developed, and then later the Two-Scale Model (TSM) \cite{wright1968new, valenzuela1978theories, hasselmann1985theory}, which is based on composite surface Bragg theory \cite{bass1968very, valenzuela1968scattering, wright1968new}. The TSM is probably the most prominent simulation method today. A three-scale model \cite{romeiser1994three} has also been used, which took into account the impact of intermediate-scale waves.

In the TSM the wind wave spectrum is divided into two regions with the purpose of separating the large-scale (long waves) and small-scale (capillary waves) roughness elements. This partitions the wavefield into deterministic and statistical computations. Unlike the SSA, there is no need for calculation of highly detailed sea surface models \cite{kay2011light}. Although parallel computation solutions have been proposed that significantly accelerate the creation of electrically large sea surface models \cite{jiang2015spectral, linghu2018gpu, linghu2018parallel, linghu2019gpu}, these solutions require additional hardware devices, such as a graphics processing unit (GPU), combined with specialist platforms like the Compute Unified Device Architecture (CUDA). The TSM method avoids these requirements and has the advantage of simplicity in its realization, and a considerably lower computational cost in both processing time and computer memory. For example, in one study \cite{hwang2015surface} it was shown that the TSM is much more computationally efficient than SSA-2, in a ratio of approximately 1:800.

As the TSM has gained prominence many different improvements and additions have been introduced: simulation of multiview SAR wave synchronization data for azimuth cutoff wavelength compensation \cite{wan2019research}, polarimetric TSM \cite{di2019closed}, sea surface velocity of wind retrieval \cite{ye2020optimized}, improved TSM \cite{li2019improved, wang2018improved}, facet-based modification of TSM \cite{zhang2011facet}, application of modified TSM to breaking waves simulation \cite{li2017facet}, and finally interesting approaches which, although not directly related to TSM, use the same scale-splitting principle to generate sea surface for infrared imaging \cite{he2018polarimetric} and for improving SSA calculations \cite{jiang2015spectral, wei2018improvement, li2019effective}.

The various radar imaging mechanisms and scattering methods mentioned above form the theoretical basis for SAR imagery simulation of complex ambient sea wave and ship wake scenes. The first group of simulation studies focused on the sea wave field \cite{alpers1983monte, lyzenga1986numerical, harger1988comparisons, vachon1989simulation, bruning1990monte, shemer1991simulation, bruning1994estimation,  zurk1996comparison, bao1997simulation, franceschetti1998ocean, schulz2001ocean, nunziata2008educational, yoshida2013sar, liu2016sar, santos2021simulator}, while the second included both sea and/or ship waves \cite{tunaley1991simulation,shemer1996simulation,oumansour1996multifrequency, hennings1999radar,wang2004simulation, arnold2007bistatic,sun2011electromagnetic, chen2012facet, sun2013scattering, zilman2014detectability, li2015study, zhao2015bistatic, jiang2016ship, fujimura2016coupled,wang2016application, wang2017sar}.

Many authors have compared simulated with actual image spectra \cite{alpers1983monte, lyzenga1986numerical, harger1988comparisons, bruning1994estimation, zurk1996comparison}. In \cite{vachon1989simulation} the simulations are relevant only to spaceborne SAR imagery. In \cite{bruning1990monte} attention was paid to the features of SAR image formation through a combination of nonlinear and linear mechanisms. Simulations of interferometric SAR (InSAR) were proposed in other works \cite{shemer1991simulation, bao1997simulation, schulz2001ocean}. Nunziata et al. \cite{nunziata2008educational} designed a simulator for educational purposes, which, although simplified, still contains all the SAR image formation mechanisms. Time-domain SAR simulations of moving ocean waves were presented in \cite{yoshida2013sar}. Simulators to produce SAR raw data, rather than actual image simulation, have also been developed, see \cite{franceschetti1998ocean, liu2016sar}. Recently, a simulator of SAR image spectra of ocean swell waves was presented \cite{santos2021simulator}. One of the earliest works on sea and ship wake radar image simulation was developed in \cite{tunaley1991simulation}. In \cite{shemer1996simulation} along-track interferometric SAR image simulation of ship wakes was proposed. The simulation of the NRCS of Kelvin arms was compared with real SAR images for different spaceborne platforms, in \cite{hennings1999radar}. The bistatic configuration of the SAR image formation models was also simulated in \cite{arnold2007bistatic, chen2012facet, zhao2015bistatic}. Some studies focused in particular on ship wakes modeling. This was the case with \cite{li2015study} and, \cite{wang2017sar} where EM scattering simulations from the ship-generated internal wave wake were presented and for turbulent wake SAR imaging in \cite{tunaley1991simulation,wang2004simulation,sun2011electromagnetic}. Linear and nonlinear sea surface models were considered and compared in \cite{sun2013scattering}. The polarized Doppler spectra of dynamic ship wakes was investigated in \cite{jiang2016ship}. In another work \cite{fujimura2016coupled} a model of far wakes was used. More general SAR simulators of ship wakes were designed in \cite{oumansour1996multifrequency, zilman2014detectability, wang2016application}. In \cite{oumansour1996multifrequency} two frequencies (X and L-band) are considered, while in \cite{wang2016application} the visualization of ship wake features in SAR images was evaluated. Zilman et al. \cite{zilman2014detectability}, investigated the detection of Kelvin wakes in simulated SAR images, and assessed the influence of significant wave height on their detectability. Finally, other original modeling studies of radar scattering from the ocean surface and ship wake include \cite{romeiser1997improved, plant2002stochastic, yang2016polarimetric, qiao2018sea, sun2018ship, zhang2019numerical}.

Although all the works mentioned above, as a whole, form an important theoretical basis, individually they are somewhat incomplete, each concerning particular aspects of RAR/SAR imaging, for example: (i) focusing only on spectra comparison of simulated and real images, or (ii) simulating only one type of vessel wave wake. The various works were also based on particular SAR configurations, for example monostatic, bistatic, or interferometric. Further, it is important to note that most of these simulation methods have been proposed for fixed modeling parameters only, for example, a single frequency, single scanning platform type, or single sea wave spectrum. Also, not all studies include the velocity bunching effect, which is crucial in forming SAR images of waves in motion. Further, some important modeling issues such as radar scene size or azimuth cut-off effects are not always addressed. The reproducibility of many of the published methods, and their integration, is also very laborious. 
All of these factors provide a strong motivation for the design of a universal, relatively simple simulation system, which can deal with the majority of the SAR imaging phenomena that arise when imaging moving waves.

In this paper, we present an extended study for modeling and simulation of SAR imagery corresponding to both ambient sea waves and superimposed ship wakes. The simulation method is based on the linear theory and stochastic concept of sea surface modeling \cite{holthuijsen2010waves, ochi2005ocean}, while the Kelvin wake model adopted here is based on Michell thin-ship theory, as previously considered in \cite{arnold2007bistatic, zilman2014detectability}.

The fundamental contribution is to integrate and present, for the first time, the majority of the known phenomena associated with SAR imaging of Kelvin ship wakes. These include the effects of wind state and Froude number, and also the effects of the SAR imaging characteristics of signal frequency, incidence angle, polarization, the spatial resolution of image cell, velocity bunching factors (image shifting and smearing), and heading of moving ship relative to the radar line-of-sight. The numerical simulation results obtained are in good agreement with the theory of SAR image formation for moving waves.

The second original contribution is in the comparison of different sea surface spectra by visibility of ship wake in the resulting SAR images. The idea here is that an understanding of the 'borderline condition' when wakes are maximally sensitive for imaging will make it possible to estimate the contribution of various spectra to the SAR image. This provides a greater understanding of how the SAR imaging of ship wakes will vary for different geographical areas. The simplest examples are coastal areas (fetch limited) or open waters (fully developed seas). To the best of our knowledge, an analysis in this respect has not been performed before. There are limited studies where various spectra are compared, but then only in terms of ensuing absolute NRCS values \cite{zheng2018sea, ryabkova2019review, xie2019effects, zhang2020modeling}. By contrast we employ five well-known sea wave spectra including Pierson-Moskowitz \cite{pierson1964proposed}, JONSWAP \cite{hasselmann1973measurements}, Fung and Lee \cite{fung1982semi}, Elfouhaily et al. \cite{elfouhaily1997unified} and Romeiser et al. \cite{romeiser1997improved}, and compare the contribution of these spectra to ship wake visualization in SAR images.  Additionally, the comparison of the utilized spectra in our work was also performed with Cox and Munk's probability density function (PDF) \cite{cox1954measurement, cox1956slopes}.

The structure of this paper is as follows: Section \ref{sec:surfaceModelling}, starts with the generic presentation of the modeling of the sea surface waves, the governing equations are presented and special attention is paid to the main modeling parameters. In Section \ref{sec:wakeModelling}, the ship wake modeling approach is presented and common SAR vessel signatures are discussed. The SAR image formation within the TSM framework is formulated in Section \ref{sec:seaSAR}. Simulation experiments with discussions are included in Section \ref{sec:results}. Conclusions and potential directions for future research are described in Section \ref{sec:conc}.

\section{Sea waves modeling}\label{sec:surfaceModelling}

\subsection{The random deep-sea wave elevation model}
The linear theory of sea surface gravity waves has been used for describing ocean waves for more than 150 years \cite{holthuijsen2010waves}. In this theory, water is assumed to be inviscid and incompressible while its motion is irrotational. The basis for describing three-dimensional moving sea waves relies on random phase modeling with summation of many independent harmonic waves, whereby they propagate within $x$-$y$ and $t$ spaces in direction $\theta$, with Rayleigh distributed amplitude $A$, and a uniformly distributed random phase $\epsilon \in (0, 2\pi)$. 

Within the linear theory of surface waves, the sea wave elevation model is related to the fluid velocity potential $\Phi_{sea}$ at free surface through
\begin{align}\label{equ:1}
    \left.{{Z}_{sea}}=-\frac{1}{g}\frac{\partial {{\Phi }_{sea}}}{\partial t}\right\vert_{z=0}
\end{align}
where
\begin{dmath}\label{equ:2}
    {{\Phi }_{sea}}(x,y,z,t)=g\sum\limits_{i}{\sum\limits_{j}{\frac{{{A}_{ij}}}{{{\omega }_{i}}}{{e}^{{{k}_{i}}z}}\sin \left[ {{k}_{i}}\left( x\cos {{\theta }_{j}}+y\sin {{\theta }_{j}} \right)-{{\omega }_{i}}t+{{\epsilon }_{ij}} \right]}}
\end{dmath}

Then, the irregular sea surface elevation model can be described using the double superimposition model as

\begin{dmath}\label{equ:3}
{{Z}_{sea}}(x,y,z,t)=\sum\limits_{i}{\sum\limits_{j}{{{A}_{ij}}\cos \left[ {{k}_{i}}\left( x\cos {{\theta }_{j}}+y\sin {{\theta }_{j}} \right)-{{\omega }_{i}}t+{{\epsilon }_{ij}} \right]}}
\end{dmath}
where $k$ and $\omega$ are the wave wavenumber and wave circular (radian) frequencies, respectively. The general expression of the dispersion in deep water includes capillary and gravity-capillary waves as

\begin{align}\label{equ:4}
    {{\omega }^{2}}=gk\left( 1+{{{k}^{2}}}/{{{k}_{m}}^{2}}\; \right)
\end{align}
with

\begin{align}\label{equ:5}
    {{k}_{m}}^{2}={g\rho }/{\tau }\;
\end{align}
where $\tau$ is the surface tension of water (N/m), $\rho$ is the sea water density (kg/m$^3$), and $g$ is the gravitational acceleration. For the gravity range (short-gravity and gravity waves), the expression in (\ref{equ:4}) is simplified as

\begin{align}\label{equ:6}
    \omega =\sqrt{gk}
\end{align}

The amplitude $A$ is expressed as:

\begin{align}\label{equ:7}
    {{A}_{ij}}=\sqrt{2S\left( {{k}_{i}} \right)D\left( {{k}_{i}},{{\theta }_{j}} \right)\Delta {{k}_{i}}\Delta {{\theta }_{j}}}
\end{align}
where $S(k)$ is the wave spectrum as a function of angular spatial frequency (it can also be expressed as a function of frequency $S(f)$ or circular frequency $S(\omega)$), $D(k, \theta)$ is the angular spreading function, while $\Delta k$, and $\Delta\theta$ are the sampling intervals of the wavenumbers and propagation angles.

In simulations, two main parameters should be well-thought-out: (i) the minimum size of the model and (ii) the size of the scatterers (or facets) of the sea model. The minimum size of the sea surface model affects the wavelength and the accuracy of modeled large gravity waves, specifically the amplitude of large waves for the appropriate wind velocity $V_w$, while the facet size affects the proportion of large and small-scale waves. Some studies \cite{arnold2007bistatic, wang2015doppler, mobley2015polarized, mobley2016modeling} have focused on determining the minimum size of the model for correctly transferring the energy of the spectrum to the sea surface model. In \cite{wang2015doppler} it was argued that the size should be larger than the dominant wavelength of the sea waves to reflect their modulation effects. In \cite{arnold2007bistatic}, using Fung and Lee’s spectrum, it has been determined that the minimum size, $L_{min}$, of the scene representing the sea surface can be approximated as $L_{min} = 3.28\times V_{w19.5}^2$, where $V_{w19.5}$ is the wind velocity at 19.5m above the mean sea level. In another study \cite{mobley2016modeling}, $L_{min}$ was determined via the wavelength ($\Lambda_p$) of the spectral peak (changes depending on $V_w$) of the Pierson-Moskowitz spectrum as $L_{min} = 2\Lambda_p$. Another example based on Elfouhaily et al. spectrum \cite{elfouhaily1997unified}, demonstrates the losing effect of the energy spectrum (decreasing model heights) when the sea model size is reduced to 100 m (cf. Fig. 3 in \cite{mobley2015polarized}). In our calculations (Section \ref{sec:results}), we select $L_{min}$ as 1000 m, which satisfies the correct transferring of energy spectrum to the sea surface model in a range of $V_w = 3.5 - 11$ m/s.

The facet size determines the discretization level of the sea surface model or the high frequency component. In the simulator presented in \cite{nunziata2008educational}, the facet size is half of the SAR resolution cell. In some studies \cite{zhang2011facet}, \cite{chen2012facet}, it was experimentally determined that the average NRCS does not crucially depend on the facet size in ranges from 0.5 to 2 m \cite{zhang2011facet} and from 0.5 to 1.5 m \cite{chen2012facet}. In \cite{linghu2018gpu} the facet size was used in the range of 0.1-1 m corresponding to different SAR wave bands. However, choosing the facet size depends primarily on the modeling of the EM scattering. Thus, the SSA method requires sizes of orders less than one-eighth/one-tenth of the incident wavelength \cite{jiang2015spectral, wei2018improvement, li2019effective}. Within the two-scale model (TSM), the facet size must be large enough when compared to the wavelength of Bragg waves, but sufficiently small compared to the wavelength of long waves \cite{rufenach1981imaging}. Hence, in \cite{plant1994dependence}, the facet size was determined as 10 times the microwave wavelength at X and Ka bands. In \cite{alpers1986relative}, facet sizes of the order of 10 to 20 Bragg wavelengths were considered. However, for the TSM, there are two versions that determine division of the wavefield via separation scale (facet size) into deterministic and statistical calculations \cite{hasselmann1985theory}. For the electromagnetic-hydrodynamic (EMH) two-scale model, the separation wavenumber is approximated as $k_{EMH} \approx  k_B / 5$, where $k_B$ is Bragg wavenumber. For the SAR two-scale model, the separation wavenumber is determined as $k_{SAR} \approx 2\pi / p_{SAR}$, where $p_{SAR}$ is the size of the SAR resolution cell. In general, for SAR simulations a facet size equal to the SAR resolution cell is usually assumed \cite{rufenach1981imaging, liu2016sar, zilman2014detectability}, since it implies a lower memory cost. On the other hand, taking into account the Nyquist criterion, the choice of SAR resolution should depend on the resolvable sea wave or ship wake size. Thus, in \cite{graziano2017performance} the effect of the SAR resolution is demonstrated by considering the Kelvin wake system. Therefore, in this work we consider a facet size equal to the SAR image resolution ($k_{SAR}$ separation scale), which can be varied for different SAR configurations. In simulations, we set the facet size at 2.5 m, which is in the order of typical real SAR imagery and matches the details of both types of gravity waves: wind- and ship-induced.

\subsection{Omnidirectional sea wave spectrum}\label{sec:Spectra}
The one-dimensional, omnidirectional wave spectrum, represented in (\ref{equ:7}) is interpreted as the empirical interrelation between energy distribution and frequency. Many wind-wave models based on wind speed \cite{ryabkova2019review} have been developed and involve various additional parameters that control the energy of the spectrum. 

\begin{figure}[htbp]
\centering
\subfigure[]{\includegraphics[width=.49\linewidth]{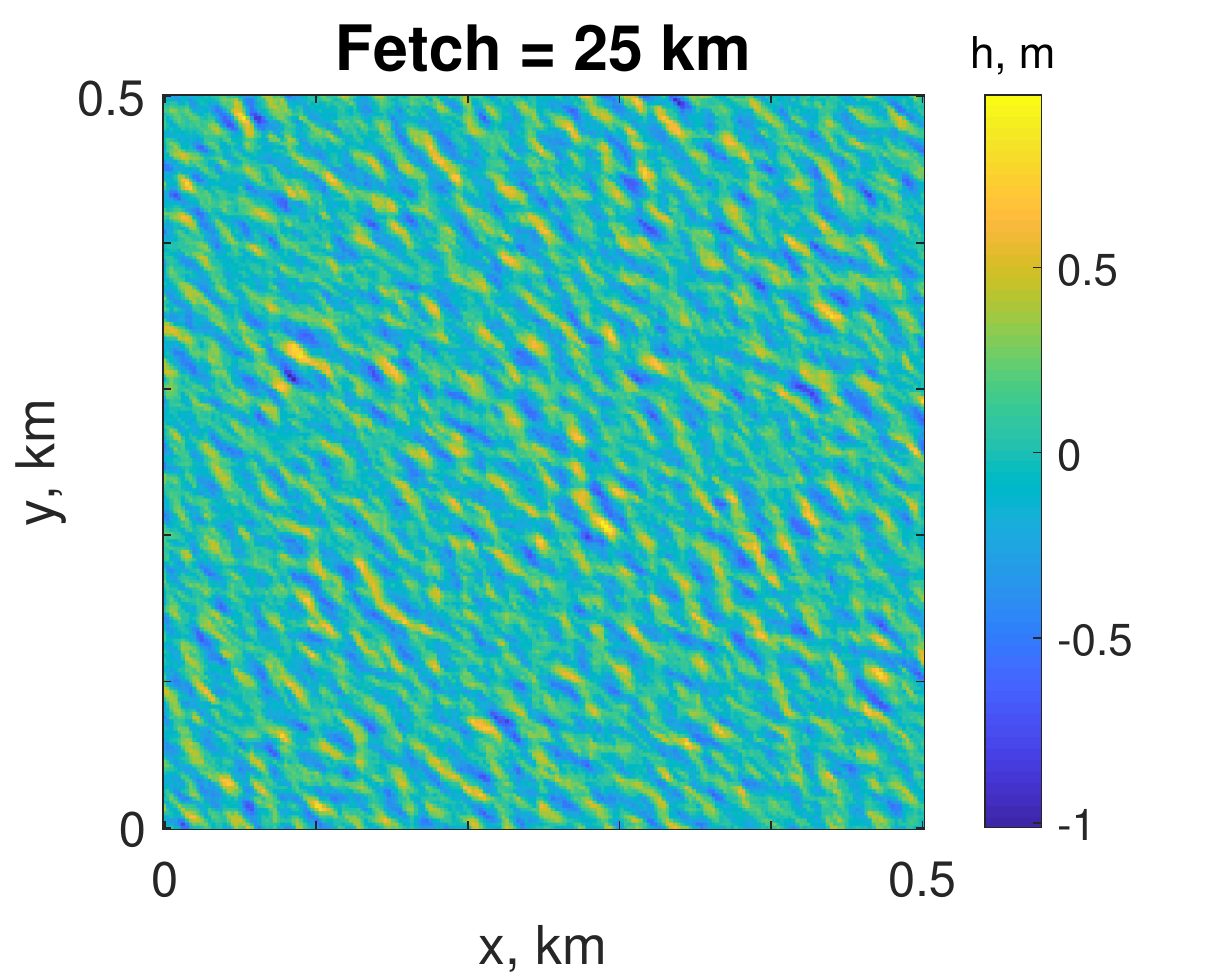}}
\subfigure[]{\includegraphics[width=.49\linewidth]{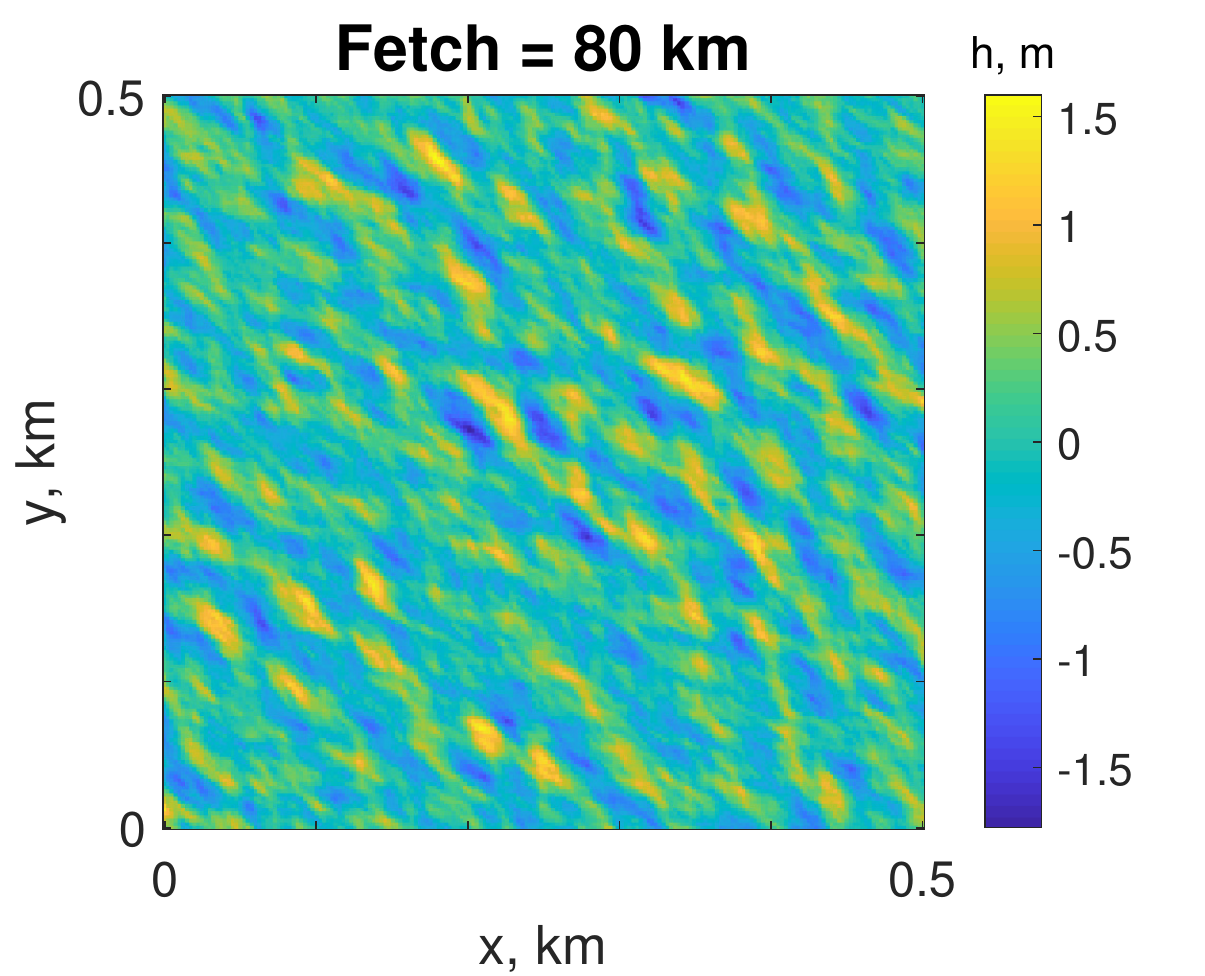}}
\caption{Effect of different fetch length (sea state) in the sea surface modeling for JONSWAP spectrum. The size of the model is 0.5$\times$0.5 km, facet size is 2.5 m, ${{V}_{w10}} = 10$ m/s, wind direction is $45^{\circ}$. (a) fetch = 25 km. (b) fetch = 80 km.}
\label{fig:fig1}
\end{figure}

Different sea states can be reproduced by these models, with the two main categories being (i) fully developed sea (waves in equilibrium with the wind) and (ii) developing or young sea (modeled using fetch and/or inverse wave age parameter). It should be noted that, for the same wind speed, the wave height or amplitude will differ for fully developed and developing sea conditions. This effect is demonstrated in Fig. \ref{fig:fig1}.  

\begin{figure}[htbp]
\centering
\subfigure[]{\includegraphics[width=.49\linewidth]{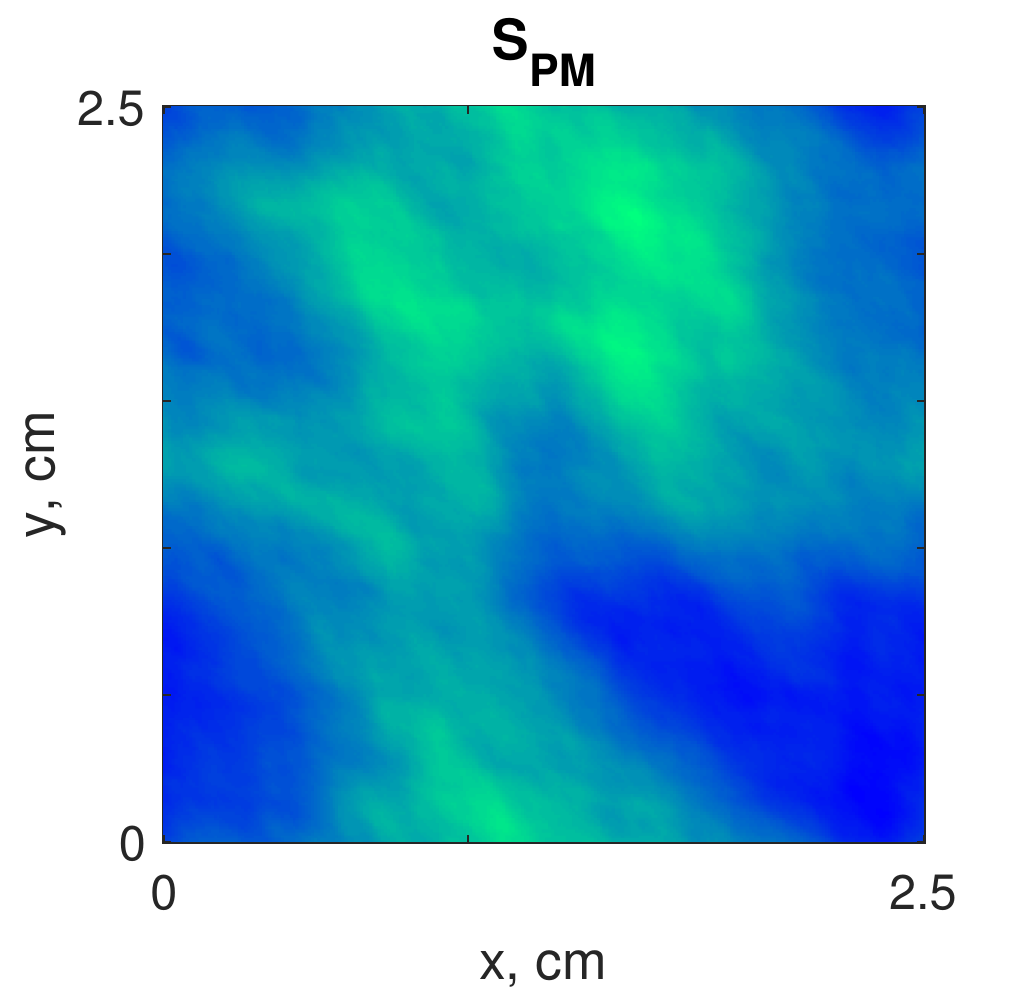}}
\subfigure[]{\includegraphics[width=.49\linewidth]{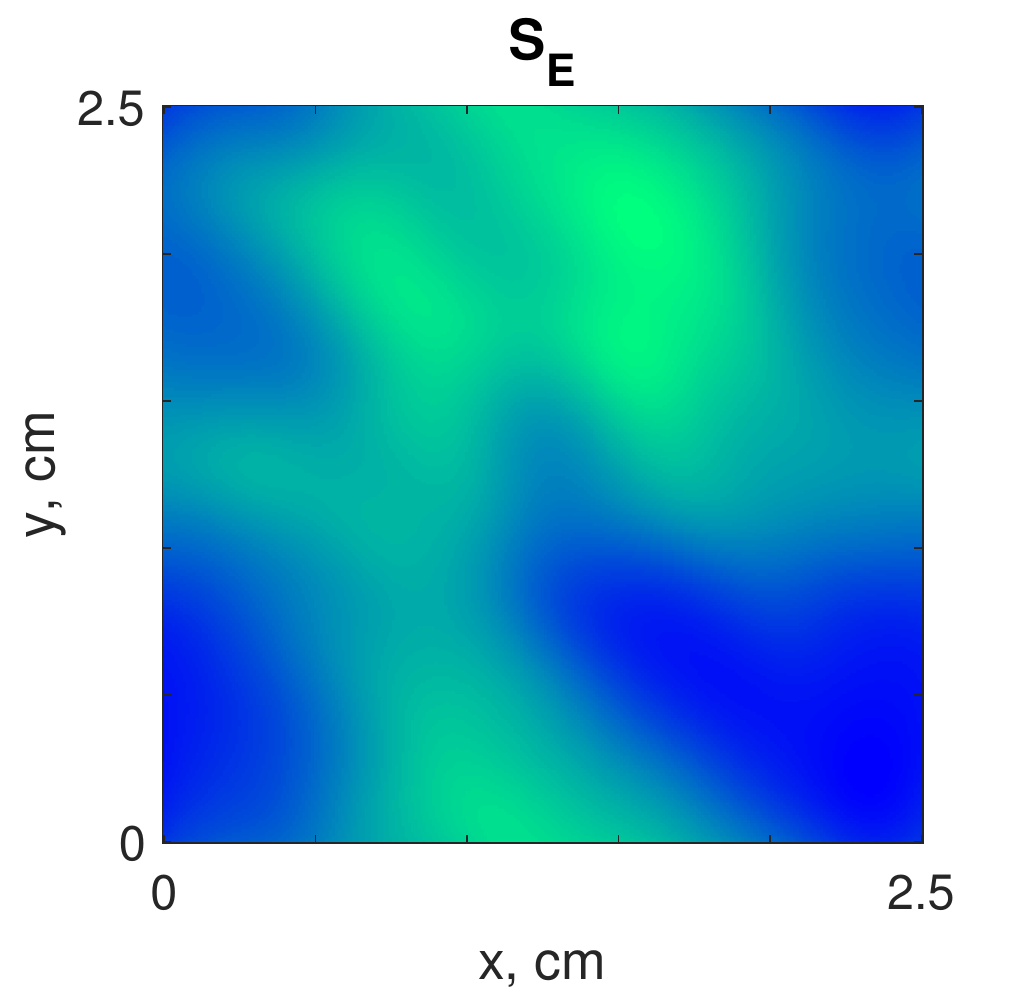}}
\caption{Effect of scale in the sea surface modeling. The size of the model is 2.5$\times$2.5 cm, facet size is 0.1 mm, ${{V}_{w10}} = 8$ m/s. (a) gravity Pierson-Moskowitz spectrum. (b) gravity-capillary Elfouhaily et al. spectrum.}
\label{fig:fig2}
\end{figure}

\begin{figure*}[htbp]
\centering
\subfigure[]{\includegraphics[width=.19\linewidth]{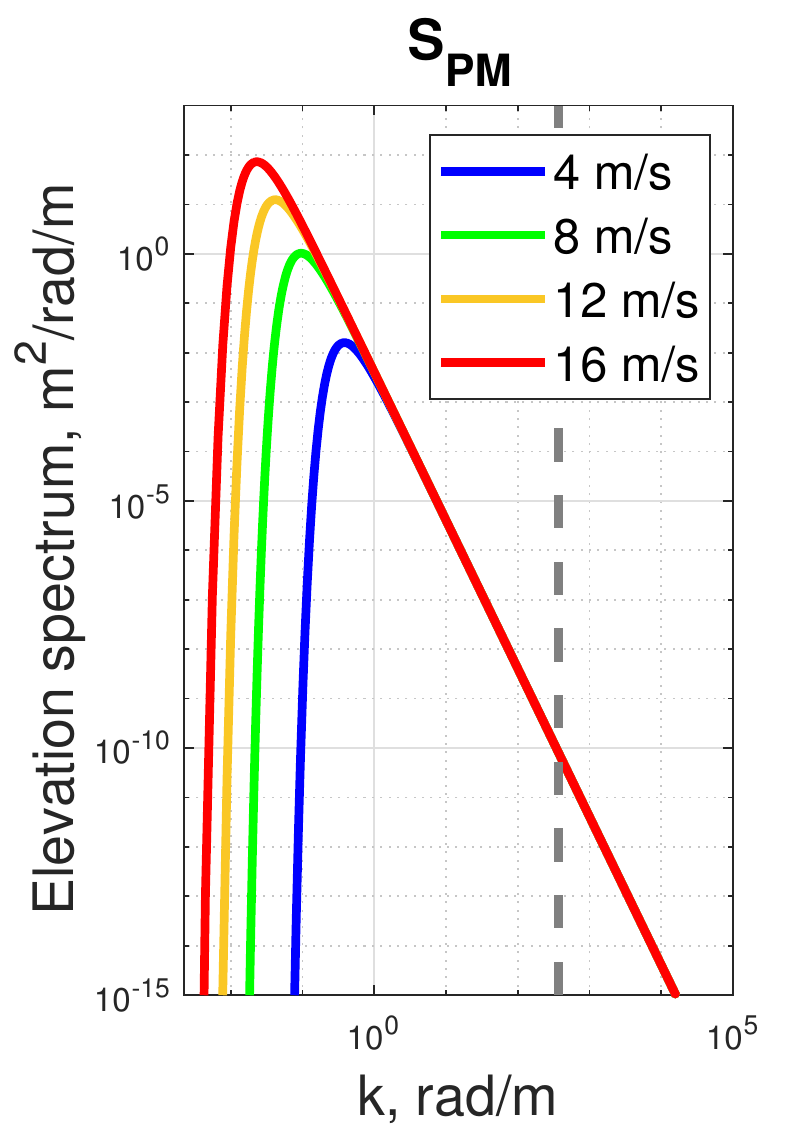}}
\subfigure[]{\includegraphics[width=.19\linewidth]{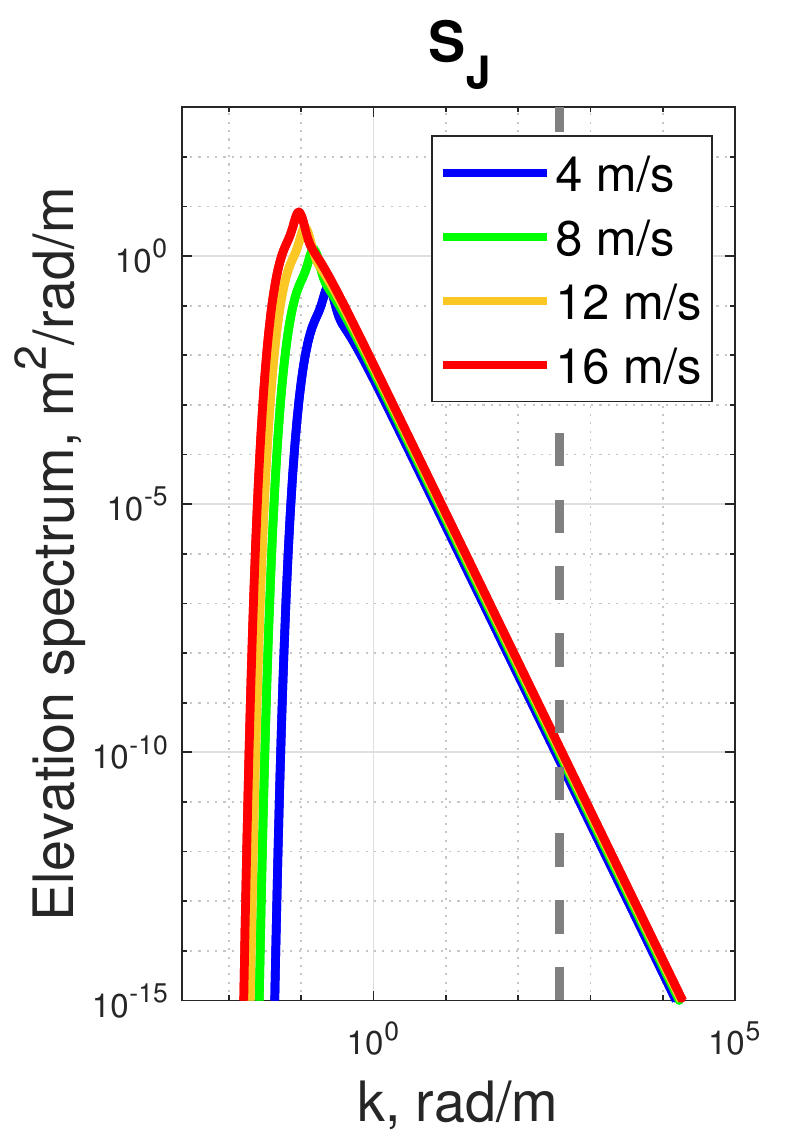}}
\subfigure[]{\includegraphics[width=.19\linewidth]{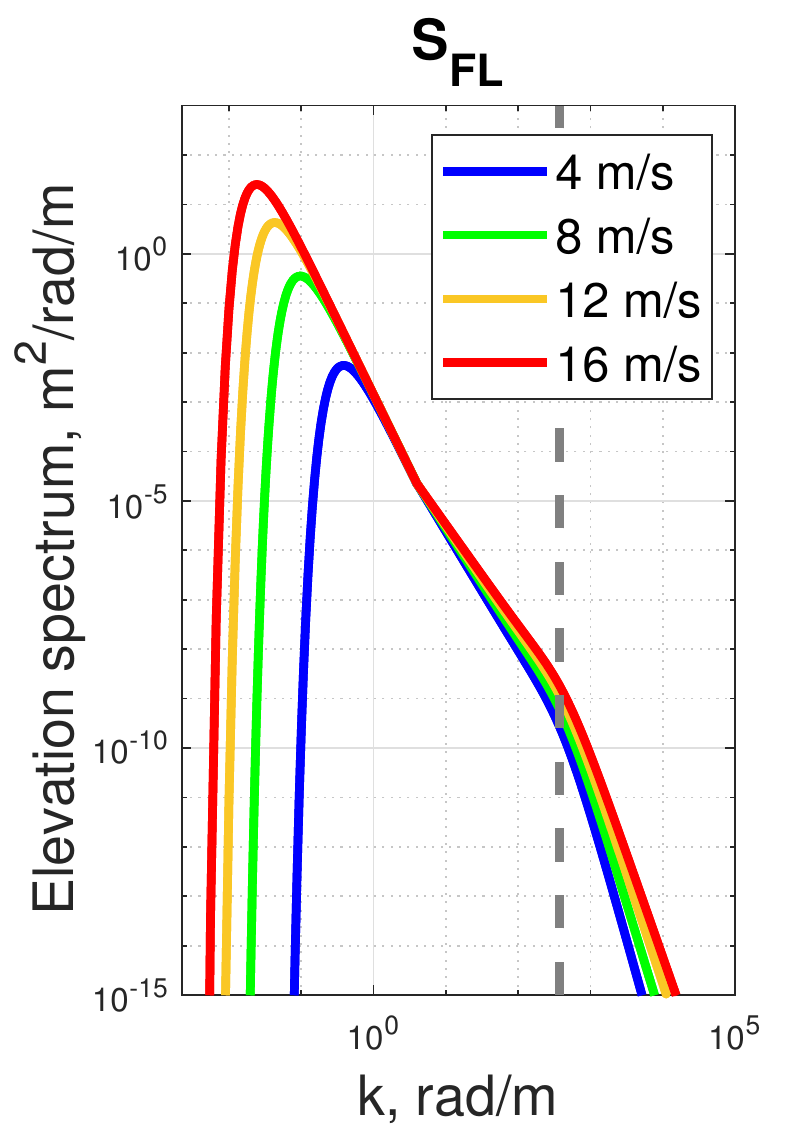}}
\subfigure[]{\includegraphics[width=.19\linewidth]{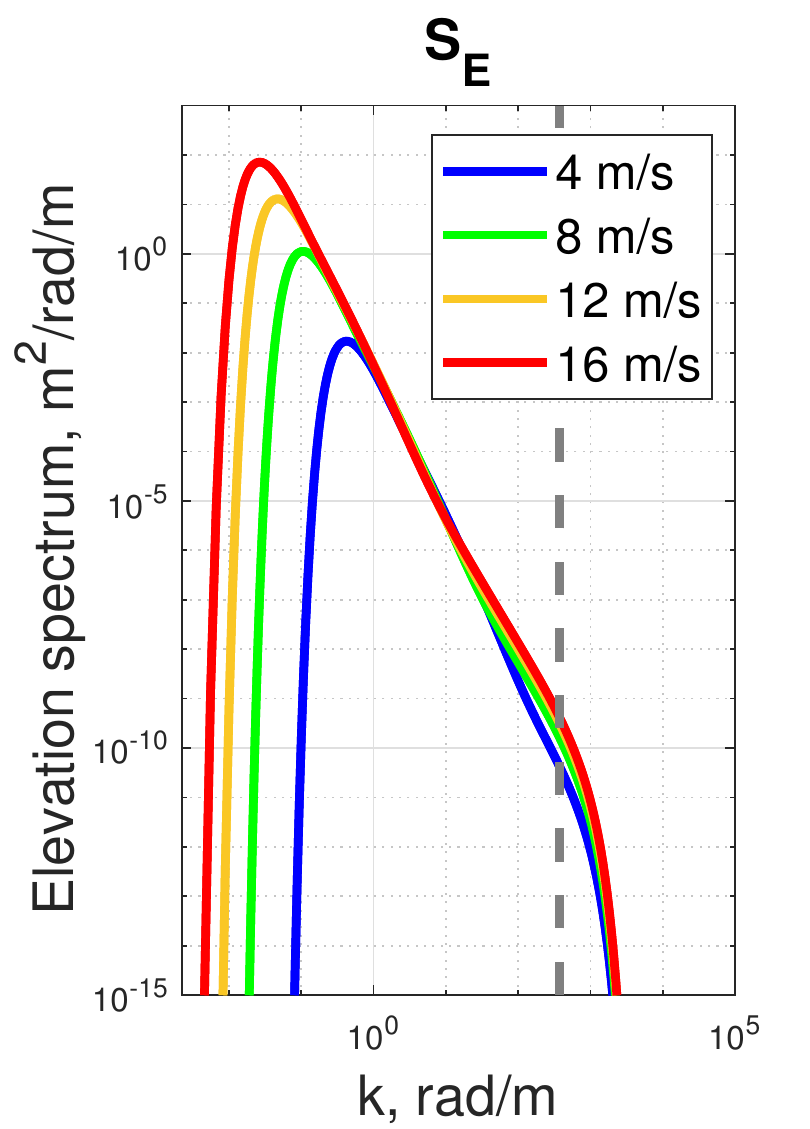}}
\subfigure[]{\includegraphics[width=.19\linewidth]{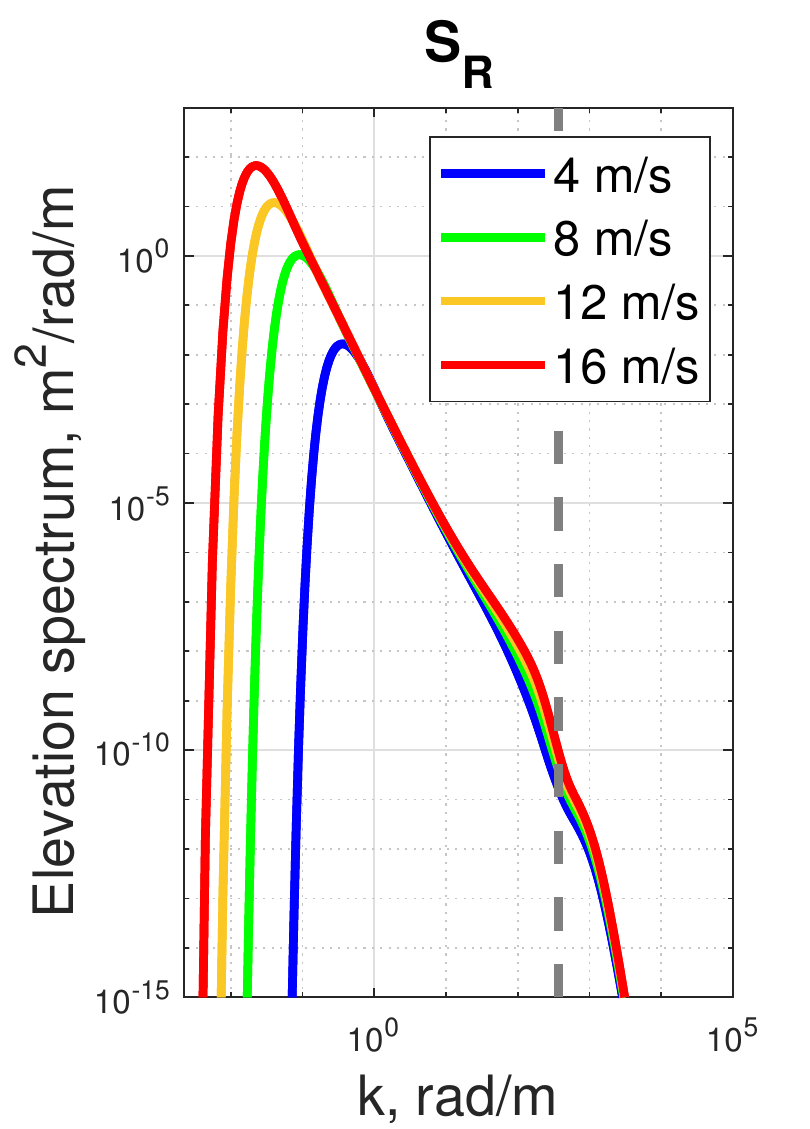}}
\caption{Omnidirectional elevation spectra for different wind speeds (${{V}_{w10}}$) corresponding to different models: (a) $S_{PM}$, (b) $S_{J}$, (c) $S_{FL}$, (d) $S_{E}$, and (e) $S_{R}$. For $S_{J}$ the fetch is set to 80 km, while for ${S_E}$ the inverse wave age $\Omega$ is 0.84. The vertical-dashed line ($k = 370$ rad/m) represents the boundary between the gravity and capillary waves.}
\label{fig:fig3}
\end{figure*}

Another general characteristic is the existence of a range of wavelengths covered by the spectrum (e.g. from gravity to short-gravity or gravity to capillary). A particular spectrum describes only gravity waves while another can go up to capillary scale. Here for the first time we depict this effect in models. Fig. \ref{fig:fig2} displays small scale models where the size for both models was chosen as $2.5 \times 2.5$ cm (gravity-capillary scale \cite{zilman2014detectability}) and the facet size is set to 0.1 mm (capillary scale). At the scale of small ripples the surface tension is the dominant restoring force which aims to stretch the surface of the water into flatness. Even with a significant change in the scale of the sea model, its patterns (but not the heights) follow that of gravitational waves, cf. Fig. \ref{fig:fig2}-(a). That is to say, in the sense of structure, the model is scale-invariant, which confirms that the Pierson-Moskowitz spectrum describes only the gravity part. This is further verified in Fig. \ref{fig:fig3}. In contrast, the Elfouhaily et al. spectrum in Fig. \ref{fig:fig2}-(b) shows smoother capillary patterns which differ from the gravity scale.

In the remainder of this section, we describe the most well-known spectra: Pierson-Moskowitz \cite{pierson1964proposed}, JONSWAP \cite{hasselmann1985theory}, Fung-Lee \cite{fung1982semi}, Elfouhaily et al. \cite{elfouhaily1997unified} and Romeiser et al. \cite{romeiser1997improved}. For ease of reference, henceforth we refer to them as $S_{PM}$, $S_J$, $S_{FL}$, $S_E$, and $S_R$, respectively. Let us note that we only emphasize key parameters and general mathematical formulations for all spectra, and we refer the readers to the original publications for further details. It must be emphasized that in the literature, different spectra are often expressed in terms of different frequency variables.

Conversion to wavenumber spatial frequency is achieved through the following expression:

\begin{align}\label{equ:9}
    S\left( k \right)=S\left( f \right)\frac{d f}{d k}
\end{align}
where $f$ and $k$ are interconnected by dispersion relation (\ref{equ:6}).

The first and second spectra considered include only the gravity component, while all others are implemented using both the gravity and the capillary regime of waves. The $S_{PM}$ spectrum is a well-known approximation for the gravity waves and, for a fully developed sea state, is formulated in terms of $k$ as \cite{pierson1964proposed}

\begin{align}\label{equ:10}
{{S}_{PM}}\left( k \right)=\frac{\alpha }{2{{k}^{3}}}\exp \left[ -\beta {{\left( \frac{g}{k} \right)}^{2}}\frac{1}{{V_{w19.5}}^{4}} \right]
\end{align}
where $\alpha = 0.0081$ is a Phillips constant, $\beta = 0.74$, $g$ is the gravitational acceleration, and $V_{w19.5}$ is the wind speed at 19.5 m above the mean sea surface.

The general form of the spectrum for developing sea state was described
in the Joint North Sea Wave Project (JONSWAP) as \cite{kanevsky2008radar}

\begin{dmath}\label{equ:11}
    {{S}_{J}}\left( k \right)=\frac{\alpha }{2}{{k}^{-3}}\exp \left[ -1.25{{\left( \frac{k}{{{k}_{p}}} \right)}^{-2}} \right]\exp \left\{ \ln \gamma \exp \left[ -\frac{{{\left( \sqrt{{k}/{{{k}_{p}}}\;}-1 \right)}^{2}}}{2{{\sigma }^{2}}} \right] \right\}
\end{dmath}
where $\alpha = 0.076(V_{w10}^2/Fg)^{0.22}$, and the parameter $\sigma$, which describes the width of the spectrum is either $\sigma = 0.07$ if $k \leq k_p$ or $\sigma = 0.09$ if $k > k_p$, where $k_p$ is the peak wavenumber \cite{hasselmann1973measurements, massel2017ocean}

\begin{align}\label{equ:12}
    {{k}_{p}}={{\left[ 7\pi \frac{g}{{{V}_{w10}}\sqrt{g}}{{\left( \frac{{{V}_{w10}}^{2}}{gF} \right)}^{0.33}} \right]}^{2}}
\end{align}

Here and above, $V_{w10}$ is the wind speed at 10 m above the mean sea surface and $F$ is the fetch length in meters.
The peak enhancement factor $\gamma$ is usually set to
3.3, although sometimes other values are employed by certain authors \cite{mitsuyasu1980observation}.

Another widely known spectrum was developed by Fung and Lee \cite{fung1982semi} specifically for the estimation of radar backscatter over L to Ku frequency bands. Their spectrum includes two parts: gravity and capillary wave ranges, which join at the separation point $k_j = 0.04$ rad/cm: $S_{FL} = S_{FLg}$ if $k < k_j$ and $S_{FLc}$ if $k > k_j$. The gravity spectrum $S_{FLg}$ is the same as in (\ref{equ:10}), except for $\alpha = 2.8 \times 10^{-3}$, while the capillary spectrum is given by

\begin{dmath}\label{equ:13}
    {{S}_{FLc}}\left( k \right)=0.875{{\left( 2\pi  \right)}^{p-1}}\left( 1+3{{{k}^{2}}}/{{{k}_{m}}^{2}}\; \right){{g}^{{\left( 1-p \right)}/{2}\;}}{{\left[ k\left( 1+{{{k}^{2}}}/{{{k}_{m}}^{2}}\; \right) \right]}^{-{\left( p+1 \right)}/{2}\;}}
\end{dmath}
with $k_m$ the same as in (\ref{equ:5}) and $p = 5 - log_{10}(V_{w*})$. The calculation of $V_{w*}$ is provided later within the paper, in (\ref{equ:23}), (\ref{equ:24}).

Probably the most well known spectrum is Elfouhaily et al. \cite{elfouhaily1997unified}, which, like the $S_{FL}$ spectrum, consists of both wave ranges. It has however an additional advantage in that, similar to the $S_{J}$ spectrum, it can also simulate different sea states. The general expression of the omnidirectional $S_{E}$ spectrum is

\begin{align}\label{equ:14}
    {{S}_{_{E}}}\left( k \right)={{k}^{-3}}\left( {{B}_{l}}+{{B}_{h}} \right)
\end{align}
where the gravity or long-wave part of the spectrum is

\begin{align}\label{equ:15}
    {{B}_{l}}=\frac{1}{2}{{\alpha }_{p}}\frac{{{c}_{p}}}{c}{{L}_{PM}}{{J}_{p}}\exp \left[ -\frac{\Omega }{\sqrt{10}}\left( \sqrt{\frac{k}{{{k}_{p}}}-1} \right) \right]
\end{align}
whereas the capillary or short-wave part of spectrum is given by

\begin{align}\label{equ:16}
    {{B}_{h}}=\frac{1}{2}{{\alpha }_{m}}\frac{{{c}_{m}}}{c}{{L}_{PM}}{{J}_{p}}\exp \left[ -\frac{1}{4}{{\left( \frac{k}{{{k}_{m}}}-1 \right)}^{2}} \right]
\end{align}

In the short-wave part, we included the shape spectrum factor $L_{PM}$ and the peak enhancement factor $J_P$ as in \cite{mobley2016modeling}. All parameters of Elfouhaily's spectrum can be found in their original article \cite{elfouhaily1997unified}.

Unlike in $S_{PM}$ and $S_{J}$, Romeiser et al. spectrum has been developed for the improved composite surface model for the radar backscattering  \cite{romeiser1997improved}, which is based on the Apel spectrum \cite{apel1994improved}. The main omnidirectional formulation is

\begin{align}\label{equ:17}
    {{S}_{R}}\left( k \right)={{k}^{-3}}{{P}_{L}}{{W}_{H}}{{\left( \frac{{{V}_{w10}}}{{{V}_{n}}} \right)}^{\beta }}
\end{align}
where $P_L$ is computed as

\begin{dmath}\label{equ:18}
    {{P}_{L}}=0.00195\exp \left[ -\frac{{{k}_{p}}^{2}}{{{k}^{2}}}+0.53\exp \left( -\frac{{{\left( \sqrt{k}-\sqrt{{{k}_{p}}} \right)}^{2}}}{0.32{{k}_{p}}} \right) \right].
\end{dmath}
The peak wavenumber, $k_p$, is

\begin{align}\label{equ:19}
    {{k}_{p}}=\frac{1}{\sqrt{2}}\frac{g}{{{V}_{w10}}^{2}}
\end{align}
The wind speed exponent $\beta$ is given by

\begin{dmath}\label{equ:20}
    \beta =\left[ 1-\exp \left( -\frac{{{k}^{2}}}{{{k}_{1}}^{2}} \right) \right]\exp \left( -\frac{k}{{{k}_{2}}} \right)+\left[ 1-\exp \left( -\frac{k}{{{k}_{3}}} \right) \right]\exp \left[ -{{\left( \frac{k-{{k}_{4}}}{{{k}_{5}}} \right)}^{2}} \right]
\end{dmath}
with the values of the constants being $k_1 = 183$ rad/m, $k_2= 3333$ rad/m, $k_3=33$ rad/m, $k_4 = 140$ rad/m, and $k_5 = 220$ rad/m. $W_H$ is the shape of Bragg wave region of the spectrum and can be expressed as

\begin{dmath}\label{equ:21}
    {{W}_{H}}=\frac{{{\left[ 1+{{\left( {k}/{{{k}_{6}}}\; \right)}^{7.2}} \right]}^{0.5}}}{\left[ 1+{{\left( {k}/{{{k}_{7}}}\; \right)}^{2.2}} \right]{{\left[ 1+{{\left( {k}/{{{k}_{8}}}\; \right)}^{3.2}} \right]}^{2}}}\exp \left( -\frac{{{k}^{2}}}{{{k}_{9}}^{2}} \right)
\end{dmath}
where $k_6 = 280$ rad/m, $k_7 = 75$ rad/m, $k_8 = 1300$ rad/m, and $k_9 = 8885$ rad/m.

Since the $S_{PM}$ and $S_{J}$ spectra do not include high-frequency parts, the short wave Phillips spectrum is employed for both spectra \cite{guinard1971variation}

\begin{align}\label{equ:22}
{{W}_{P}}\left( k \right)=\beta {{k}^{-4}}
\end{align}
where $\beta$ = $6 \times 10^{-3}$.

In order to ensure the consistency of wind speed at different altitudes in different spectrum models and angular spreading functions ($V_{w10}$, $V_{w12.5}$, and $V_{w19.5}$), we apply the Fung and Lee method \cite{fung1982semi}. The same method was used for calculation of the wind friction velocity $V_{w*}$ in (\ref{equ:13})

\begin{align}\label{equ:23}
    {{V}_{w}}=\left( {{{V}_{w*}}}/{0.4}\; \right)\ln \left( {z}/{{{Z}_{0}}}\; \right)
\end{align}
and

\begin{align}\label{equ:24}
    {{Z}_{0}}=\left( {0.684}/{{{V}_{w*}}}\; \right)+4.28\times {{10}^{-5}}{{V}_{w*}}^{2}-0.0443
\end{align}
where $V_{w}$ is the wind speed at altitude $z$ above the mean sea surface in, cm.

The sea surface elevation model is usually characterized in terms of significant wave height $H_s$ (the average of 1/3 of the highest presented waves) \cite{ochi2005ocean}, which relates to $Z_{sea}$ or the sea wave spectrum as

\begin{align}\label{equ:25}
    {{H}_{s}}=4\sqrt{\int{S\left( k \right)dk}}=4\sqrt{\operatorname{var}\left( {{Z}_{sea}} \right)}
\end{align}

In Fig. \ref{fig:fig3}, the elevation spectra for different wind speeds are shown. We further discuss in Section \ref{sec:results} the effect of ambient sea waves on the SAR image simulation.

\subsection{Directional spreading function}\label{sec:Directionalfunction}
The underlying idea for the representation of wave directionality is that gravity waves are aligned with the mean wind direction. However, shorter waves are more divergent and can propagate perpendicularly or even move against the wind direction and, then gravity-capillary waves which once again are aligned with wind direction
\cite{elfouhaily1997unified, huang19906}. In \cite{mitsuyasu1975observations}, by using buoy measurements it was concluded that the direction of waves changes when the size of the waves decreases. A detailed review on this has recently been published, see \cite{du2017improved}.

The frequency spectrum $S(k)$ alone is not sufficient to adequately describe the propagation of sea waves in a two-dimensional (2D) space. In order to represent the state of superimposed directional components of the wave energy transferring along the wind direction, the angular spreading function $D(k, \theta)$ (sometimes called angular/directional distribution function) also needs to be used.

Although many spreading functions have been proposed (see for example \cite{huang19906}), a satisfactory model of the directional energy distribution still needs to be developed. Nonetheless, all types of the spreading functions presented in (\ref{equ:7}) have to meet the condition of

\begin{align}\label{equ:26}
    \int\limits_{-\pi }^{\pi }{D\left( k,\theta  \right)}d\theta =1
\end{align}
where the sea waves direction $\theta = \theta - \theta_{w}$ is corrected by the mean wind direction $\theta_{w}$, which is taken relative to the flight direction of the SAR platform.
One of the first practical and simple forms of the directional spreading function is the cosine-squared spreading function with directional width around $30^{\circ}$ \cite{oumansour1996multifrequency}. For $\left| \theta  \right|> {{90}^{\circ}}$, $D(\theta)$ = 0, and for  $\left| \theta  \right|\leq {{90}^{\circ}}$ the function is

\begin{align}\label{equ:27}
    D\left( \theta  \right)=\frac{2}{\pi }{{\cos }^{2}}\theta 
\end{align}

\begin{figure*}[htbp]
\centering
\subfigure[]{\includegraphics[width=.19\linewidth]{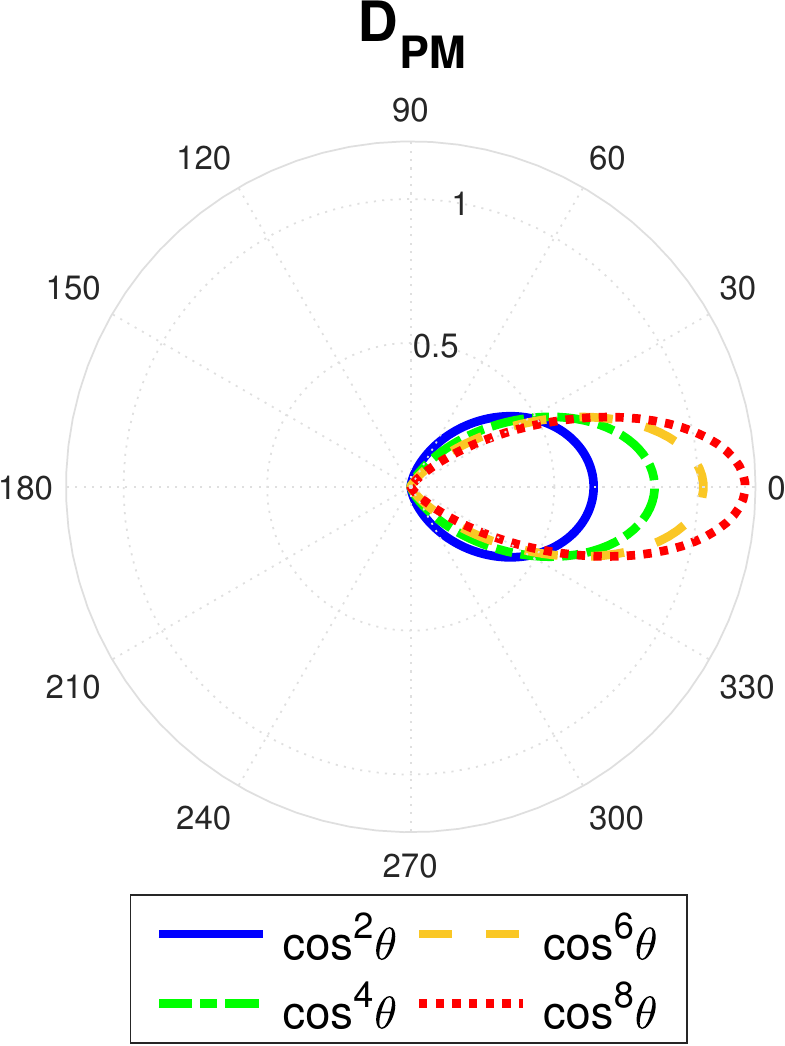}}
\subfigure[]{\includegraphics[width=.19\linewidth]{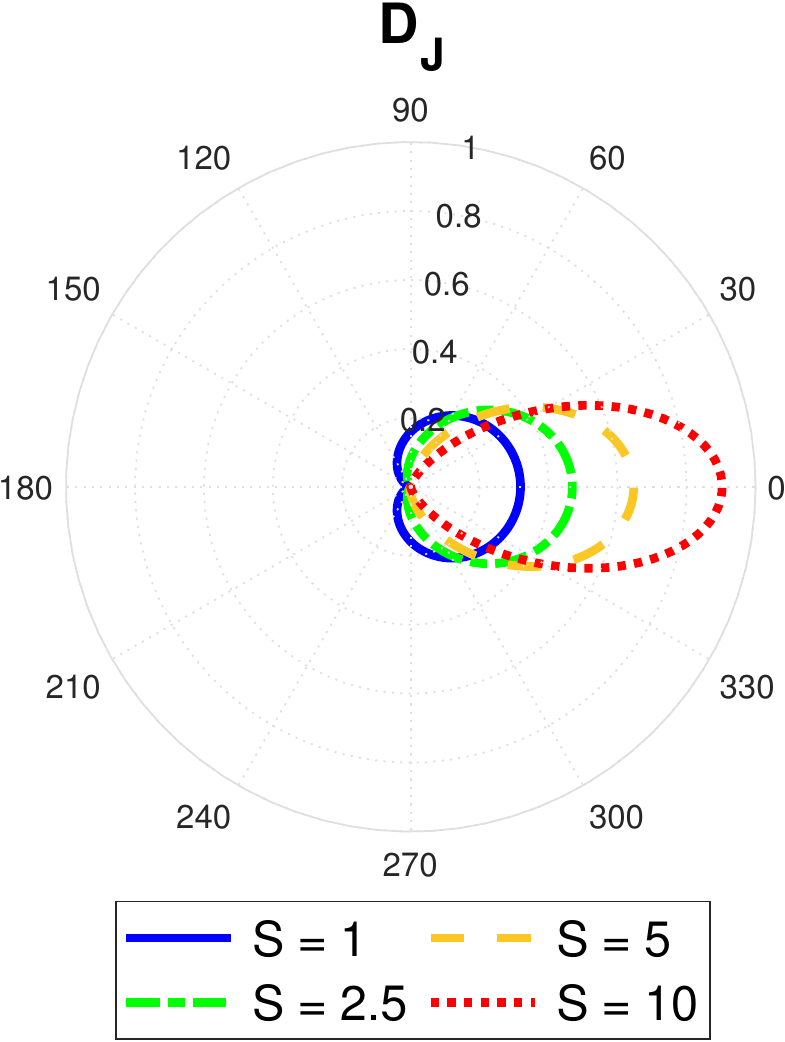}}
\subfigure[]{\includegraphics[width=.19\linewidth]{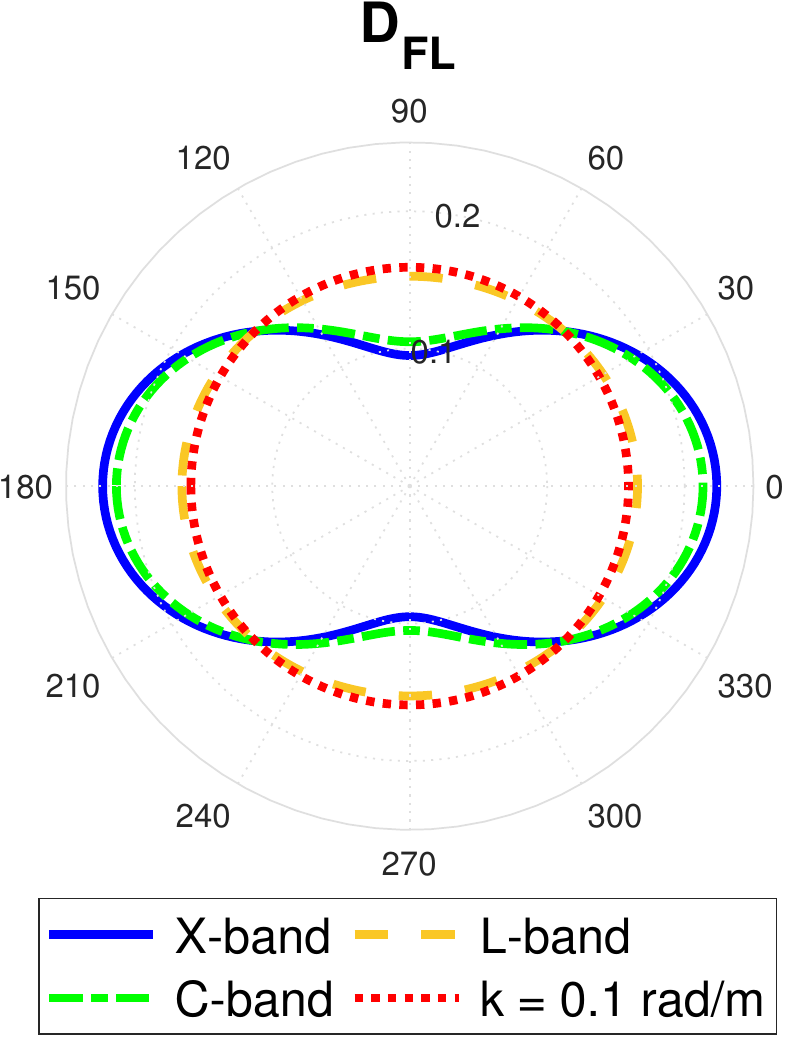}}
\subfigure[]{\includegraphics[width=.19\linewidth]{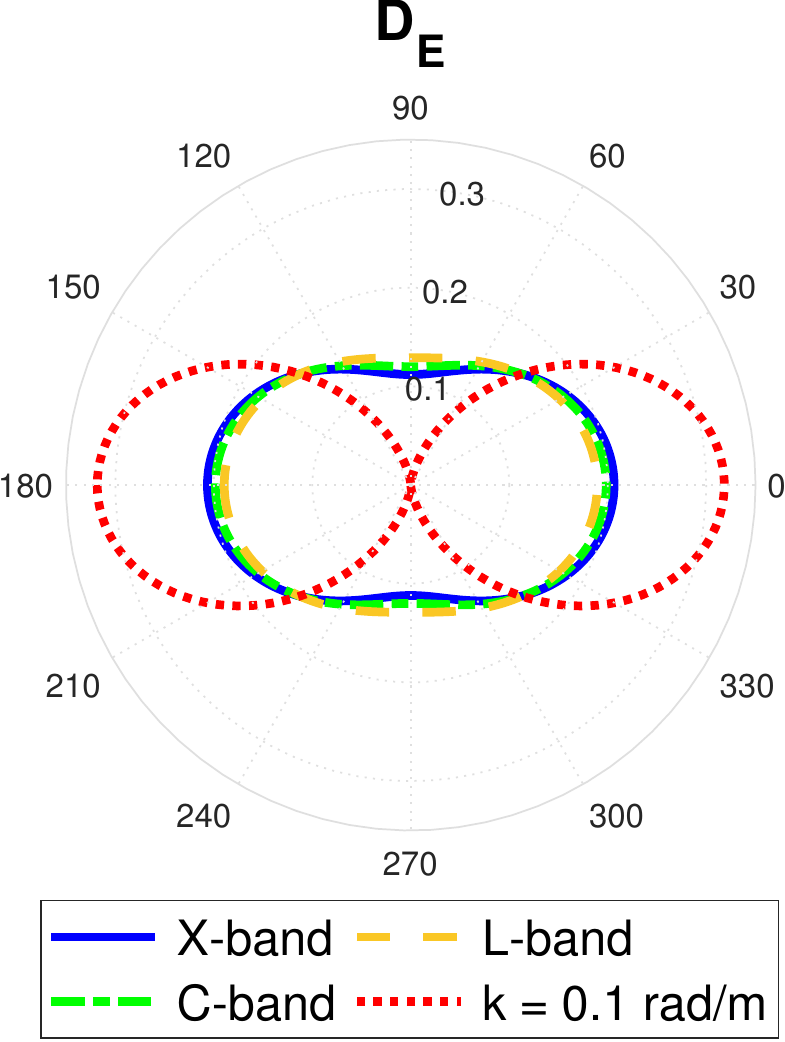}}
\subfigure[]{\includegraphics[width=.19\linewidth]{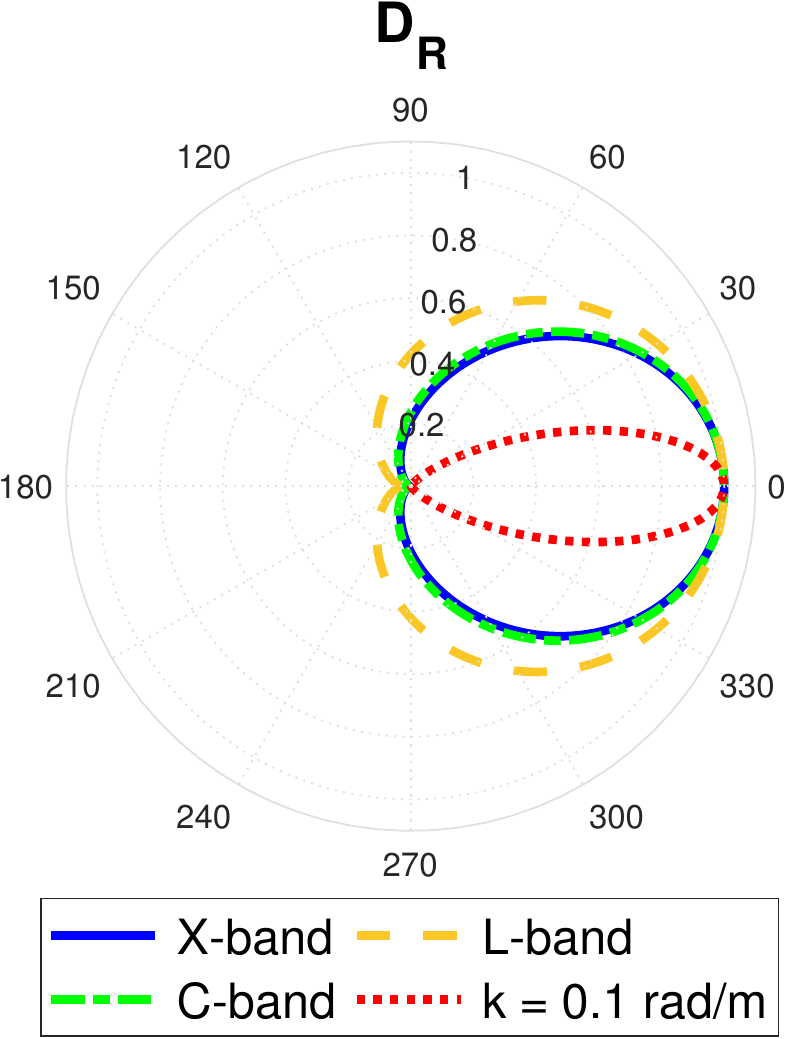}}
\caption{The angular spreading functions presented in polar coordinates at different wavenumbers: gravity-capillary waves (X-band), short gravity waves (C and L-bands), gravity waves (0.1 rad/m). For wind depended functions, the wind speed is set to ${{V}_{w10}} = 8$ m/s. 
(a) Cosine-squared. (b) Longuet-Higgins et al. with S parameter. (c) Fung and Lee. (d) Elfouhaily et al. (e) Romeiser et al.}
\label{fig:fig4}
\end{figure*}

The main limitation of this function is that it does not explicitly depend on frequency and wind speed, which means that all wave components propagate in the same direction. An alternative to the cosine-squared spreading function \cite{nagai1972computation} is presented in Fig. \ref{fig:fig4}-(a). In general, higher-order functions correspond to better directionality for wave propagation.
Probably the best known and the most widely used is the Longuet-Higgins et al. cosine type spreading function (see \cite{holthuijsen2010waves}) based on empirical field measurements of pitch-and-roll buoys

\begin{align}\label{equ:28}
    D\left( \theta  \right)=\frac{\Gamma \left( S+1 \right)}{\left[ \Gamma \left( S+0.5 \right)2\sqrt{\pi } \right]}{{\cos }^{2S}}\left( \frac{\theta }{2} \right)
\end{align}
where $\Gamma(\cdot)$ is the gamma function, and the parameter $S$ controls the width of the function and depends on $k$. This parameter was later refined in some studies \cite{mitsuyasu1975observations, hasselmann1980directional}. For simplicity, we keep $S$ as a constant value, following \cite{holthuijsen2010waves}. When $S$ is increased, the directionality of waves also increases, which is shown in Fig. \ref{fig:fig4}-(b). For example, in our simulations, we employed $S = 20$ in Fig. \ref{fig:fig11}, (a), (b) and $S = 8$ in Fig. \ref{fig:fig12}, (b) and Fig. \ref{fig:fig14}, (e), (f).

Some specific spreading functions have also been proposed together with their omnidirectional spectra $S_{FL}$, $S_{E}$, and $S_{R}$. 
Fung and Lee spreading function in Fig. \ref{fig:fig4}-(c) can be expressed as

\begin{align}\label{equ:29}
    D\left( k ,\theta \right)={{\left( 2\pi  \right)}^{-1}}+{{a}_{1}}\left( 1-{{e}^{-b{{k}^{2}}}} \right)\cos \left( 2\theta  \right)
\end{align}
where

\begin{align}\label{equ:30}
    {{a}_{1}}=\frac{{\left( 1-R \right)}/{\left( 1+R \right)}\;}{\pi \left( 1-B \right)}
\end{align}
with

\begin{align}\label{equ:31}
    B=\left( {1}/{{{\sigma }_{t}}^{2}}\; \right)\int\limits_{0}^{\infty }{{{k}^{2}}{{S}_{FL}}\left( k \right){{e}^{-b{{k}^{2}}}}dk}
\end{align}
and

\begin{align}\label{equ:32}
    {{\sigma }_{t}}^{2}=\int\limits_{0}^{\infty }{{{k}^{2}}{{S}_{FL}}\left( k \right)dk}
\end{align}
The parameter $b = 1.5$ cm$^2$ and $R$ is based on Cox and Munk’s ratio of slope variances as follows \cite{cox1954measurement}

\begin{align}\label{equ:33}
    R=\frac{0.003+1.92\times {{10}^{-3}}{{V}_{w12.5}}}{3.16\times {{10}^{-3}}{{V}_{w12.5}}}
\end{align} 
where $V_{w12.5}$ (m/s) is the wind velocity at 12.5 m above the mean sea surface.

Elfouhaily et al. spreading function in Fig. \ref{fig:fig4}-(d) takes the form:

\begin{align}\label{equ:34}
    D\left( k,\theta  \right)=\frac{1}{2\pi }\left[ 1+\Delta \left( k \right)\cos \left( 2\theta  \right) \right]
\end{align}
where the value of $\Delta(k)$ is determined as

\begin{align}\label{equ:35}
    \Delta \left( k \right)=\tanh \left[ {{a}_{0}}+{{a}_{p}}{{\left( {c}/{{{c}_{p}}}\; \right)}^{2.5}}+{{a}_{m}}{{\left( {{{c}_{m}}}/{c}\; \right)}^{2.5}} \right]
\end{align}
For the way to set the remaining parameters, we refer the reader to the original paper \cite{elfouhaily1997unified}.

The $D_{FL}$ and $D_{E}$ functions are symmetric about $\pi / 2$ (Fig. \ref{fig:fig4},-(c) and (d)), which is more realistic for electromagnetic modeling \cite{elfouhaily1997unified}. However, the $D_{FL}$ function is not realistic for the simulation of long waves \cite{elfouhaily1997unified}, which is apparent from the isotropic shape of the function at $k = 0.1$ rad/m (Fig. \ref{fig:fig4}-(c)).

Finally, Romeiser et al. \cite{romeiser1997improved} spreading function is based on a Gaussian kernel as follows

\begin{align}\label{equ:36}
    D\left( k,\theta  \right)=\exp \left( -\frac{{{\theta }^{2}}}{2{{\delta }^{2}}} \right)
\end{align} 
with			 

\begin{dmath}\label{equ:37}
    \frac{1}{2{{\delta }^{2}}}=0.14+0.5\left[ 1-\exp \left( -\frac{k{{V}_{w10}}}{{{c}_{1}}} \right) \right]+5\exp \left[ 2.5-2.6\ln \left( \frac{{{V}_{w10}}}{{{V}_{n}}} \right)-1.3\ln \left( \frac{k}{{{k}_{n}}} \right) \right]
\end{dmath} 
where $c_1 = 400$ rad/s, and $k_n = 1$ rad/m. This function has maximal directionality at wavenumber 0.1 rad/m as opposed to other $k$-depended functions (cf. Fig. \ref{fig:fig4}-(e)).

The effect of wind speed on spreading functions here has been omitted. It should be noted that when wind speed is increased, the differences between gravity-capillary and short gravity waves are minimized. An example of this effect can be found in \cite{zhou2017directional}, p. 10.

All the spreading functions presented in this section can potentially be applied in conjunction with different omnidirectional spectra in order to offer exhaustive modalities of simulating SAR images. This could also broaden their applicability (e.g. some spectra may represent better real conditions in some geographical areas compared to others), for mitigating the distortions introduced during the SAR image simulation process for both sea waves and ship wakes. However, in our study, unless specified otherwise in a particular experiment, for the $S_{FL}$, $S_{E}$ and $S_{R}$ spectra we kept their original $D(k, \theta)$ functions, whereas $S_{PM}$ was combined with the cosine-squared function $D_{PM}$ while $S_{J}$ with Longuet-Higgins cosine  type  spreading  function $D_{J}$.

\section{Ship wake modeling}\label{sec:wakeModelling}
A moving vessel in the sea generates different types of wave patterns called – wakes, which are usually classified into Kelvin wakes, turbulent wake, internal-wave wakes and narrow-V wakes \cite{lyden1988synthetic, pichel2004ship}. It is not always possible to simultaneously identify all these wake types in a SAR image. This is due to variation in the SAR parameters, for example the orientation of the antenna, ship parameters, and background sea condition or wind speed.

It is well known that the most distinguishable feature of ship signatures in SAR images is the turbulent wake \cite{graziano2017performance}. Although some simulation approaches have been performed \cite{wang2004simulation, arnold2007bistatic, sun2011electromagnetic, liu2020sar, ren2021sar}, there is still no complete understanding of this phenomenon in SAR images. This is due to turbulent wakes' complicated nature and complex geometry\cite{ren2021sar}. Approaches to the modeling of ship-generated internal wakes have also been formulated in some papers \cite{wang2017sar, li2017facet}, but these wakes have only been successfully visualized in SAR images where there is a shallow thermocline in the sea \cite{lyden1988synthetic}. In real SAR images, observing the presence of the narrow-V wave wakes (bright lines also known as ship-generated Bragg waves \cite{lyden1988synthetic}) is challenging and there is no clear mechanism explaining the reason for their appearance. It is usually suggested that the reason is associated with the incidence angle and radar frequency. For example, depending on the radar frequency, these wakes have been considered as either narrow-V wakes (L-band) or as components of a turbulent system (X-band) \cite{graziano2017performance}. A further complexity in the modeling of narrow-V wakes results from the fact that they are only visible on SAR images at low wind speed ($V_{w}<3$ m/s) \cite{soomere2007, lyden1988synthetic}. However, most sea spectra models are theoretically valid only when the wind friction velocity $V_{w*} > 12$ cm/s ($ V_{w10} > 3.3$ m/s) \cite{fung1982semi, arnold2007bistatic1}. Therefore, the possibilities for the simulation of SAR images with narrow-V wakes may be limited.

The Kelvin wake is a ship wake structure which is (along with the turbulent wake) most often visible in satellite SAR images \cite{pichel2004ship}.
The classical Kelvin wave system of a ship includes transverse and divergent waves, and the cusp waves which are formed by interference at the wake edges (see further in \cite{arnold2007bistatic1, pichel2004ship}) and which produce cusp lines \cite{zilman2014detectability} on the SAR images. These are also known as Kelvin envelope lines \cite{lyden1988synthetic} and Kelvin arms after their discoverer Sir W. Thomson (Lord Kelvin) \cite{thomson1887ship}, and form opening angles of about $\pm$ 19.50 degrees relative to the ship track. However, a recent study \cite{rabaud2013ship} based on the analysis of optical images has improved our understanding of Kelvin wakes by showing that at large Froude numbers (Section \ref{subsec:FroudeNumber}) the opening angles differ.

Therefore, we recommend using a well known Kelvin wake modeling methodology that has been employed in a wide range of SAR simulation works \cite{jiang2016ship, zilman2014detectability, oumansour1996multifrequency}. The limitation of this methodology is that it assumes that ships only move linearly, while in reality they also move along curves.

Assuming that water is inviscid and incompressible, and with the ship’s hull presented as a Wigley parabolic shape, the general asymptotic expression of the Kelvin wake based on the Michell thin ship theory is \cite{zilman2014detectability}

\begin{align}\label{equ:38}
    {{Z}_{ship}}\left( x,y \right)=\operatorname{Re}\int\limits_{{-\pi }/{2}\;}^{{\pi }/{2}\;}{A\left( \theta  \right)}{{e}^{i{{k}_{0}}\left( x\cos \theta +y\sin \theta  \right)}}d\theta 
\end{align} 
where ${{k}_{0}}=\nu {{\sec }^{2}}\theta $, $\nu ={g}/{{{V}_{s}}^{2}}\;$, $V_{s}$ is the velocity of the ship, and $A(\theta)$ is a function of the ship’s shape with parameters: $B$ – beam (m), $L$ – length at the waterline (m), $D_{t}$ – draft (m).

By analogy with a sea surface elevation model (\ref{equ:1}), and replacing ${\partial \Phi }/{\partial t=-{{V}_{s}}{\partial \Phi }/{\partial x}}$, the Kelvin wake elevation surface is related to the fluid velocity potential as:

\begin{align}\label{equ:39}
    {{Z}_{ship}}=\frac{{{V}_{s}}}{g}\frac{\partial {{\Phi }_{ship}}}{\partial x}
\end{align}
Here, Zilman et al.'s \cite{zilman2014detectability} approximated form of fluid velocity potential is employed

\begin{align}\label{equ:40}
    {{\Phi }_{ship}}\left( x,y,z \right)=-16BL{{\pi }^{-1}}F{{r}^{6}}\operatorname{Re}\int\limits_{0}^{\infty }{C\left( \tau ,x,z \right)}{{e}^{iy\tau }}d\tau 
\end{align}
where

\begin{dmath}\label{equ:41}
C\left( \tau ,x,z \right)=\frac{\left( 1-{{e}^{-\nu \alpha {{D}_{t}}}} \right)\sin \left( \beta -\beta \cos \beta  \right)}{{{\alpha }^{{3}/{2}\;}}\sqrt{{1}/{4+{{{\tau }^{2}}}/{{{\nu }^{2}}}\;}\;}}\cos \left( x\nu \sqrt{\alpha } \right){{e}^{z\nu \alpha }}
\end{dmath}
with

\begin{align}\label{equ:42}
\alpha ={\left( 1+\sqrt{1+{4{{\tau }^{2}}}/{{{\nu }^{2}}}\;} \right)}/{2}\; \quad \text{and} \quad \beta ={\sqrt{\alpha }}/{2F{{r}^{2}}}\;
\end{align}
and

\begin{align}\label{equ:43}
\tau =\nu \sqrt{{{\sec }^{2}}\theta -\sec \theta }
\end{align}

The Froude number $Fr$ appearing in (\ref{equ:40}) is introduced later in (\ref{equ:58}). Since the secant function in (\ref{equ:43}) is undefined for angles of $-\pi / 2$ and $\pi / 2$, in implementation, a range for $\theta$ excluding the borders $(-\pi / 2, \pi / 2)$ is used.

Other numerical methods for Kelvin wake simulation have been described in various studies \cite{oumansour1996multifrequency, shemer1996simulation, rabaud2014narrow}.

\section{SAR imaging of the sea surface}\label{sec:seaSAR}
The scattering of the microwave radiation from the disturbed sea surface is a complex process which includes: the physical properties of the surface, scanning platform geometry and roughness conditions (e.g. local slopes), as well as microwave signal properties. As mentioned in the Introduction \ref{sec:introduction}, there are two main scattering mechanisms: Bragg scattering and non-Bragg scattering (from breaking waves and specular reflection) \cite{kudryavtsev2003semiempirical}. SAR works mainly at wavelengths in the centimeters to decimeters range, and covers moderate incidence angles of $20^{\circ}$ $\sim$ $70^{\circ}$ for VV polarization (for HH polarization $20^{\circ}$ $\sim$ $60^{\circ}$), which is related to Bragg scattering. In this study we only consider the Bragg scattering region, which is determined by the sea surface roughness at the scale of the radar signal's wavelength. These short-scale waves are modulated in motion and orientation by long-scale waves (tilt and hydrodynamic modulations), thus allowing the real aperture radar (RAR) to image wind- and ship-driven waves. Also, the rationale for considering only VV and HH polarizations and not cross-polarization is the better and clearer imaging of sea/ship waves in co-polarized SAR images. The two-scale model (TSM) employed here is based on resonant Bragg scattering theory \cite{wright1968new, valenzuela1978theories, hasselmann1985theory} and is a good compromise between the calculation time and accuracy for approximation of scattering. The Bragg scattering solution for the ensemble-averaged normalized radar cross-section (NRCS) with VV and HH polarizations is represented as \cite{zilman2014detectability, romeiser1997improved}

\begin{align}\label{equ:44}
    {{\sigma }_{0}}\left( x,y \right)=8\pi {{k}_{e}}^{4}{{\cos }^{4}}{{\theta }_{l}}W\left( {{k}_{Bx}},{{k}_{By}} \right){{\left| T \right|}^{2}}
\end{align}
where $k_e = 2\pi / \lambda$ is the radar electromagnetic wavenumber and $\lambda$ is the wavelength of the radar signal, $\theta_l = \cos^{-1} [\cos(\theta_r - s_p) \cos(s_n)]$ is the local radar incidence angle, where $\theta_r$ is the nominal radar incidence angle, while $s_n = \tan^{-1} (dZ/dx)$ and $s_p = \tan^{-1} (dZ/dy)$ are local slopes normal and parallel to the radar look direction respectively, $W(\cdot)$ is the 2D wavenumber spectral density of the sea surface roughness determined as the short wave spectrum, with Bragg scattering components $k_{Bx}$, $k_{By}$ \cite{zilman2014detectability}, $T(\cdot)$ is the complex scattering function which controls polarization of the radar signal and depends on the relative dielectric constant $e$ of the sea water. Here, $e = 49-35.5i$ for X-band, $e = 60-36i$ for C-band and $e = 72-59i$ for L-band according to \cite{donelan1987radar}.

Taking into account the tilt and hydrodynamic modulations, NRCS is then defined by \cite{alpers1981detectability, schulz2001ocean}

\begin{align}\label{equ:45}
\sigma \left( x,y \right)={{\sigma }_{0}}\left[ 1+\int{\left( M\left( k \right)\widehat{Z}\left( k \right){{e}^{ikx}}+c.c. \right)dk} \right]
\end{align}
where $\widehat{Z}(k)$ is the 2D Fourier transform of the sea surface model $Z$. The complex RAR modulation transfer function (MTF) is presented as a sum $M(k) = M_{t}(k) + M_{h}(k)$, where for the left-looking SAR antenna \cite{lyzenga1986numerical, schulz2001ocean}

\begin{align}\label{equ:46}
    {{M}_{t}}(k)=\frac{4\cot {{\theta }_{r}}}{1\pm {{\sin }^{2}}{{\theta }_{r}}}i{{k}_{y}}
\end{align}
with the plus sign related to VV and minus sign to HH polarization, and

\begin{align}\label{equ:47}
    {{M}_{h}}\left( k \right)=-4.5\omega \frac{{{k}_{y}}^{2}}{\left| k \right|}\frac{\omega -i\mu }{{{\omega }^{2}}+{{\mu }^{2}}}
\end{align}
where $\mu$ is the hydrodynamic relaxation rate which depends on $V_w$ and signal frequency and was set according to \cite{bruning1994estimation, nunziata2008educational}: when $V_w \leq 5$ m/s $\Rightarrow$ $\mu = 0.24$ s$^{-1}$ for X-band, $\mu = 0.1$ s$^{-1}$ for C-band, and $\mu = 0.01$ s$^{-1}$ for L-band, and when $V_w > 5$ m/s $\Rightarrow$ $\mu = 1.7$ s$^{-1}$ for X-band, $\mu = 0.7$ s$^{-1}$ for C-band, and $\mu = 0.1$ s$^{-1}$ for L-band. Hence, the NRCS with the MTF represents the RAR imaging, where MTF is at maximum for the sea waves traveling in the range direction, and insignificant for the azimuth moving direction.

Once the RAR image is calculated, it can be used for additional correction (specific SAR imaging mechanism) of the nonuniform displacements of water surface facets in the azimuthal direction via a velocity bunching (VB) mechanism \cite{alpers1979effect, lyzenga1985sar}. This leads to shifted imaging of moving waves by the Doppler effect \cite{bruning1990monte, bruning1994estimation}

\begin{dmath}\label{equ:48}
    I({{x}_{i}},{{y}_{i}})=\iint{\delta \left( {{y}_{i}}-y \right)\frac{\sigma \left( x,y \right)}{{{p}_{a}}^{\prime }\left( x,y \right)}}\times \exp \left\{ -{{\pi }^{2}}{{\left[ \frac{{{x}_{i}}-x-\frac{R}{V}{{{\bar{U}}}_{r}}\left( x,y \right)}{{{p}_{a}}^{\prime }\left( x,y \right)} \right]}^{2}} \right\}dxdy
\end{dmath}
where $I(x_i,y_i)$ is a speckle-free intensity SAR image, $\delta(\cdot)$ is the SAR impulse response function in range direction, approximated by the Dirac delta function, $R={H}/{\cos {{\theta }_{r}}}$ is the range distance between the antenna and the surface facets, $H$ is the platform flight height, $V$ is the platform velocity (it should be noted that there is a difference between the platform velocity and the footprint velocity \cite{vachon1994airborne}), while $R / V$ is the range-to-velocity ratio. ${{\bar{U}}_{r}}$ is the mean radial velocity of the surface facets in range direction and is written as

\begin{align}\label{equ:49}
    {\bar{U}}_r=Re\left\{{\mathcal{F}}^{-1}\left[\mathcal{F}\left(U_r\right)B_f\right]\right\}
\end{align}
where
$\mathcal{F}^{-1}\left(\cdot\right)$ and $\mathcal{F}\left(\cdot\right)$ refer to inverse and forward Fourier transforms, respectively, and

\begin{dmath}\label{equ:50}
    B_{f}=\frac{2}{{{k}_{x}}\Delta x}\sin \left( \frac{{{k}_{x}}\Delta x}{2} \right)\frac{2}{{{k}_{y}}\Delta y}\sin \left( \frac{{{k}_{y}}\Delta y}{2} \right) \frac{2}{\omega {{T}_{i}}}\sin \left( \frac{\omega {{T}_{i}}}{2} \right)
\end{dmath}
The radial orbital velocity field is given by \cite{zilman2014detectability}

\begin{align}\label{equ:51}
    {{U}_{r}}={{U}_{z}}\cos {{\theta }_{r}}-\sin {{\theta }_{r}}\left( {{U}_{x}}\sin {{\theta }_{w}}+{{U}_{y}}\cos {{\theta }_{w}} \right)
\end{align}
where $\theta_w$ is the angle between flight direction and wind direction or, if replaced by $\theta_{s}$, the angle between flight direction and ship moving direction. Both angles are measured counterclockwise from the flight (azimuth) direction. The additional filtering factor $B_{f}$ in (\ref{equ:49}) averages radial velocity over the integration time $T_i$ \cite{lyzenga1986numerical}. The large-scale (long-wave) orbital velocity components $U_x$, $U_y$ and $U_z$ in (\ref{equ:51}) are calculated as the gradients of the fluid velocity potential at the free surface ($z = 0$) for both sea waves and ship wake models

\begin{align}\label{equ:52}
    {{U}_{x}}&=\frac{\partial \Phi \left( x,y,0 \right)}{\partial x} & {{U}_{y}}&=\frac{\partial \Phi \left( x,y,0 \right)}{\partial y} & {{U}_{z}}&=\frac{\partial \Phi \left( x,y,0 \right)}{\partial z}
\end{align}
The degraded azimuthal resolution $p_a'$ in (\ref{equ:48}) is equal to

\begin{align}\label{equ:53}
    {{{p}'}_{a}}\left( x,y \right)=N_l{{p}_{a}}{{\left[ 1+\frac{{{\pi }^{2}}{{T}_{i}}^{4}}{{{N_l}^{2}}{{\lambda }^{2}}}{{{\bar{A}}}_{r}}\left( x,y \right)+\frac{1}{{{N_l}^{2}}}\frac{{{T}_{i}}^{2}}{{{\tau }_{c}}^{2}} \right]}^{{1}/{2}\;}}
\end{align}
where $N_l$ is the number of incoherent looks in azimuth direction, $\lambda$ is the wavelength of the radar signal, $p_a$ is the nominal single-look azimuthal resolution as follows

\begin{align}\label{equ:54}
    {{p}_{a}}=\frac{\lambda R}{2V{{T}_{i}}}
\end{align}

Since the integration time $T_i$ is one of the smearing factors in SAR images and relates to the full azimuthal resolution $p_a$, it follows that with decreasing azimuthal resolution the integration time also decreases, which reduces the smearing effect \cite{schulz2001ocean}.

The mean radial acceleration of the surface facets ${{\bar{A}}_{r}}$ is calculated the same way as ${{\bar{U}}_{r}}$, but by inserting the large-scale acceleration components $A_x$, $A_y$, $A_z$ into (\ref{equ:51}), which in turn is calculated using (\ref{equ:52}) by replacing  $\Phi(x,y,0)$ with $\partial \Phi(x,y,0) / \partial t$.

The scene coherence time $\tau_c$ is related to the spreading of the facets motion within a SAR resolution cell \cite{alpers1986relative}. However, for simplicity, and assuming a Pierson-Moskowitz type wave spectrum, the coherence time is approximated as \cite{frasier2001dual}

\begin{align}\label{equ:55}
    {{\tau }_{c}}\approx 3\frac{\lambda }{{{V}_{w19.5}}}er{{f}^{{-1}/{2}\;}}\left( 2.7\frac{{p}}{{{V}_{w19.5}}^{2}} \right)
\end{align}

It is worth noting that the velocity bunching method is not valid for large values of the integration time $T_i$ \cite{zurk1996comparison}, however, a recent idea about the time-divided velocity bunching model can help to overcome this limitation \cite{rim2019sar}.

Finally, the intensity of the SAR image which includes a multiplicative noise component can be presented as \cite{zilman2014detectability}:

\begin{align}\label{equ:56}
    {{I}_{n}}\left( {{x}_{i}},{{y}_{i}} \right)=I\left( {{x}_{i}},{{y}_{i}} \right)N({{x}_{i}},{{y}_{i}})
\end{align}
where the noise sequence can be expressed with the exponential distribution with a PDF of $P(N) = \exp(-N)$. Please also note that for some specific applications, various other advanced intensity speckle models, such as Gamma, $K$, Gen-Rayleigh \cite{kuruoglu2004modeling}, Gen-Rician \cite{karakucs2020generalized}, Gen-Gamma \cite{li2010efficient} can be used.

The important limitation of SAR imaging of waves moving in flight direction which is associated with the velocity bunching is the azimuthal cut-off effect \cite{pleskachevsky2016meteo}. The minimal detectable wavelength of the surface waves can be approximated as \cite{beal1983large, vachon2004ocean}

\begin{align}\label{equ:57}
    {{\lambda }_{\min }}=C_{0}\frac{R}{V}\sqrt{{{H}_{s}}}
\end{align}
where $C_{0}$ is a constant of order 1 ($m^{1/2}$ $s^{-1}$). Here an increase in the $R/V$ ratio and significant wave height $H_s$, reduces the ability for the azimuth-traveling wave to be imaged. In order to illustrate the azimuthal cut-off effect we assume the $S_{PM}$ spectrum where the significant wave height can be approximated \cite{ochi2005ocean} as $H_{s}=0.21V_{w19.5}^2/g$ and peak wavenumber $k_{p}=0.16/H_{s}$, whence the dominant wavelength of the surface waves $\lambda_{d}=2\pi/k_{p}$. The resulting graph-map is shown in Fig. \ref{fig:fig5}-(a) which is calculated for a range of $V_{w19.5}=3.5-13.5$ m/s ($H_{s}=0.26 - 3.9$ m or $\lambda_{d} = 10.3 - 153.2$ m) and SAR ratios $R/V = 10 - 250$ s. Although the original relationship (\ref{equ:57}) is based on significant wave height, we use the dominant wavelength of the surface waves, because when analyzing the waves from the SAR images it is more convenient to use $\lambda_{d}$ (as this can be easily estimated via the spatial or spectral domain) than $H_s$. For example, in Fig. \ref{fig:fig5}-(b) and (c) two SAR images for $R/V=23.1$ s (b) and $R/V=107.1$ s (c) with $V_{w19.5}\approx10.7$ m/s or $\lambda_{d}\approx 95.5$ m are presented. Using the presented graph-map, for the first SAR image in Fig. \ref{fig:fig5}-(b) the minimal detectable wavelength $\lambda_{min}=36.2$ m and for the second image in Fig. \ref{fig:fig5}-(c) $\lambda_{min}=167.7$ m. Thus, in the first case (b) the waves are clearly detectable because the condition $\lambda_{min}<\lambda_{d}$ is satisfied, while in the second case (c) where $\lambda_{min}>\lambda_{d}$ it is not. A similar example for the ship wake is presented in Fig. \ref{fig:fig10}-(a) and (f)). It is worth mentioning that the proposed graph-map (Fig. \ref{fig:fig5}) is suitable for a fully developed sea state.

\newsavebox{\bigleftbox}

\begin{figure*}[ht]
\centering
\sbox{\bigleftbox}{%
  \begin{minipage}[b]{.55\textwidth}
  \centering
  \vspace*{\fill}
    \subfigure[]{\includegraphics[width=12cm,height=11cm]{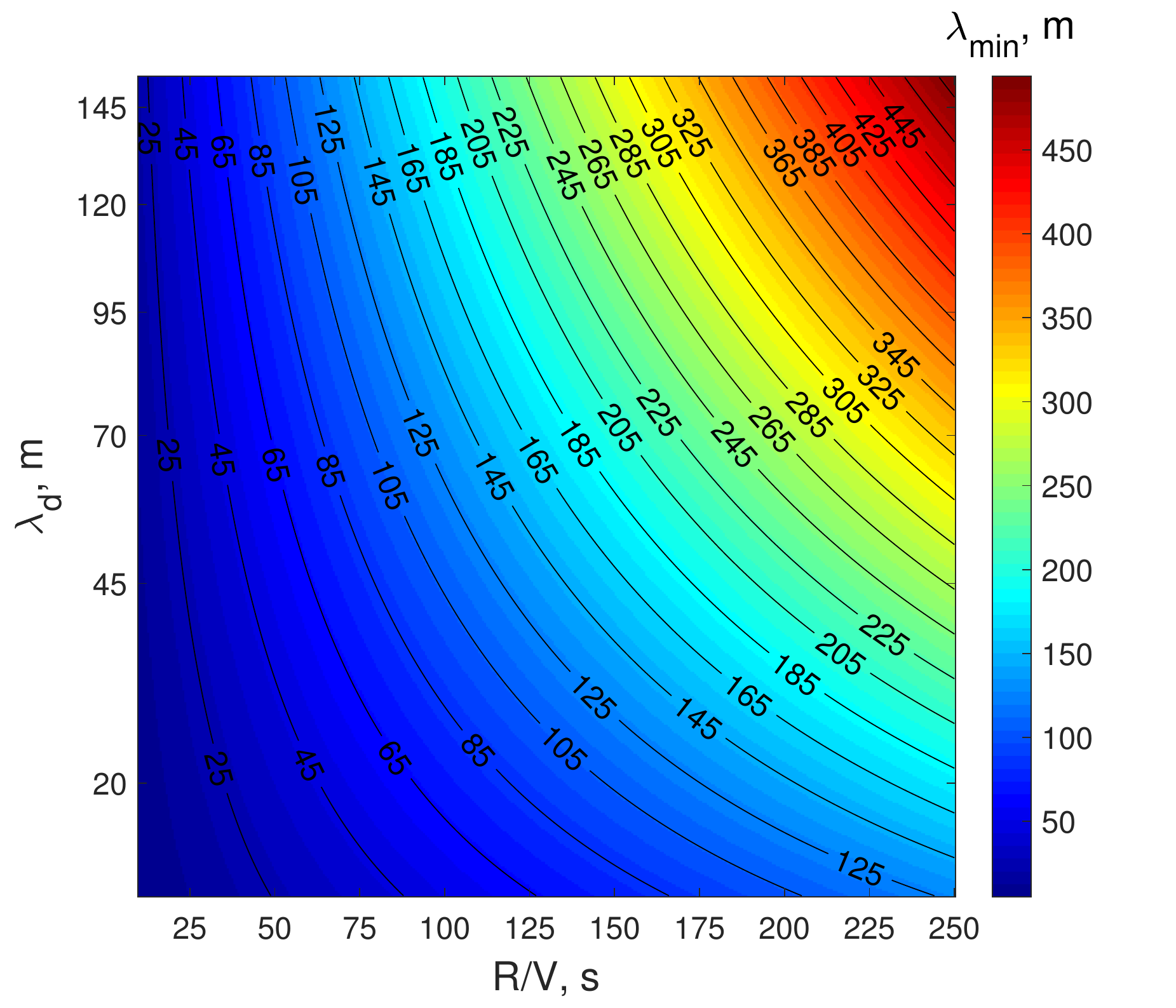}}
  \end{minipage}%
}\usebox{\bigleftbox}%
\begin{minipage}[b][\ht\bigleftbox][s]{.65\textwidth}
\centering
  \subfigure[]{\includegraphics[width=.43\linewidth]{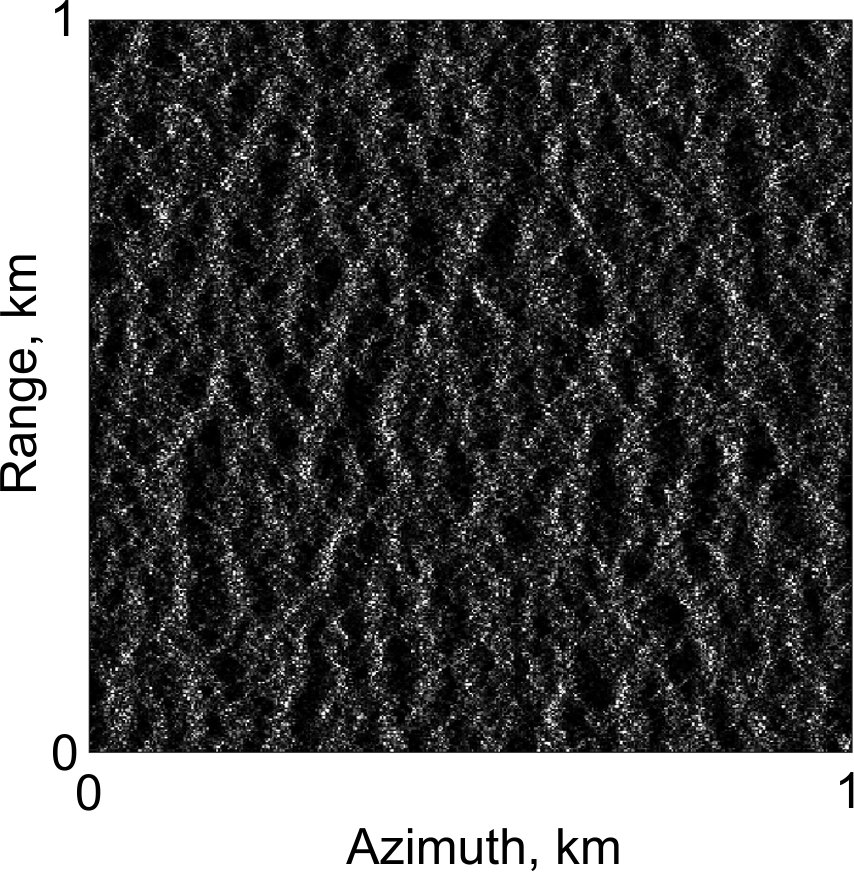}}
\vfill
  \subfigure[]{\includegraphics[width=.43\linewidth]{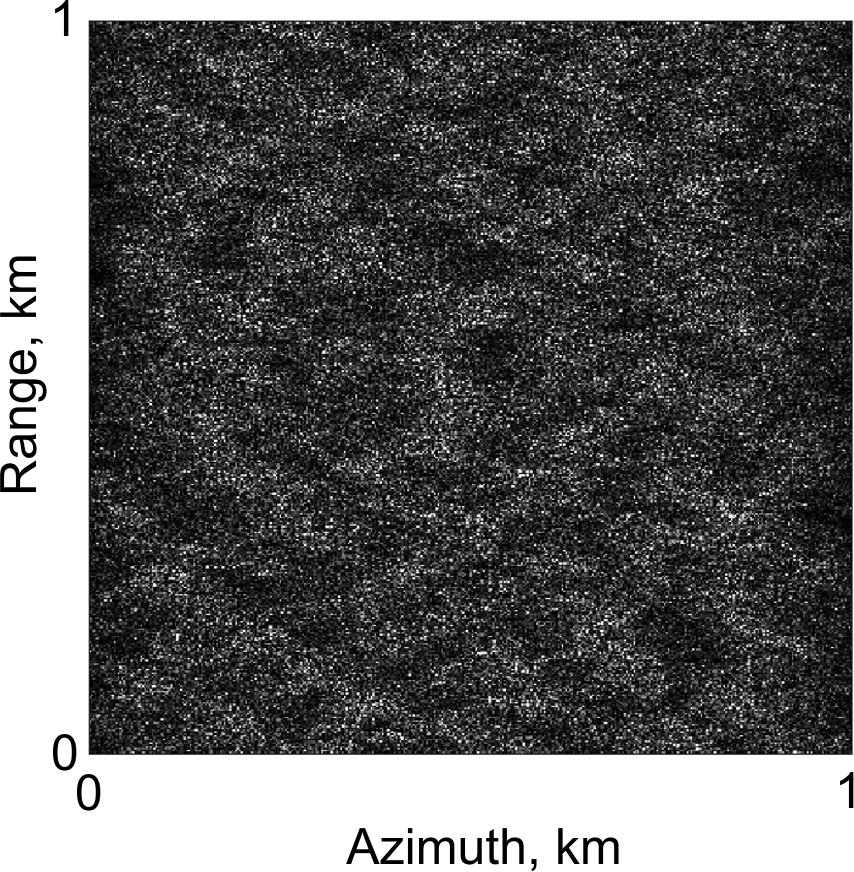}}
\end{minipage}
\hspace*{\fill}
\caption{The SAR azimuthal cut-off effect whith the minimal detectable  wavelength of the sea surface waves $\lambda_{min}$ presented in terms of the dominant wavelength of waves $\lambda_{d}$ and the $R/V$ ratios. (a) The graph-map is based on the $S_{PM}$ spectrum and is valid for the range $V_{w19.5} = 3.5 - 13.5$ m/s to $R/V = 10 - 250$ s. Simulated SAR images (X-band, $\theta_r = 30^{\circ}$, HH polarization) of the sea surface with $V_{w19.5}\approx10.7$ m/s and $\lambda_{d}\approx 95.5$ m ($S_{PM}$ and $D_{J}$ = $0^{\circ}$ with S = 20) for (b) airborne ($R/V$ = 23.1 s) and (c) satellite ($R/V$ = 107.1 s) platforms.}
\label{fig:fig5}
\end{figure*}

\section{Results and Discussions}\label{sec:results}
In this section, we show the results of our simulation experiments involving SAR image formation, including for different wave spectra. Detailed attention has been paid to specific parameters of the radar image formation. Firstly, we consider the factors that affect the imaging of both types of waves, sea and ship. In general, these factors can be reduced to hydrodynamic (surface modeling) and SAR imaging constituents.
There are hydrodynamic effects due to waves superposition: various sea surfaces and ship wakes with different Froude numbers are examined. The numerical simulation analysis was performed in two steps. First, the geometry of the SAR platforms and scanning parameters was considered: different platforms, $R/V$ ratios, resolutions, incidence angles, signal frequencies, and polarizations were utilized. Second, the specific radar mechanisms in terms of ship wakes and sea waves imaging were evaluated.

The comparison of different spectra has been performed in two aspects: sea surface roughness validation and ship wake visibility evaluation. Since the Bragg scattering directly depends on surface roughness, we compared simulated spectra with the well-known Cox and Munk’s probability density function \cite{cox1956slopes} of surface slopes. The visibility evaluation of the ship wake was also performed in two stages: determination of the boundary condition, and qualitative and quantitative assessment.

Simulated SAR images were created corresponding to both airborne and satellite platforms, with details shown in Table \ref{tab:table1}. Airborne platforms were presented at low altitude (AI) and high altitude (AII). Satellite platforms were also separated into low altitude (SI) and high altitude (SII) types, with main parameters similar to the TerraSAR-X and Sentinel-1 respectively. The terms low and high should be understood as relative to conventional Earth observation SAR scanners. It must be noted that the range-to-velocity ratio $R/V$ was determined in accordance with the specified incidence angle, which may vary for some calculations. For simplicity, we neglected the differences of SAR resolution in range and azimuth, which usually exist for real images, and used 2.5 m for both resolutions. Also, only single-look processing is considered ($N_l$ = 1). For the simulations, we utilized four different ship models with varying Froude numbers of $Fr = 0.15-0.5$
\begin{itemize}
    \item  \textbf{Ship-I} $\rightarrow$ $L = 35$ m, $B = 5$ m, $D_{t} = 2.5$ m,
    \item \textbf{Ship-II} $\rightarrow$ $L = 50$ m, $B = 6.5$ m, $D_{t} = 3.5$ m,
    \item \textbf{Ship-III} $\rightarrow$ $L = 65$ m, $B = 10$ m, $D_{t} = 4.6$ m,
    \item \textbf{Ship-IV} $\rightarrow$ $L = 135$ m, $B = 25$ m, $D_{t} = 10$ m.
\end{itemize}

\begin{table*}[t]
  \centering
  \caption{Main parameters of scanning platforms for the simulated SAR images.}
  \renewcommand{\arraystretch}{0.75}
    \resizebox{\linewidth}{!}{\begin{tabular}{lcccc}
    \toprule
    Parameter & \textbf{AI} & \textbf{AII} & \textbf{SI} & \textbf{SII} \\
    \toprule
    Frequency $f$, [GHz] & \multicolumn{4}{c}{9.65 (X-band), 5.3 (C-band), 1.275 (L-band)} \\
    \midrule
    Wavelength $\lambda$, [m] & \multicolumn{4}{c}{0.031 (X-band), 0.057 (C-band), 0.235 (L-band)} \\
    \midrule
    Incidence angle $\theta_r$, [deg] & \multicolumn{4}{c}{20-70} \\
    \midrule
    Polarisation & \multicolumn{4}{c}{VV, HH} \\
    \midrule
    Platform altitude $H$, [km] & 2.5   & 7     & 514   & 705 \\
    \midrule
    Platform velocity $V$, [m/s] & 125   & 160   & 7600  & 7600 \\
    \midrule
    $R/V$, [s] & \multirow{2}[2]{*}{21-59} & \multirow{2}[2]{*}{47-128} & \multirow{2}[2]{*}{72-198} & \multirow{2}[2]{*}{99-271} \\
    ($\theta_r=20^{\circ}-70^{\circ}$) &       &       &       &  \\
    \midrule
    Integration time $T_i$, [s] & 0.13-0.36 (X-band) & 0.29-0.79 (X-band) & 0.45-1.23 (X-band) & 0.61-1.68 (X-band) \\
\cmidrule{2-5}    ($\theta_r=20^{\circ}-70^{\circ}$) & 0.24-0.66 (C-band) & 0.53-1.45 (C-band) & 0.82-2.24 (C-band) & 1.12-3.07 (C-band) \\
\cmidrule{2-5}          & 1-2.75 (L-band) & 2.19-6.01 (L-band) & 3.38-9.29 (L-band) & 4.64-12.75 (L-band) \\
    \midrule
    Coherence  time $\tau_c$, [s] & \multicolumn{4}{c}{0.035-0.034 (X-band)} \\
\cmidrule{2-5}    ( $V_{w10} = 3.5-11$ m/s) & \multicolumn{4}{c}{0.064-0.062 (C-band)} \\
\cmidrule{2-5}          & \multicolumn{4}{c}{0.266-0.256 (L-band)} \\
    \midrule
    Azimuth resolution, [m] & \multicolumn{4}{c}{2.5} \\
    \midrule
    Range resolution, [m] & \multicolumn{4}{c}{2.5} \\
    \bottomrule
    \end{tabular}}%
  \label{tab:table1}%
\end{table*}%

\subsection{Effect of different wind state}\label{subsec:windstate}
Two time-frozen elevation models, the sea surface and Kelvin wake surface are based on a superposition of wind- and ship-generated waves and form the total sea wave-wake model as $Z = Z_{sea} + Z_{ship}$. The same summation principle is related to all derivatives of these models: surface slopes, fluid velocity, and acceleration components. When considering superposition, it is important to select the correct wind velocity value for modeling. In many studies \cite{hennings1999radar, reed2002ship, pichel2004ship, panico2017sar, zilman2014detectability, wang2016application, graziano2017performance, tings2018comparison, tings2019extension, tings2021non}, it is postulated that wind velocity (or consequently significant wave height) is one of the main factors in ship wake visualization in SAR imagery. The Kelvin wake system (divergent and transverse waves) can be best observed when the wind velocity is about 3 m/s or less, as these levels give a very calm sea surface  \cite{panico2017sar}. However, visualization of the cusp waves is actually more stable for relatively high wind speeds ($6 \sim 10$ m/s) \cite{hennings1999radar, panico2017sar, graziano2017performance}.

\begin{figure*}[htbp]
\centering
\subfigure[]{\includegraphics[width=.23\linewidth]{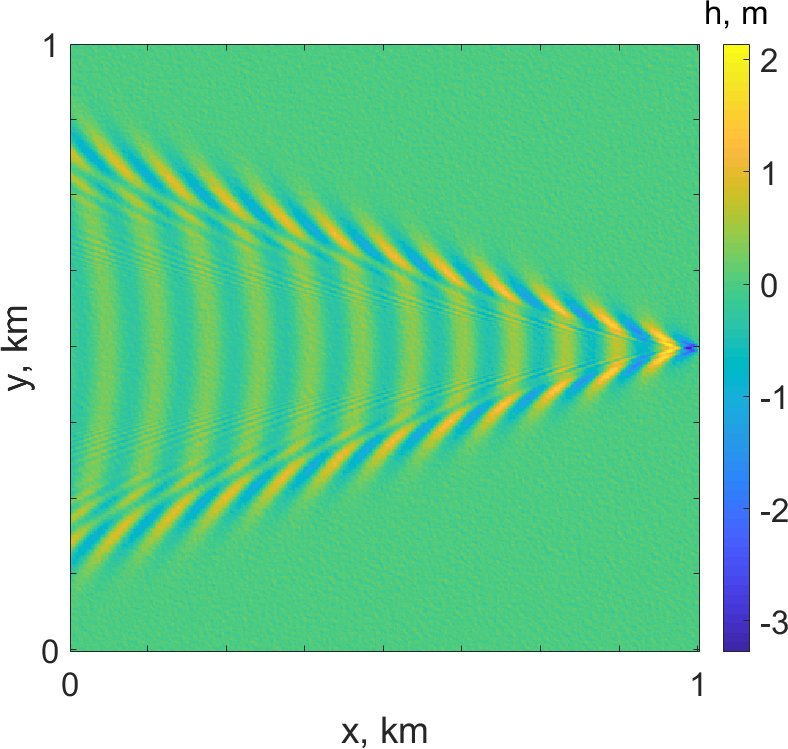}}
\hspace{0.1 cm}
\subfigure[]{\includegraphics[width=.23\linewidth]{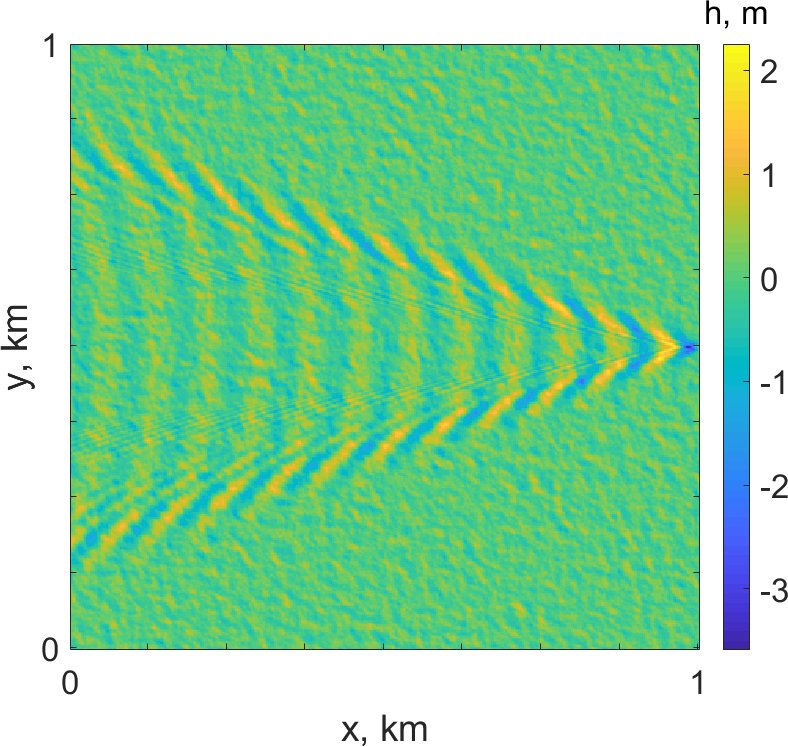}}
\hspace{0.1 cm}
\subfigure[]{\includegraphics[width=.23\linewidth]{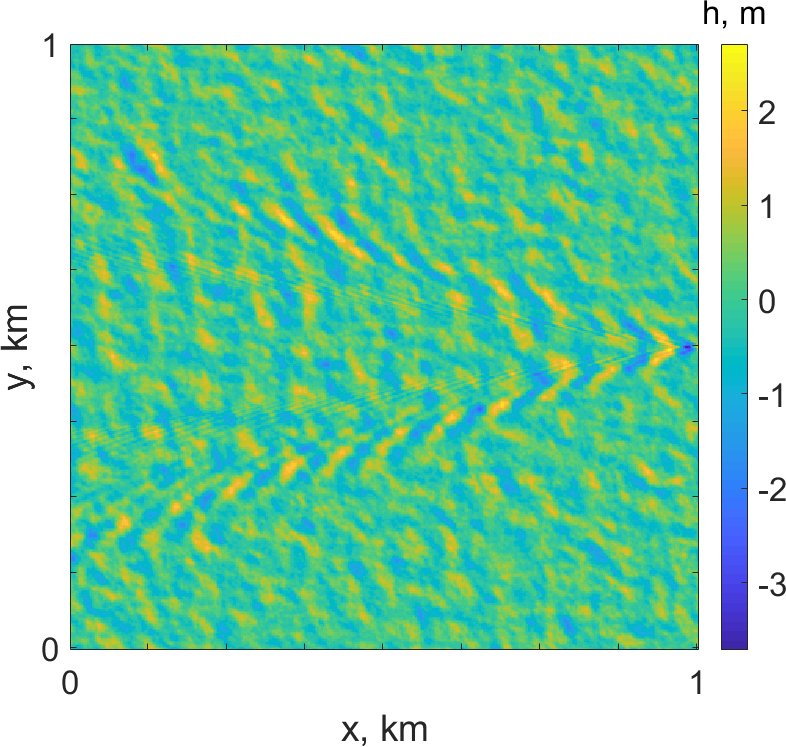}}
\hspace{0.1 cm}
\subfigure[]{\includegraphics[width=.23\linewidth]{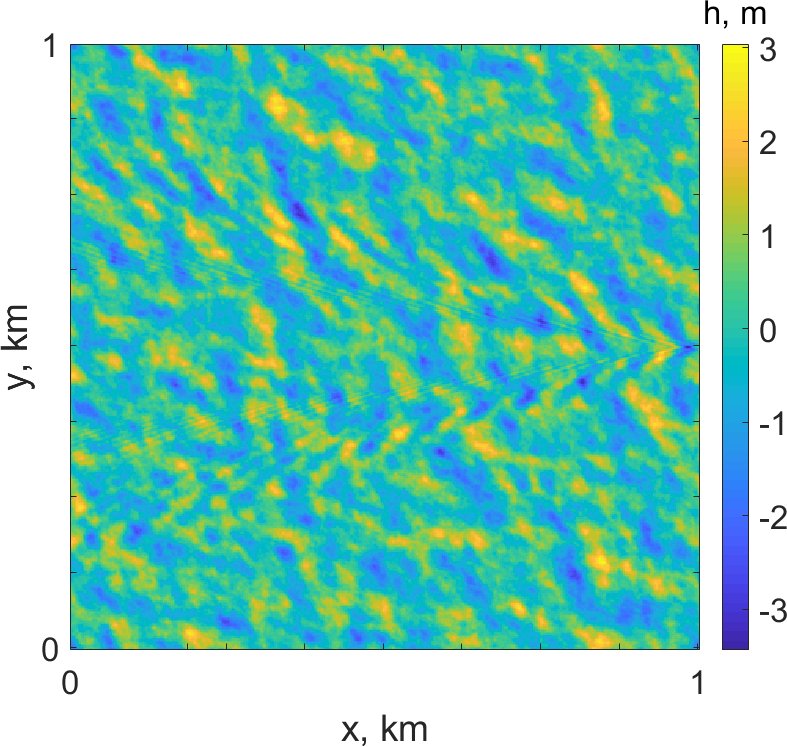}}
\caption{Composite sea-ship surface elevation models $Z$ for different velocities of wind ${{V}_{w10}}$ as follows: (a) 3.5 m/s. (b) 6 m/s. (c) 8.5 m/s. (d) 11 m/s. \textbf{Ship II}, Fr = 0.5. Sea surface parameters: $S_{E}$ spectrum with $D_{E}$ = $45^{\circ}$.}
\label{fig:fig6}
\end{figure*}

In  Fig. \ref{fig:fig6}, our modeling results clearly demonstrate that at higher wind velocity the ship wake visualization is significantly decreased, with larger sea waves and ship wakes canceling each other out. It is also important to note that with increasing amplitude of the sea waves, radial velocity and acceleration are also increased, which leads to greater smearing and shifting in the resulting SAR image.

\subsection{Effect of the Froude number}\label{subsec:FroudeNumber}
From Section \ref{sec:wakeModelling} it follows that the main parameters which influence the pattern and amplitude of the Kelvin wake are length $L$, beam $B$, and draft $D_{t}$ of the modeled ship, as well as its velocity $V_s$. When modeling the Kelvin wake, it is useful to apply a dimensionless quantity, the Froude number, which is used to determine the resistance of a submerged moving vessel and is defined by

\begin{align}\label{equ:58}
    Fr=\frac{{{V}_{s}}}{\sqrt{gL}}
\end{align} 

\begin{figure*}[htbp]
\centering
\subfigure[]{\includegraphics[width=.23\linewidth]{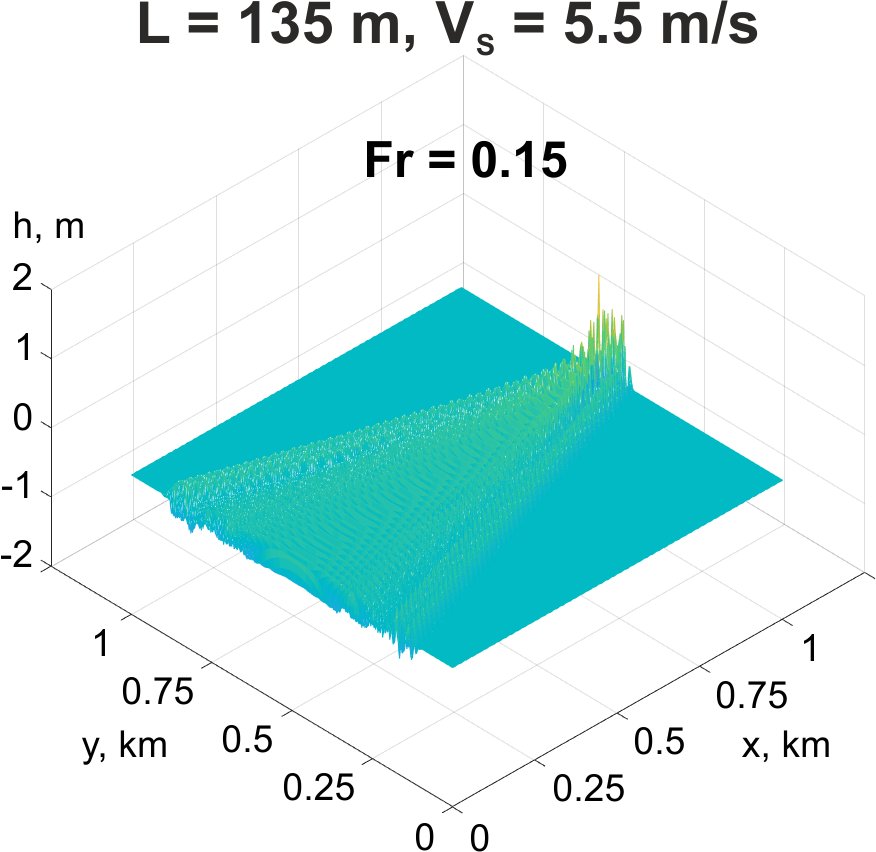}}
\subfigure[]{\includegraphics[width=.23\linewidth]{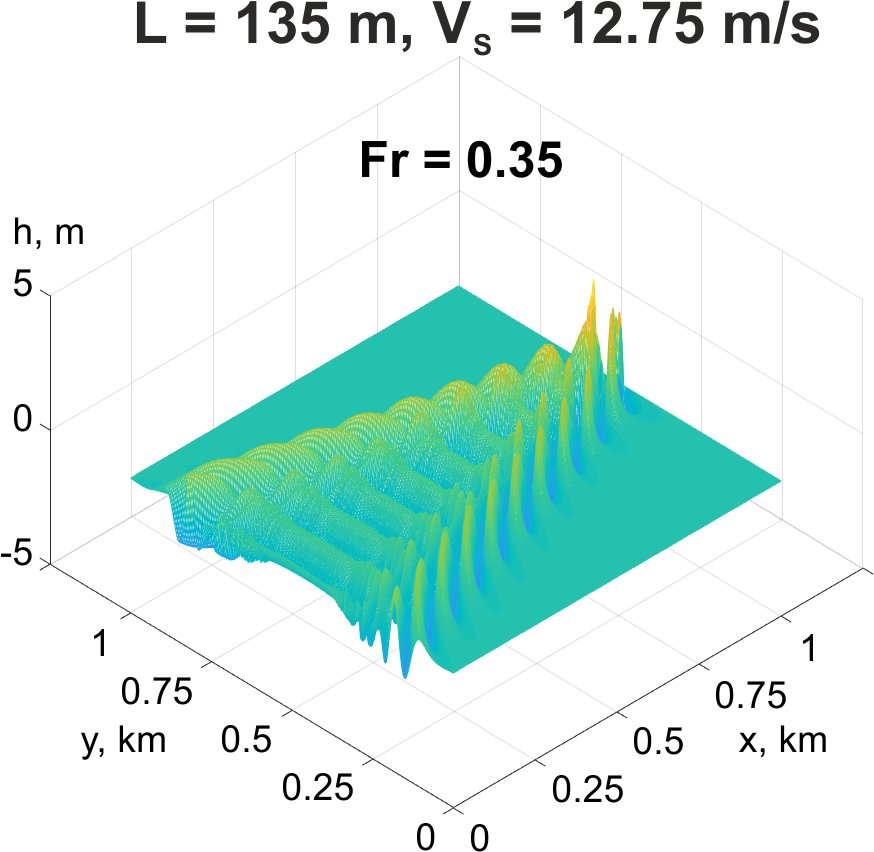}}
\subfigure[]{\includegraphics[width=.23\linewidth]{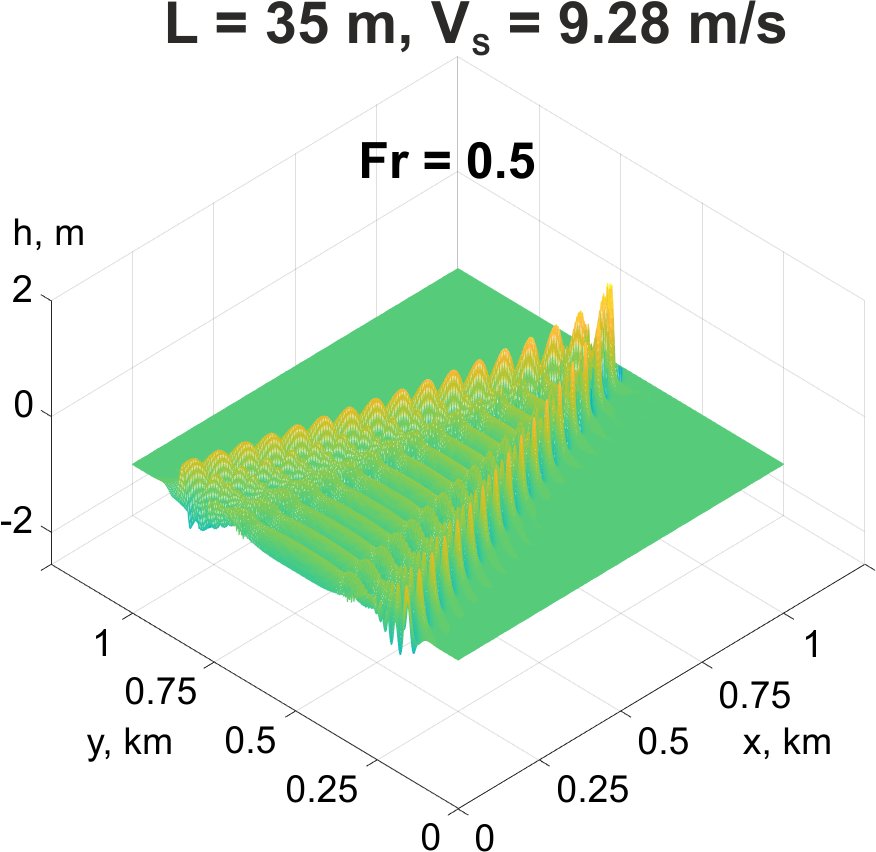}}
\subfigure[]{\includegraphics[width=.23\linewidth]{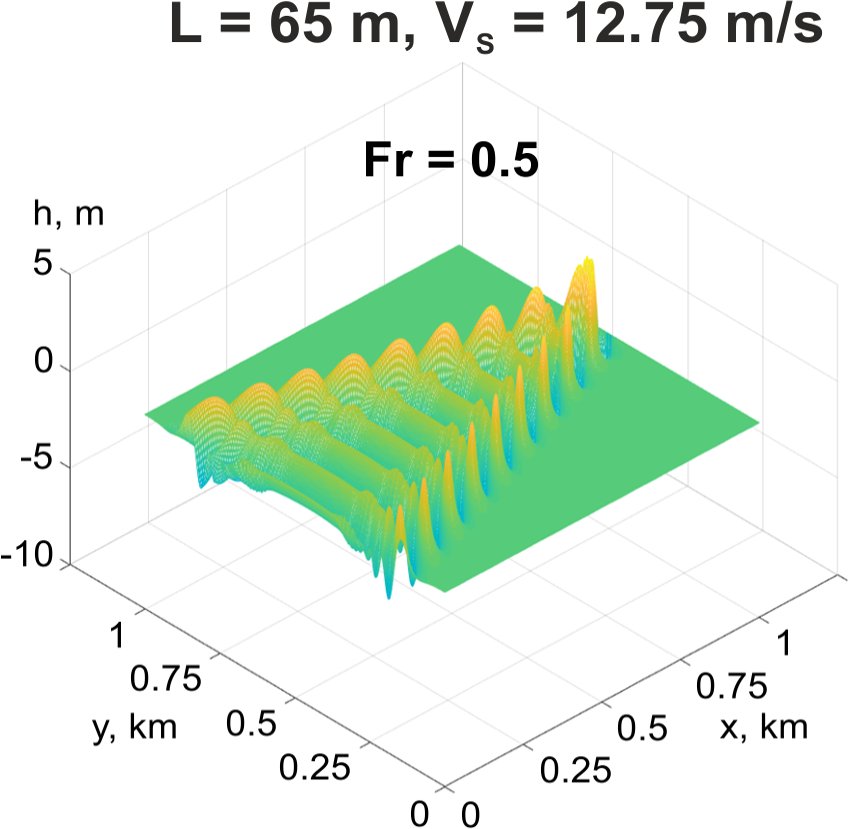}}
\subfigure[]{\includegraphics[width=.24\linewidth]{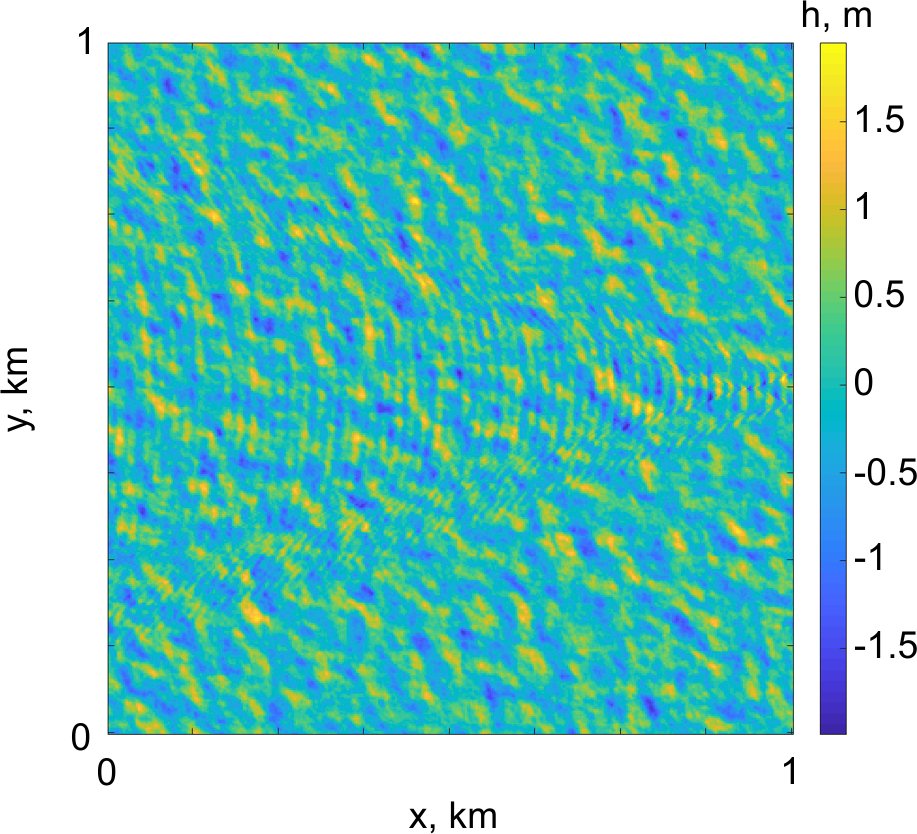}}
\subfigure[]{\includegraphics[width=.23\linewidth]{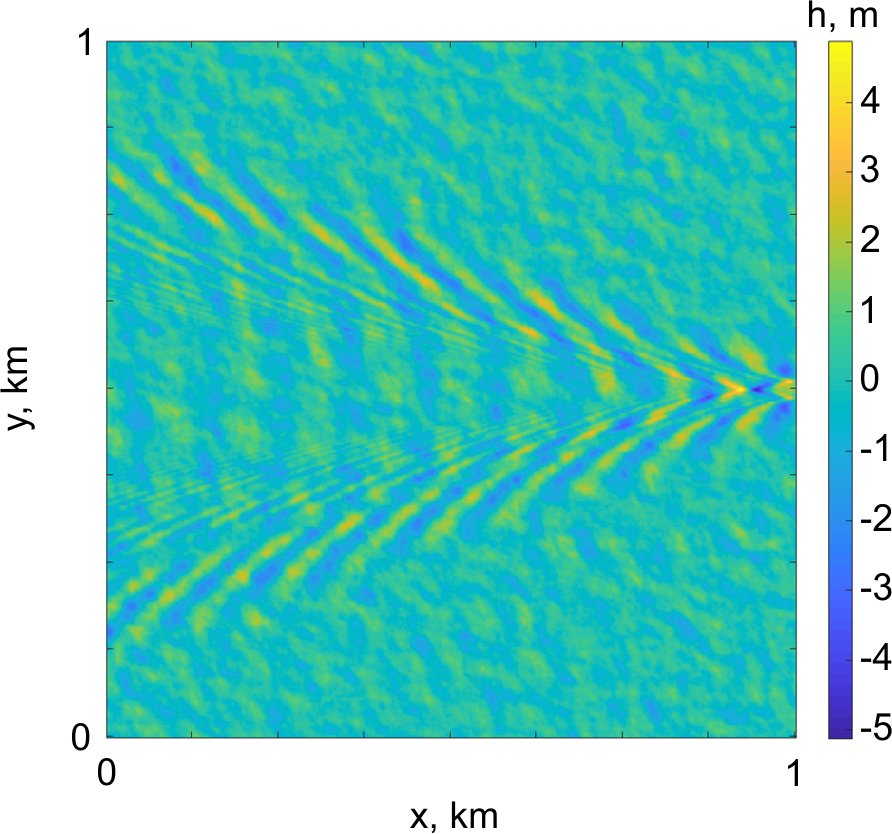}}
\subfigure[]{\includegraphics[width=.24\linewidth]{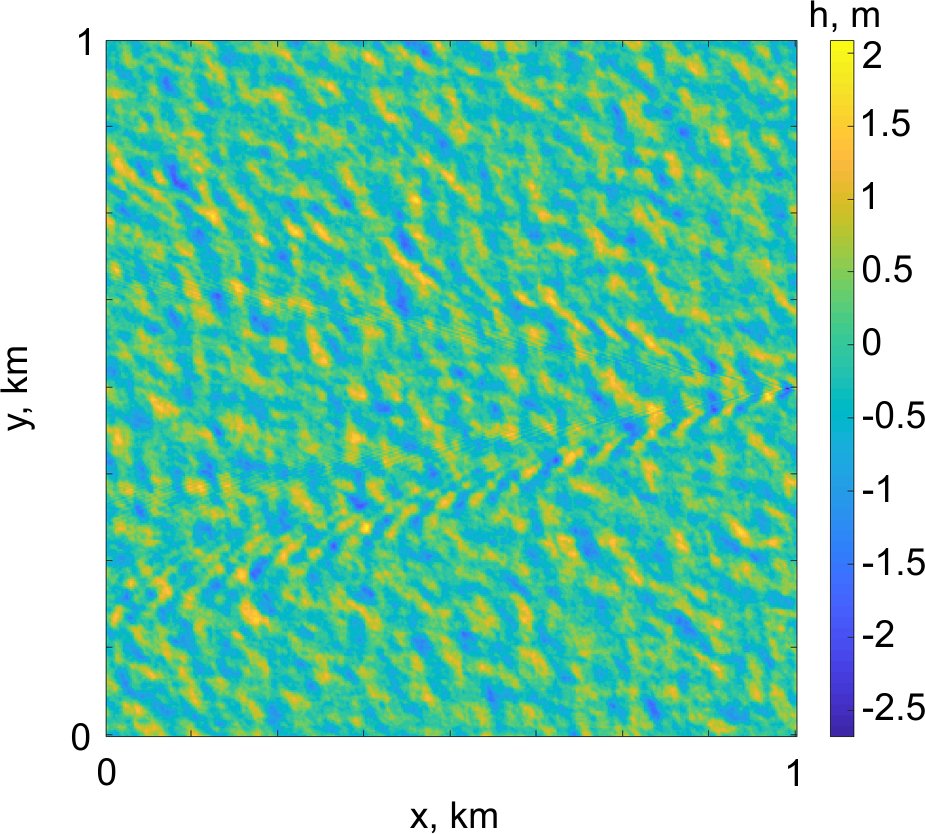}}
\subfigure[]{\includegraphics[width=.23\linewidth]{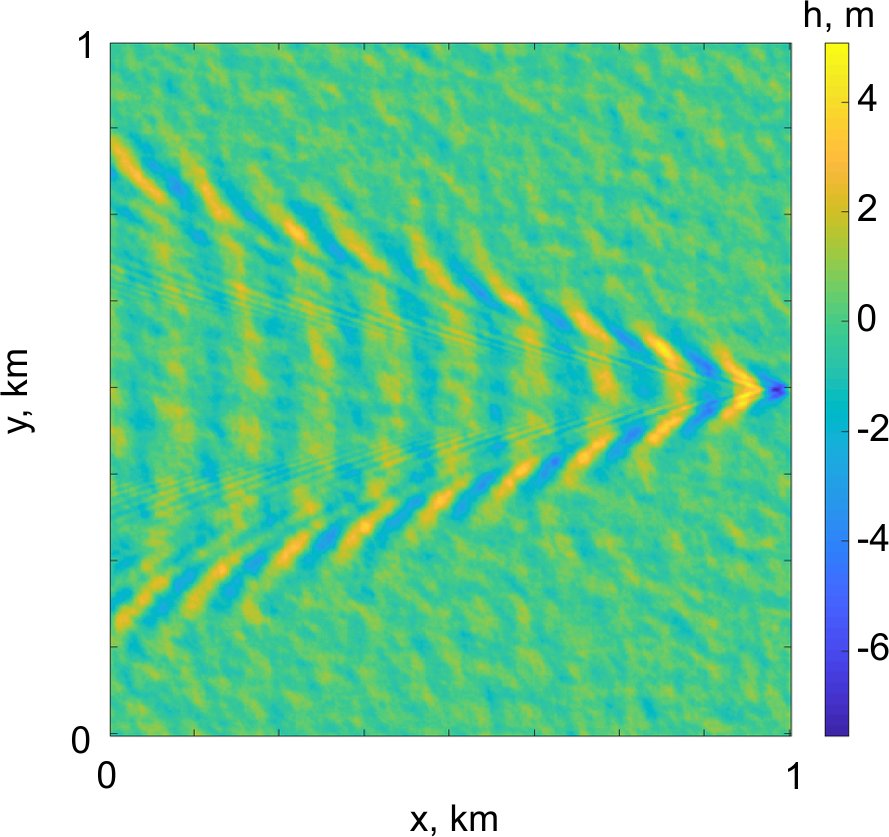}}
\caption{Simulated Kelvin wake elevation models for the \textbf{Ship I} (c), (g); \textbf{Ship III} (d), (h) and \textbf{Ship IV} (a), (b), (e), (f) with different Froude numbers. Upper images 3D representation of wake models. Lower images superposition of ship wake models with ambient sea waves under constant ${{V}_{w10}}$ = 8.5 m/s.}
\label{fig:fig7}
\end{figure*}

From Fig. \ref{fig:fig7}-(a), (b) and (e), (f) it can be easily seen that with a change in $Fr$ for a ship of the same size, the pattern and amplitude of wakes also change. However, it is interesting to note that if $Fr$ is the same (Fig. \ref{fig:fig7}-(c), (d) and (g), (h)) but the ships differ in length and velocity, then the vessels will produce an identical wave pattern (but of different amplitude). Therefore, wakes of different ships with similar $Fr$ have fractal (or scalable) properties. For example, this property has been applied in work \cite{hennings1999radar}, where the wave pattern of a modeled ship was scaled to the wave pattern of a real ship. However, a study \cite{zilman2014detectability} has also shown that a small ship with high Froude number and a bigger ship with low Froude number may produce Kelvin wakes with the same wave amplitude. All of this demonstrates that the Froude number for ships with different lengths and velocities is irrelevant for predicting wake visibility in SAR images. In addition, the wake angle between cusp waves may decrease for large ship velocities or when $Fr>0.5$ as shown in \cite{rabaud2013ship}, with the consequence that the transverse waves become invisible \cite{rabaud2014narrow}. In this case, the opening angle (rad) can be estimated using the following expression

\begin{align}\label{equ:59}
    {{\alpha }_{w}}\approx \frac{1}{2\sqrt{2\pi }Fr}
\end{align}

\subsection{Effect of SAR imaging}
\begin{figure*}[htbp]
\centering
\subfigure[]{\includegraphics[width=.23\linewidth]{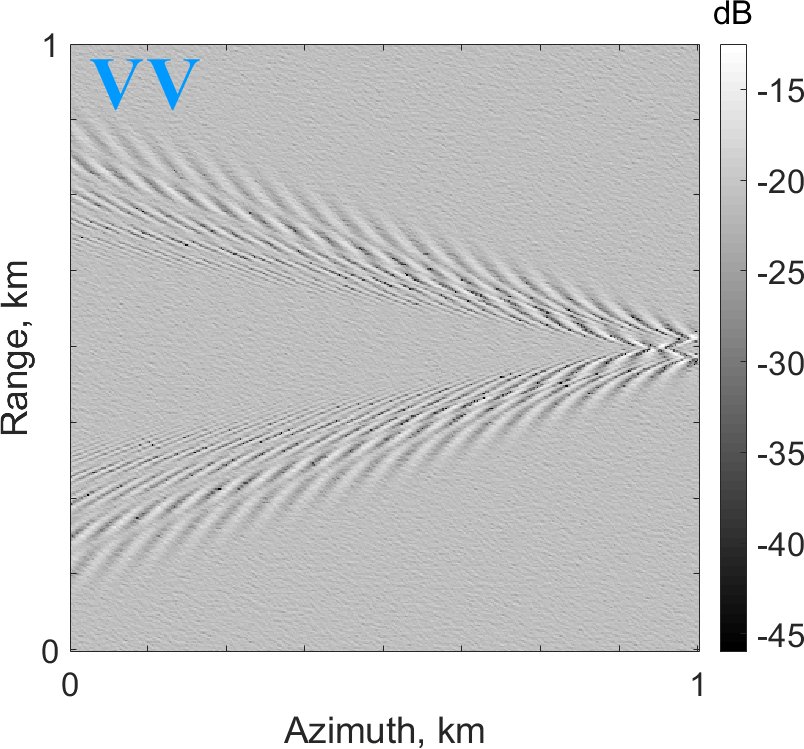}}
\subfigure[]{\includegraphics[width=.23\linewidth]{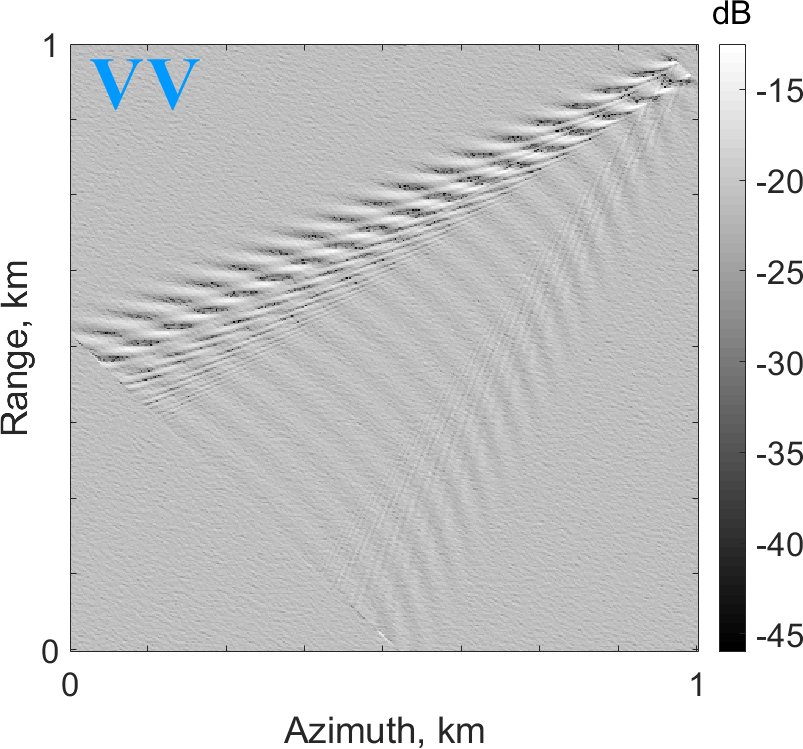}}
\subfigure[]{\includegraphics[width=.23\linewidth]{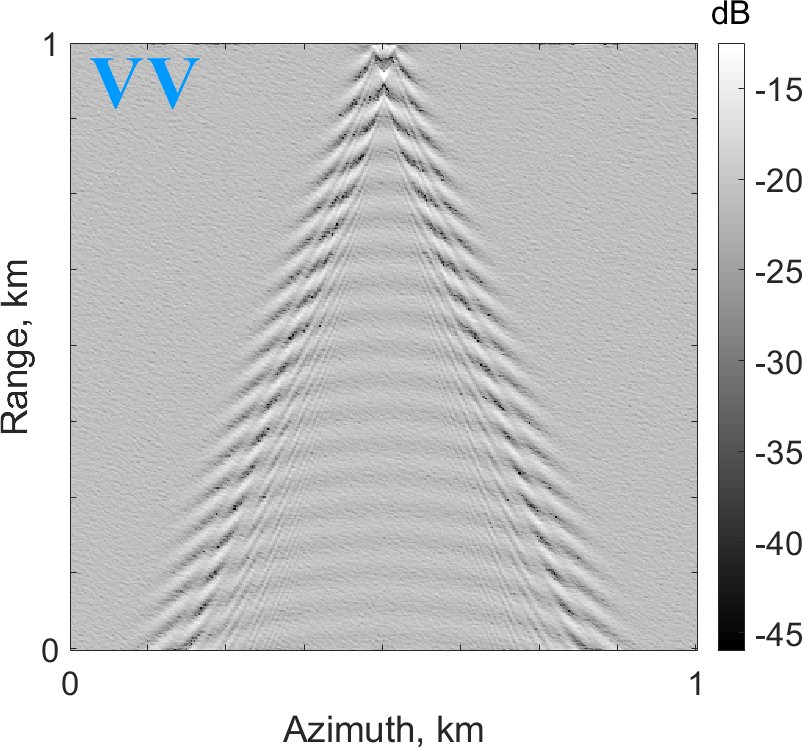}}
\subfigure[]{\includegraphics[width=.23\linewidth]{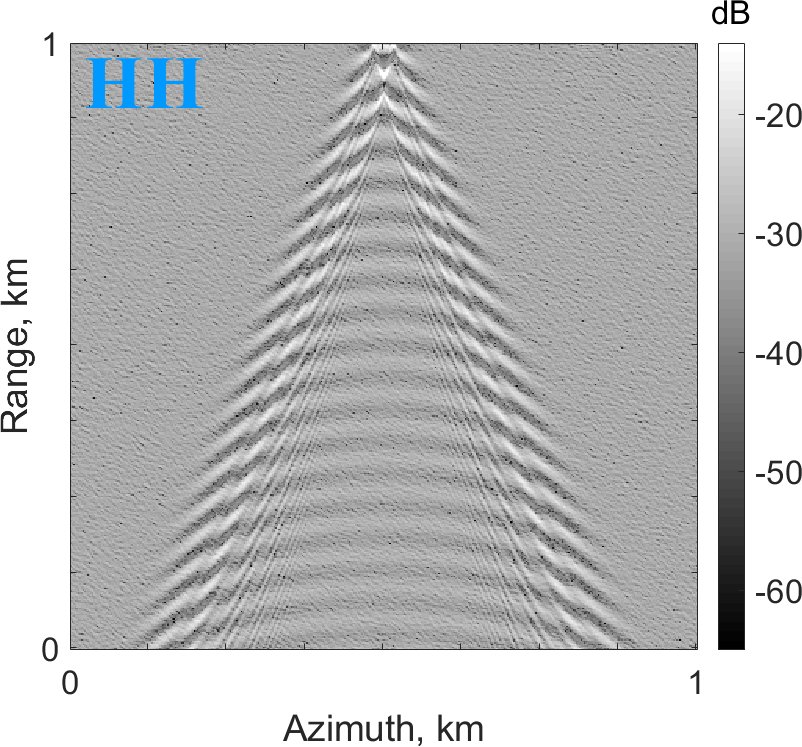}}
\caption{Simulated NRCS images (X-band, $\theta_r = 50^{\circ}$) with RAR modulations included for different ship heading direction relative to flight direction: (a) $0^{\circ}$. (b) $45^{\circ}$. (c), (d) $90^{\circ}$. \textbf{Ship IV} with Fr = 0.25; ${{V}_{w10}}$ = 3.5 m/s and $D_{E}$ = $45^{\circ}$.}
\label{fig:fig8}
\end{figure*}

\begin{figure*}[htbp]
\centering
\subfigure[]{\includegraphics[width=.19\linewidth]{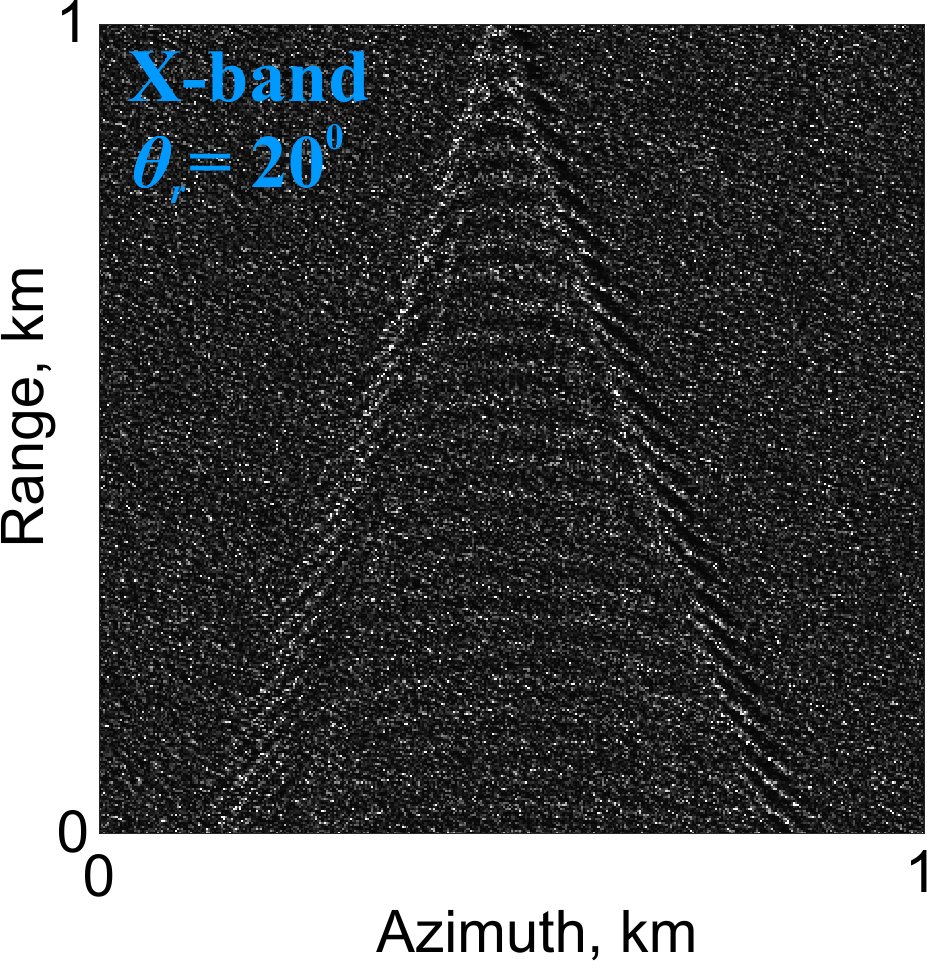}}
\subfigure[]{\includegraphics[width=.19\linewidth]{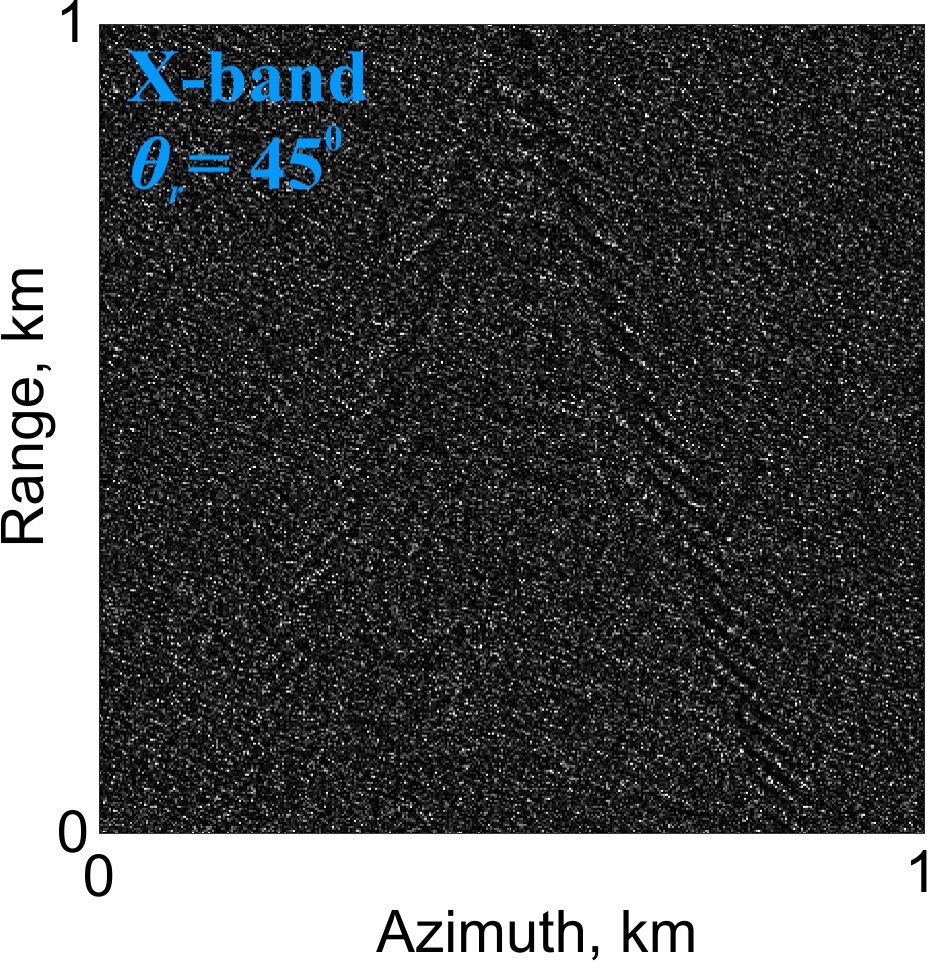}}
\subfigure[]{\includegraphics[width=.19\linewidth]{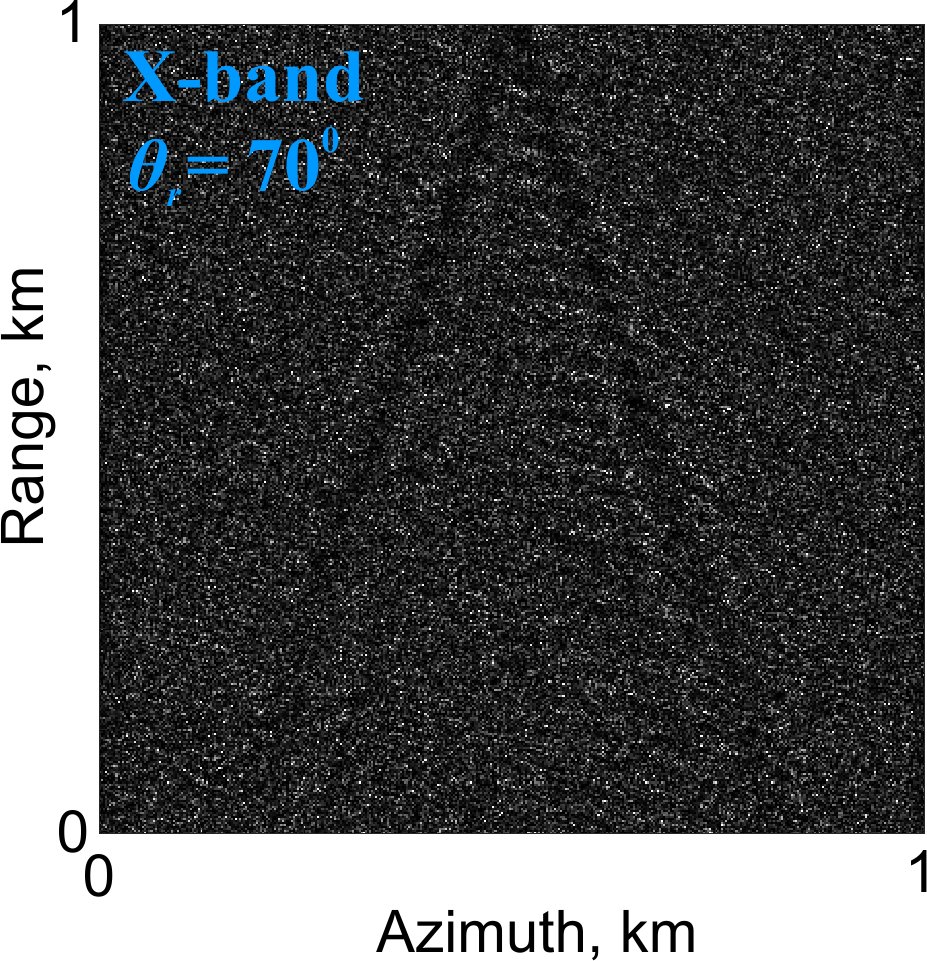}}
\subfigure[]{\includegraphics[width=.19\linewidth]{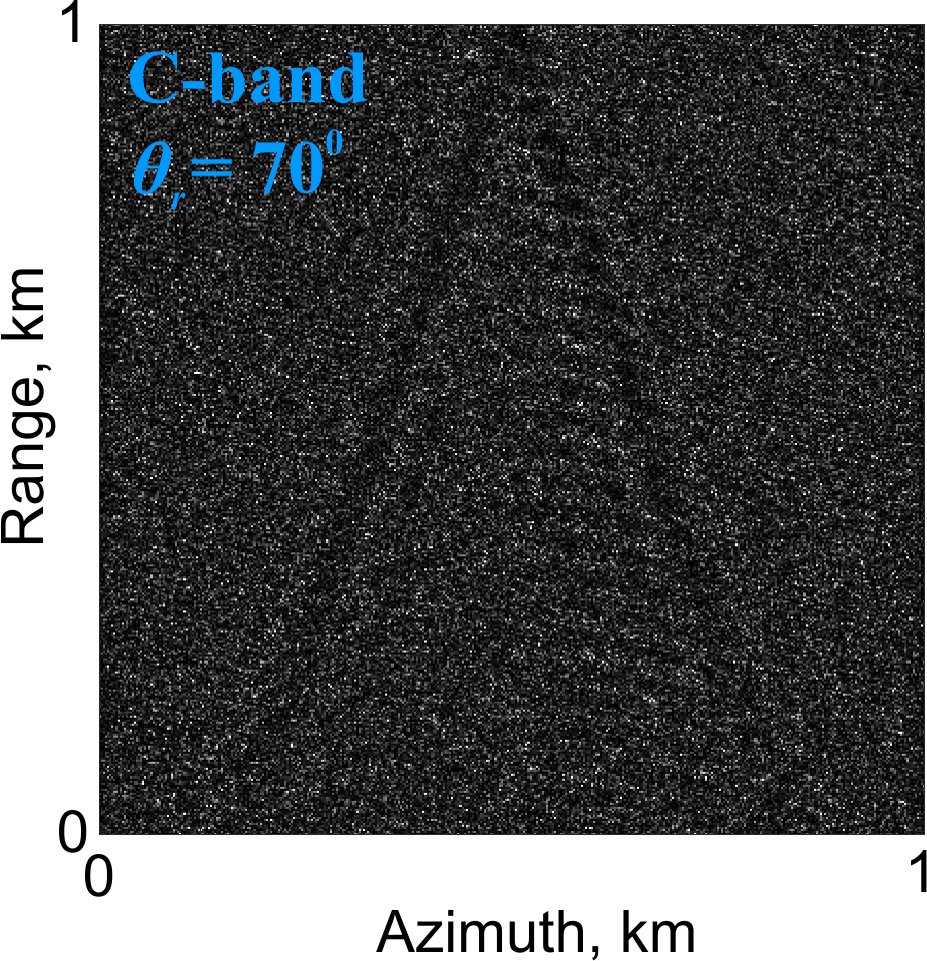}}
\subfigure[]{\includegraphics[width=.19\linewidth]{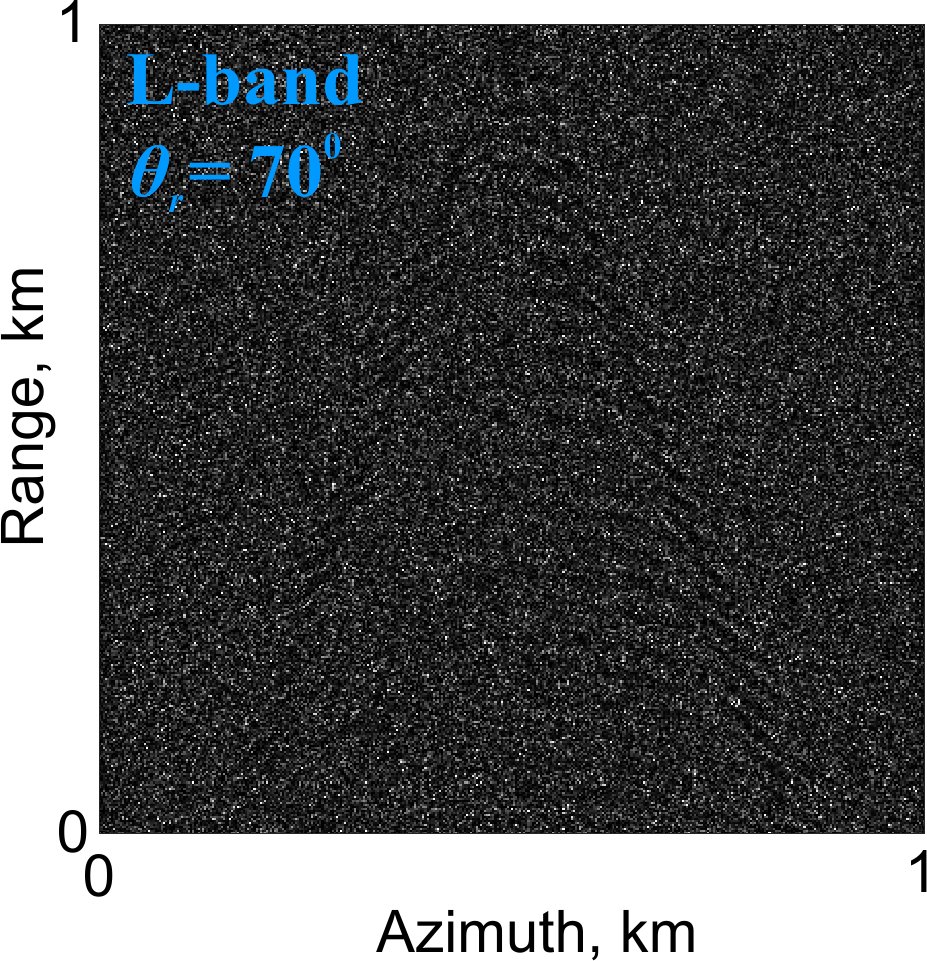}}
\caption{Effect of the incidence angle $\theta_r$ on the SAR imaging (VV polarization) of the ship wake at different bands for AI platform: (a), (b), (c) X-band. (d) C-band. (e) L-band. \textbf{Ship I} with Fr = 0.35. ${{V}_{w10}}$ = 3.5 m/s and $D_{E}$ = $45^{\circ}$.}
\label{fig:fig9}
\end{figure*}

In this subsection, we highlight the main features of SAR imaging for both sea and ship waves. For most presented models, we employed the $S_E$ spectrum, except for in Fig. \ref{fig:fig11} where, in order to better illustrate the velocity bunching effect, the $S_{PM}$ is combined with the Longuet-Higgins directional function. Since a lower wind velocity is better for wake imaging \cite{panico2017sar}, a wind velocity of $V_w = 3.5$ m/s was applied, with wind direction of $D_E = 45^{\circ}$ (except where other parameters are specified). 
The specific visualization of a Kelvin wake pattern has been studied before, and it is well known that better imaging of the ship wake results can be obtained when local waves travel along the radar line-of-sight. Fig. \ref{fig:fig8} illustrates RAR images for three main ship heading directions relative to the platform flight direction, and the visualization results match the results obtained in previous studies \cite{hennings1999radar, panico2017sar}. It has typically been thought \cite{alpers1998radar, hennings1999radar, panico2017sar, graziano2017performance} that HH polarization improves the detection of wakes (includes the effect of a larger magnitude of the tilt modulation transfer function for the HH polarization), though the larger difference in signal intensity between VV and HH polarization occurs at high incidence angles and less at low incidence angles. However, since a low incidence angle generally provides much better imaging of wake, this reduces the importance of the difference in signal intensity between VV and HH polarization.
The next illustration in Fig. \ref{fig:fig9} shows the effect of the different radar incidence angles $\theta_r$ on the visibility of the ship wake. It can be seen that the best visualization of ship wake (divergent and transverse waves) occurs when the incidence angle is smaller (Fig. \ref{fig:fig9}-(a)) and that cusp waves are mainly visible at high incidence angles (Fig. \ref{fig:fig9}-(c)). This effect has previously been noted in studies \cite{tings2018comparison, hennings1999radar, graziano2017performance}.
Although much research has been done in the field of SAR imaging of vessel signatures, there is no clear understanding of the effect of radar frequency on the imaging of ship wakes. In some studies \cite{karakus2020tgrs, tings2018comparison, velotto2016first}, real SAR data for different bands were analyzed, and an overall conclusion has been reached that ship wakes are better imaged in the X-band SAR images (e.g. TerraSAR-X) instead of the C/L-bands. It is clear that the best visibility of ship wakes is associated with the lower altitude of TerraSAR-X, compared with other satellites, which reduces the $R/V$ ratio. It is important to note that for an objective comparison, it would be necessary that all parameters be fixed, i.e. the same satellite platform, the same ship parameters, the same ambient sea waves amplitude, and direction, etc., which in practice is very hard to achieve. The comparison of different frequencies in \cite{alpers1998radar} has shown that Kelvin arms are best visible at high radar frequencies. In contrast, the simulation results in \cite{oumansour1996multifrequency} have shown better observation of ship wake at L-band rather than X-band. In studies such as \cite{hennings1999radar, pichel2004ship}, it is stated that variation in wake visualization had little dependence on radar frequency. According to our results, there is no significant difference in wake visualization for changes in radar frequency, except for at high incidence angle for L-band in Fig. \ref{fig:fig9}-(e) where the visibility of divergent and transverse waves is a little better than for X-band and C-band (but not cusp waves) at the equivalent angle (Fig. \ref{fig:fig9}-(c), (d)) and is close to X-band at lower incidence angle in Fig. \ref{fig:fig9}-(b). However, this result is based on the simulation parameters used and is not a general conclusion, so further attention is needed here.

\begin{figure*}[ht]
\centering
\subfigure[]{\includegraphics[width=.3\linewidth]{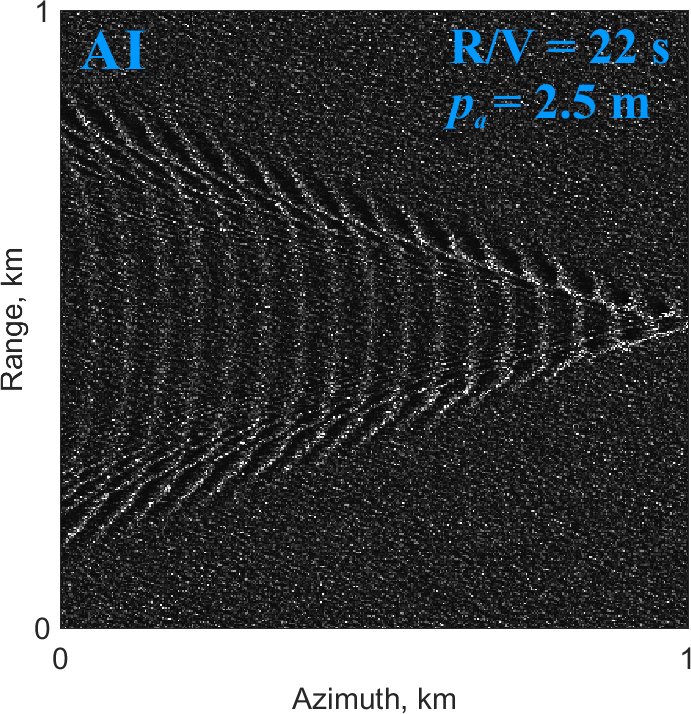}}
\subfigure[]{\includegraphics[width=.3\linewidth]{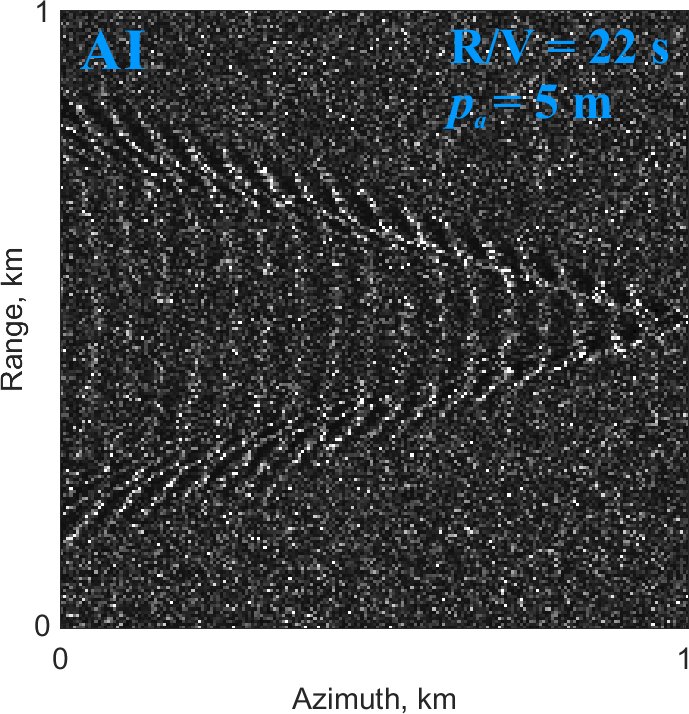}}
\subfigure[]{\includegraphics[width=.3\linewidth]{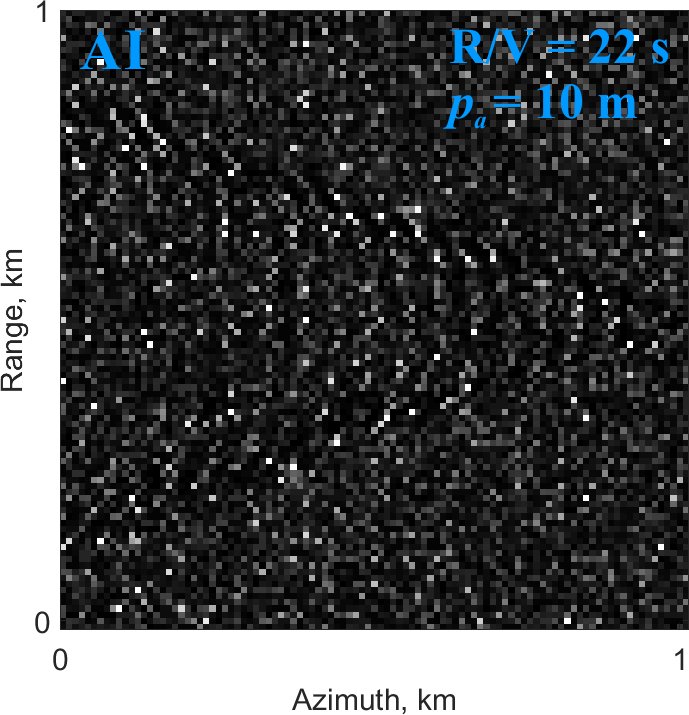}}
\subfigure[]{\includegraphics[width=.3\linewidth]{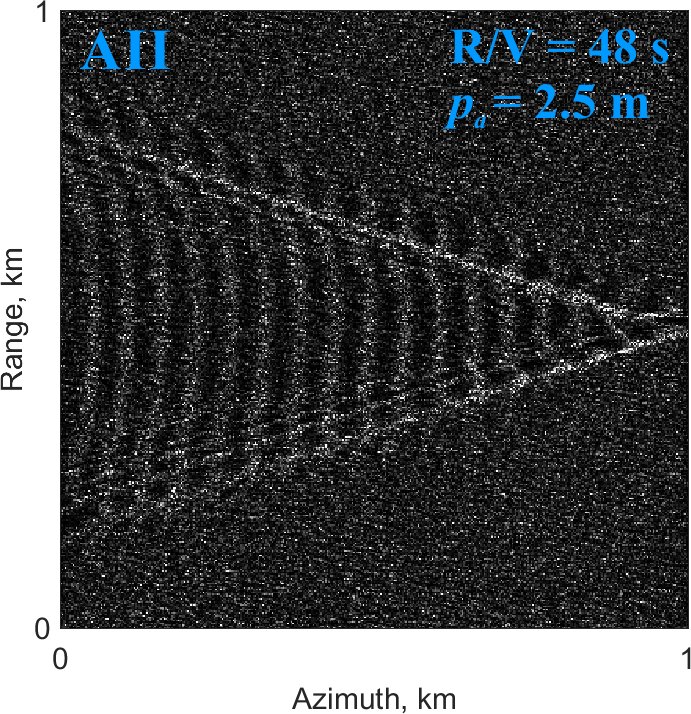}}
\subfigure[]{\includegraphics[width=.3\linewidth]{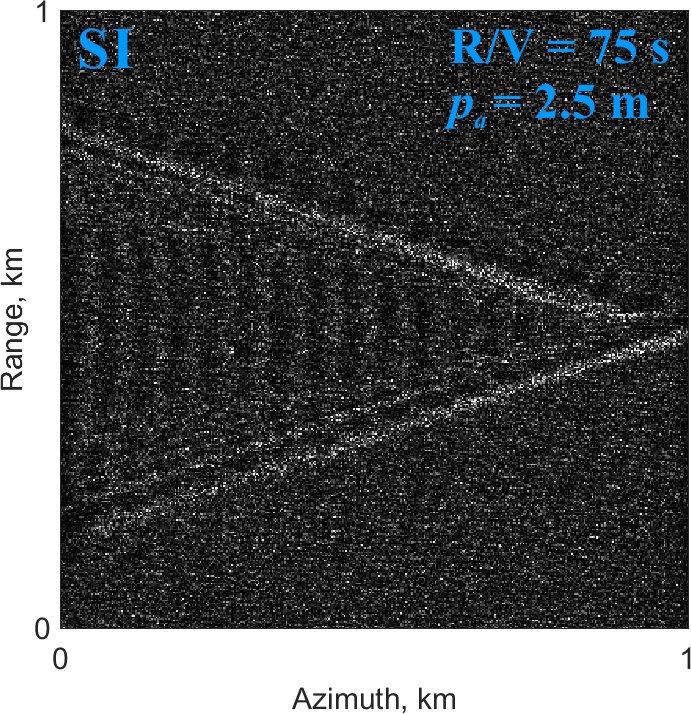}}
\subfigure[]{\includegraphics[width=.3\linewidth]{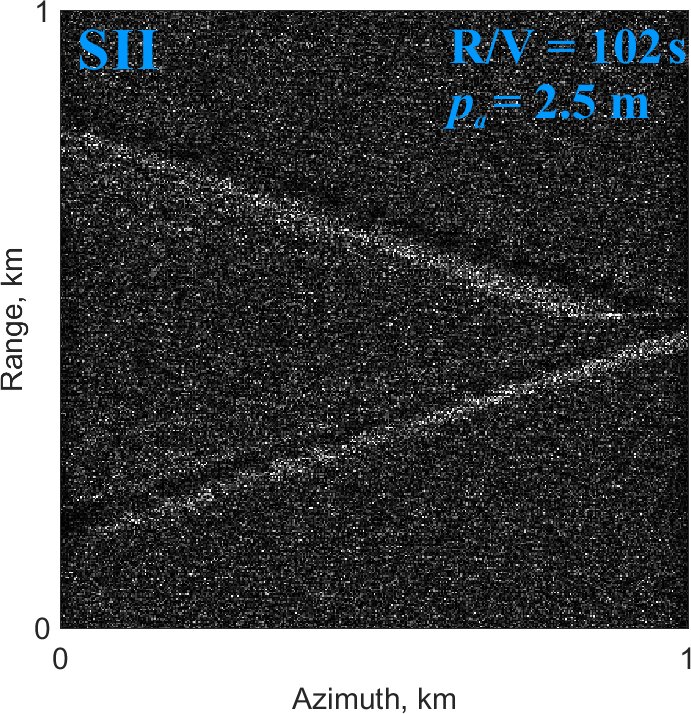}}
\caption{Effect of the different SAR image resolutions $p_a$ (a), (b), (c) and $R/V$ ratios (a), (d), (e), (f) on the SAR imaging (X-band, $\theta_r = 25^{\circ}$, HH polarization) of the ship wake. \textbf{Ship I} with Fr = 0.5. ${{V}_{w10}}$ = 3.5 m/s and $D_{E}$ = $45^{\circ}$.}
\label{fig:fig10}
\end{figure*}

\begin{figure*}[htbp]
\centering
\subfigure[]{\includegraphics[width=.23\linewidth]{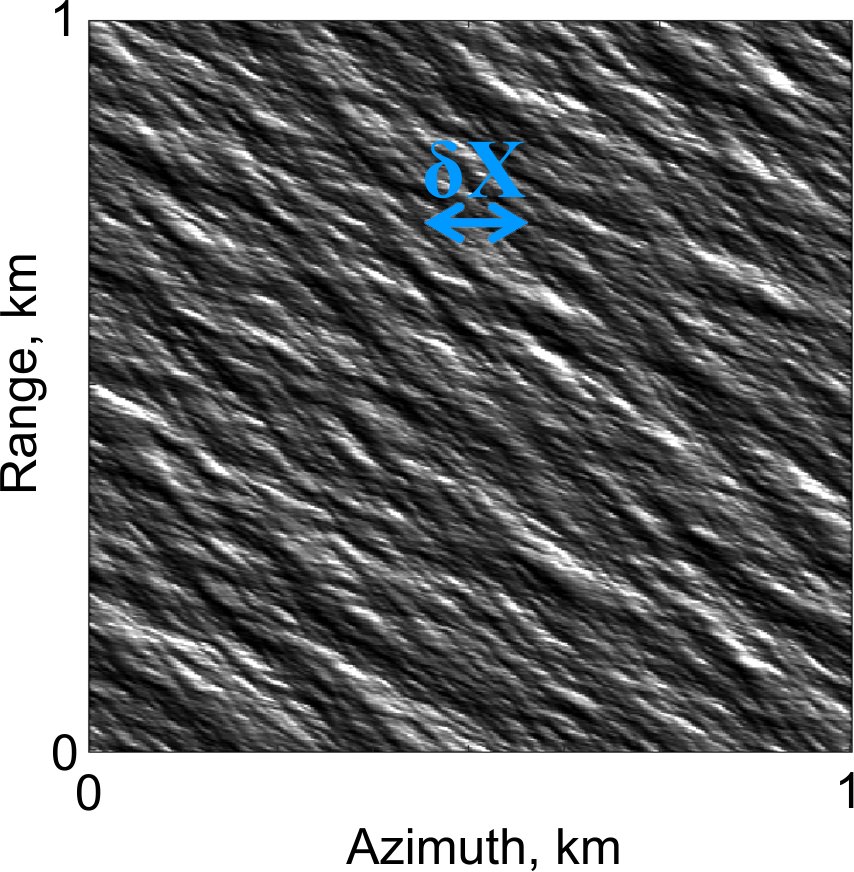}}
\subfigure[]{\includegraphics[width=.23\linewidth]{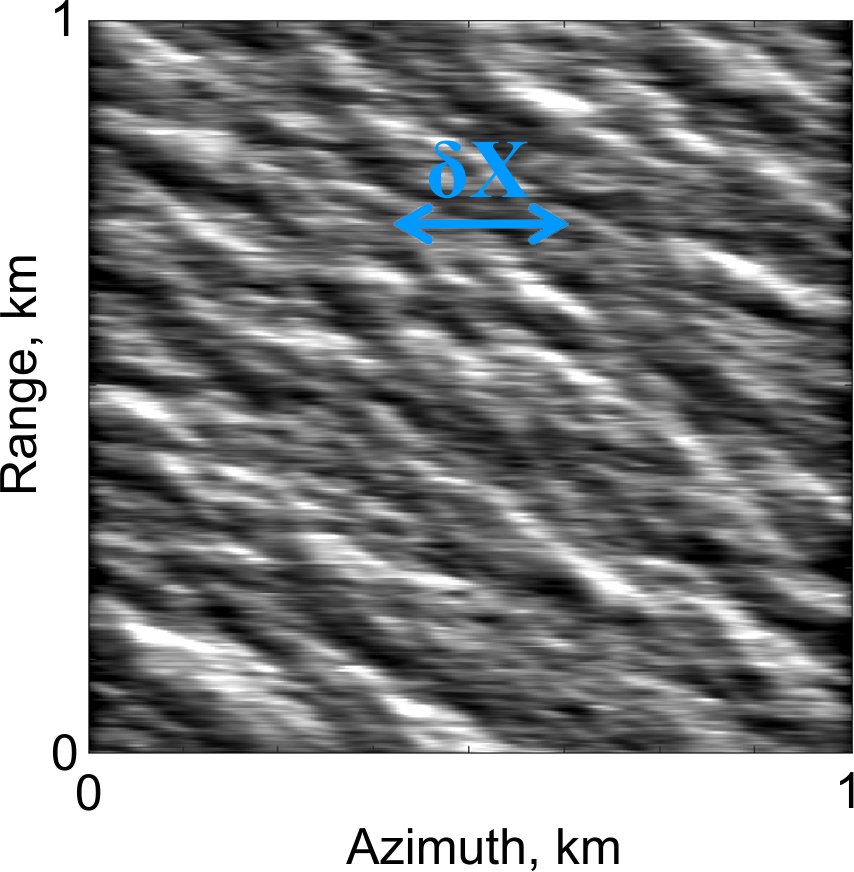}}
\subfigure[]{\includegraphics[width=.23\linewidth]{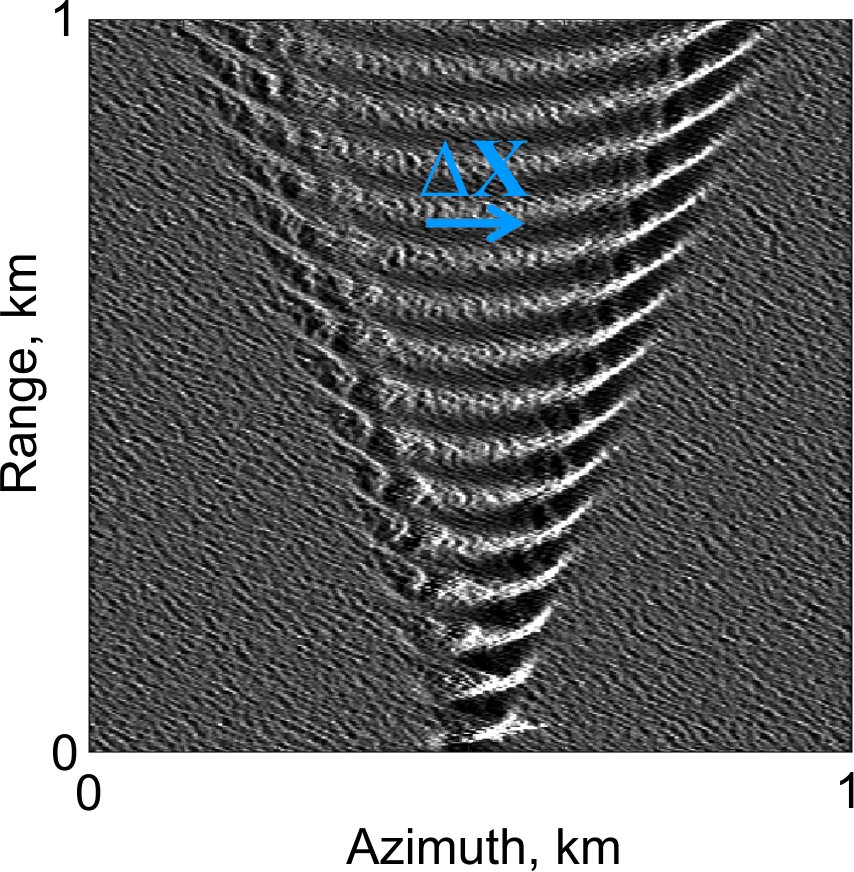}}
\subfigure[]{\includegraphics[width=.23\linewidth]{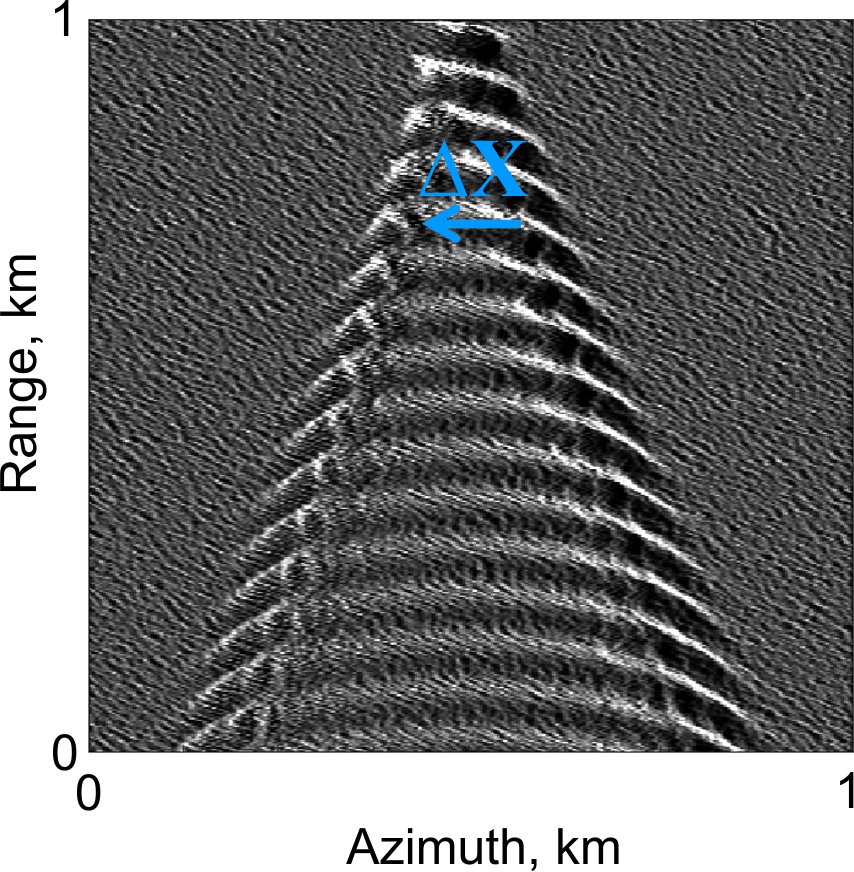}}
\caption{The smearing $\delta X$ and shifting $\Delta X$ effects in speckle-free simulated SAR images (X-band, $\theta_r = 35^{\circ}$. VV polarization). (a) AI platform and (b) SII platform with ${{\bar{U}}_{r}}$ = 0, ${{\bar{A}}_{r}}\neq 0$, ${{V}_{w10}}$ = 13.5 m/s, $S_{PM}$ with $D_{J}$ = $45^{\circ}$ where parameter S = 20. (c) and (d) AI platform with ${{\bar{U}}_{r}}\neq 0$, ${{\bar{A}}_{r}}$ = 0, ${{V}_{w10}}$ = 3.5 m/s, $S_{PM}$ with $D_{J}$ = $45^{\circ}$, and \textbf{Ship III} with Fr = 0.4.}
\label{fig:fig11}
\end{figure*}

In order to clearly demonstrate the effect of the SAR resolution cell, the SAR parameters were fixed and realized for the airborne (AI) scanning platform. It can be seen that at the detailed resolution in Fig. \ref{fig:fig10}-(a) all wake components are easily distinguishable. However, when the resolution is degraded (Fig. \ref{fig:fig10}-(c)), the wake becomes difficult to distinguish.

For the RAR imaging (Fig. \ref{fig:fig8}) the transverse and divergent waves are most visible when the ship is moving in the radar range direction, and only divergent waves exist for the azimuth direction. This effect generally also holds for the SAR imaging, although it is sometimes modified according to the SAR parameters relating to the properties of the sea and ship waves. For example, the transverse waves may still appear in radar images for a ship heading parallel to azimuth direction, at the small values of $R/V$ ratio and lower $V_w$ (Fig. \ref{fig:fig10}-(a)), mainly for airborne SAR platforms. This is due to the contribution of the orbital motion of the gravity waves to the VB imaging, and particularly where the minimum discernible azimuthal wavelength $\lambda_{min}$ in eq. (\ref{equ:57}) is less than the wavelength of transverse waves. However, for the satellite platforms, transverse waves begin to show smearing, for instance, in Fig. \ref{fig:fig10}-(e), and are completely invisible at the large $R/V$ ratio in Fig. \ref{fig:fig10}-(f). This also well explains why for the Sentinel-1 satellite, which has a large $R/V$ ratio, the images mainly display cusp waves only. This had previously been ascribed to the lower SAR resolution of the Sentinel-1 images, but our results show that this is not the sole contributor. Looking ahead, the contribution of the sea wave amplitude factor can also influence visualization of ship wake, which is demonstrated in Fig. \ref{fig:fig14}-(a), (e) and (b), (f).

Importantly, the SAR imaging of moving waves is related to the velocity bunching mechanism. The main image degradation effects consist in azimuthal shifting determined as \cite{lyzenga1986numerical}

\begin{align}\label{equ:60}
    \Delta X={R{{U}_{r}}}/{V}
\end{align} 
and smearing given as

\begin{align}\label{equ:61}
    \delta X={2R{{\sigma }_{u}}}/{V}
\end{align} 
where $\sigma_u$ is the standard deviation of the radial velocities within a SAR resolution cell.

In order to better demonstrate these effects, the simulations are presented as speckle-free images in Fig. \ref{fig:fig11}. Two different sea states with low $V_{w10} = 3.5$ m/s and relatively high $V_{w10} = 13.5$ m/s with waves traveling at $45^{\circ}$ are presented in Fig. \ref{fig:fig11}-(c), (d) and (a), (b), respectively, based on the $S_E$ spectrum and Longuet-Higgins directional function. While keeping the same sea surface model ($V_{w10} = 13.5$ m/s), the smearing is minimal for AI platform (a), and increased for SII platform (b). Next, the shifting is associated with the contribution of the radial velocity of the model's facets, and shifting direction can be easily determined \cite{lyzenga1985sar}. Here an important role is played by whether one has a right or left looking antenna SAR configuration. When facets move toward the SAR platform, then they are imaged as shifted in the flight direction (Fig. \ref{fig:fig11}-(c)), and reversely (Fig. \ref{fig:fig11}-(d)). It should be noted that shifting and smearing effects exist for all simulated scenarios, and a separate illustration of these effects in Fig. \ref{fig:fig11} is made to facilitate understanding.

\subsection{Spectra comparison: sea surface slopes vs PDF}
Many studies devoted to sea surface modeling and SAR imagery simulation do not pay due attention to the correctness of the sea surface modeling step. It is assumed, for example, that the real sea surface roughness is well enough approximated by a spectrum model. Therefore, we present a basic approach to validate the spectra models. A comparison of the probability distribution of sea surface slopes with the well known Cox and Munk probability density function (PDF) \cite{cox1954measurement, cox1956slopes} was performed. This PDF is based on the Gram-Charlier distribution as:

\begin{dmath}\label{equ:62}
p={{\left( 2\pi {{\sigma }_{c}}{{\sigma }_{u}} \right)}^{-1}}\exp \left[ -\frac{1}{2}\left( {{\xi }^{2}}+{{\eta }^{2}} \right) \right]\\\bigg\{ 1-\frac{1}{2}{{c}_{21}}\left( {{\xi }^{2}}-1 \right)\eta  -\frac{1}{6}{{c}_{03}}\left( {{\eta }^{3}}-3\eta  \right)+\\\frac{1}{24}{{c}_{40}}\left( {{\xi }^{4}}-6{{\xi }^{2}}+3 \right)+\frac{1}{4}{{c}_{22}}\left( {{\xi }^{2}}-1 \right)
 \left( {{\eta }^{2}}-1 \right)+\frac{1}{24}{{c}_{04}}\left( {{\eta }^{4}}-6{{\eta }^{2}}+3 \right) \bigg\}
\end{dmath}
where $\xi = Z_y / \sigma_c$ and $\eta = Z_x / \sigma_u$ are normalized crosswind and upwind slope components, respectively (up to $\xi = \eta = 2.5$), with local surface slopes $Z_x$ and $Z_y$ (here we assume $x$ is upwind, and $y$ is crosswind, direction as in \cite{chan1977theory}), and rms slope components are $\sigma_u$ and $\sigma_c$. The latter are expressed as

\begin{align}\label{equ:63}
    \sigma _{u}^{2}&=3.16\times {{10}^{-3}}{{V}_{w12.5}}\\
    \sigma _{c}^{2}&=0.003+1.92\times {{10}^{-3}}{{V}_{w12.5}}
\end{align} 
The skewness coefficients for the clean sea surface are presented

\begin{align}\label{equ:64}
    {{c}_{21}}&=0.01-0.0086{{V}_{w12.5}}\\
    {{c}_{03}}&=0.04-0.033{{V}_{w12.5}}
\end{align}
and the peakedness coefficients are given

\begin{align}\label{equ:65}
    {{c}_{40}}&=0.4 & {{c}_{22}}&=0.12 & {{c}_{04}}&=0.23
\end{align}

\begin{figure*}[htbp]
\centering
\subfigure[]{\includegraphics[width=.19\linewidth]{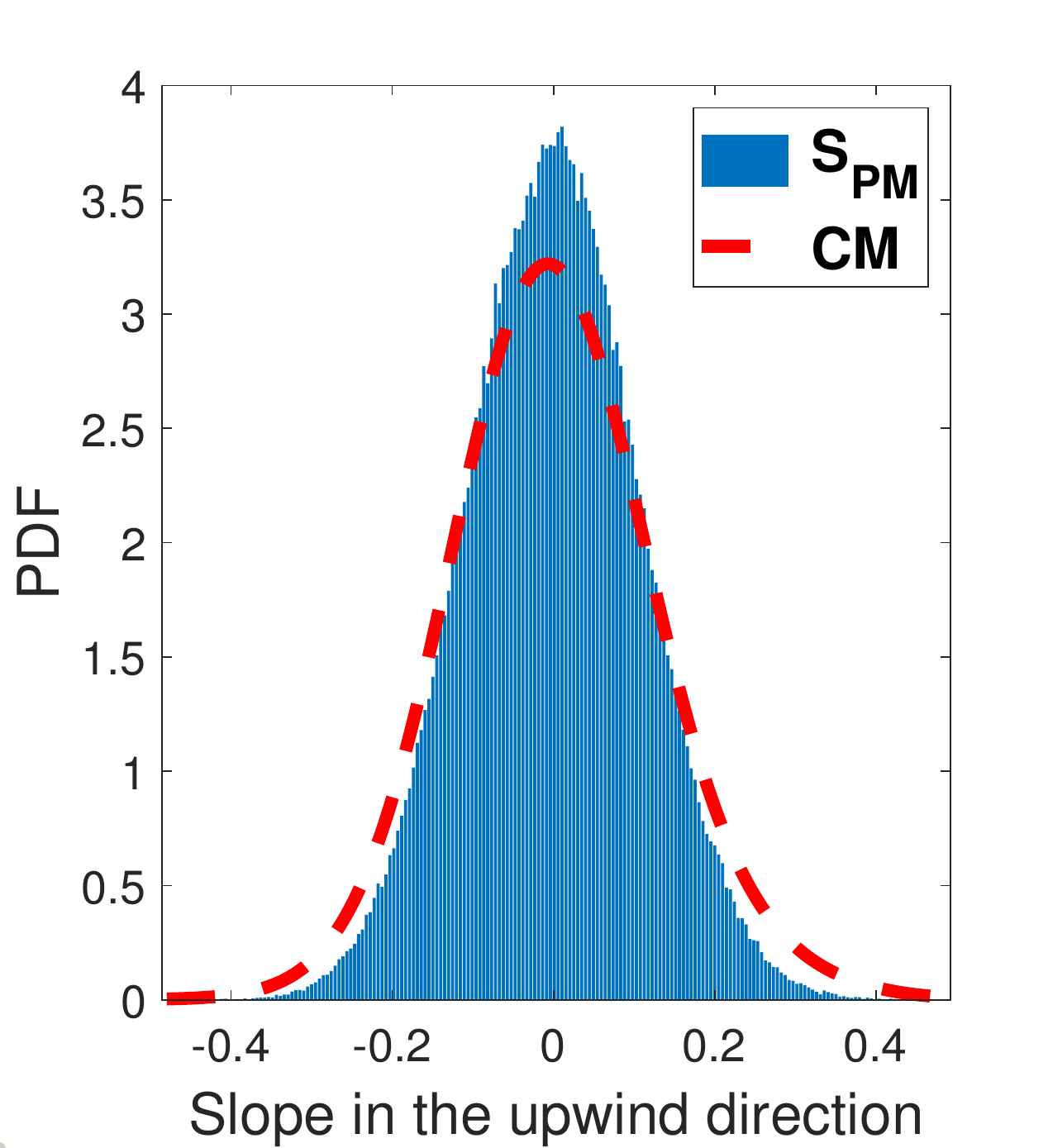}}
\subfigure[]{\includegraphics[width=.19\linewidth]{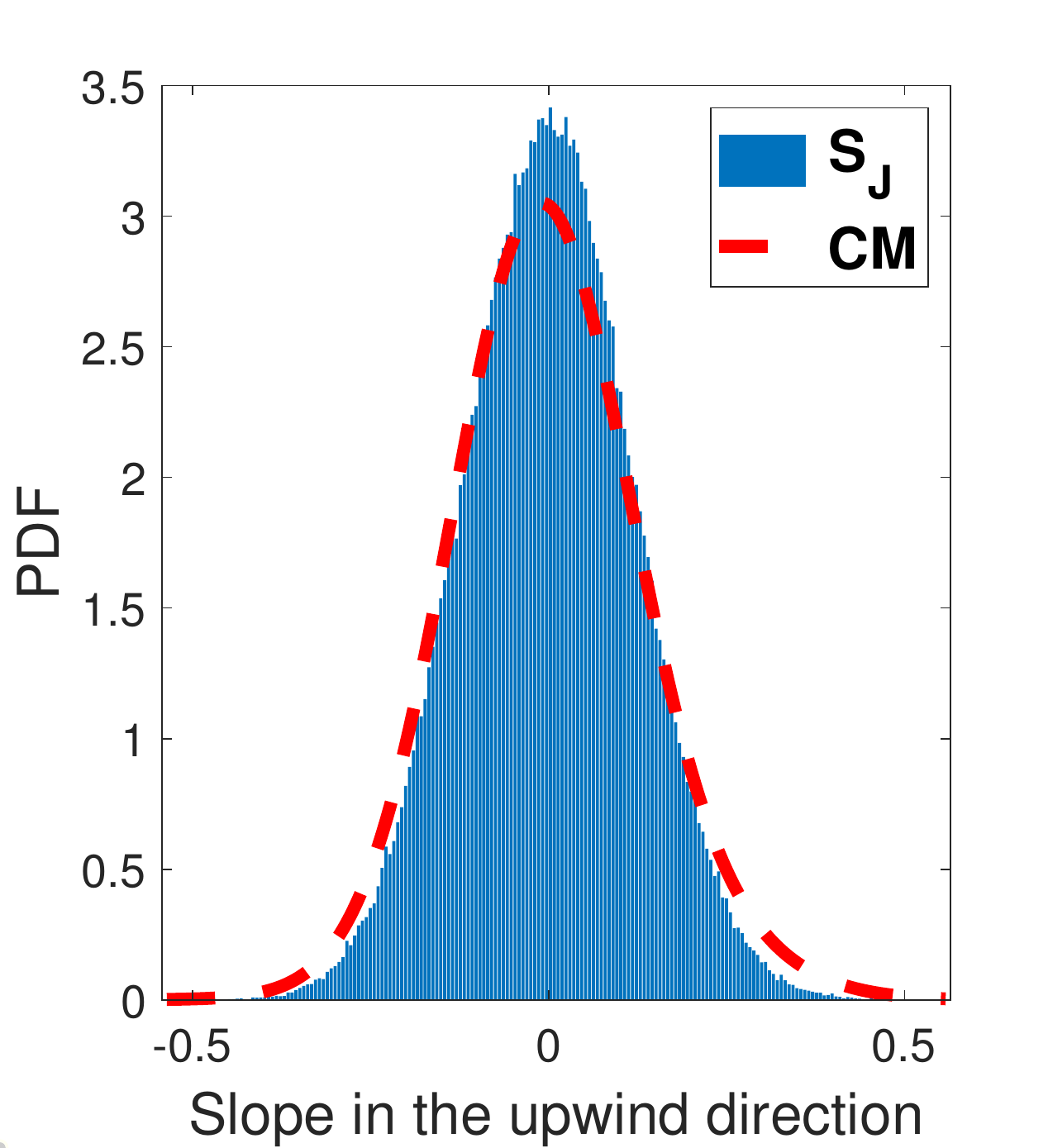}}
\subfigure[]{\includegraphics[width=.19\linewidth]{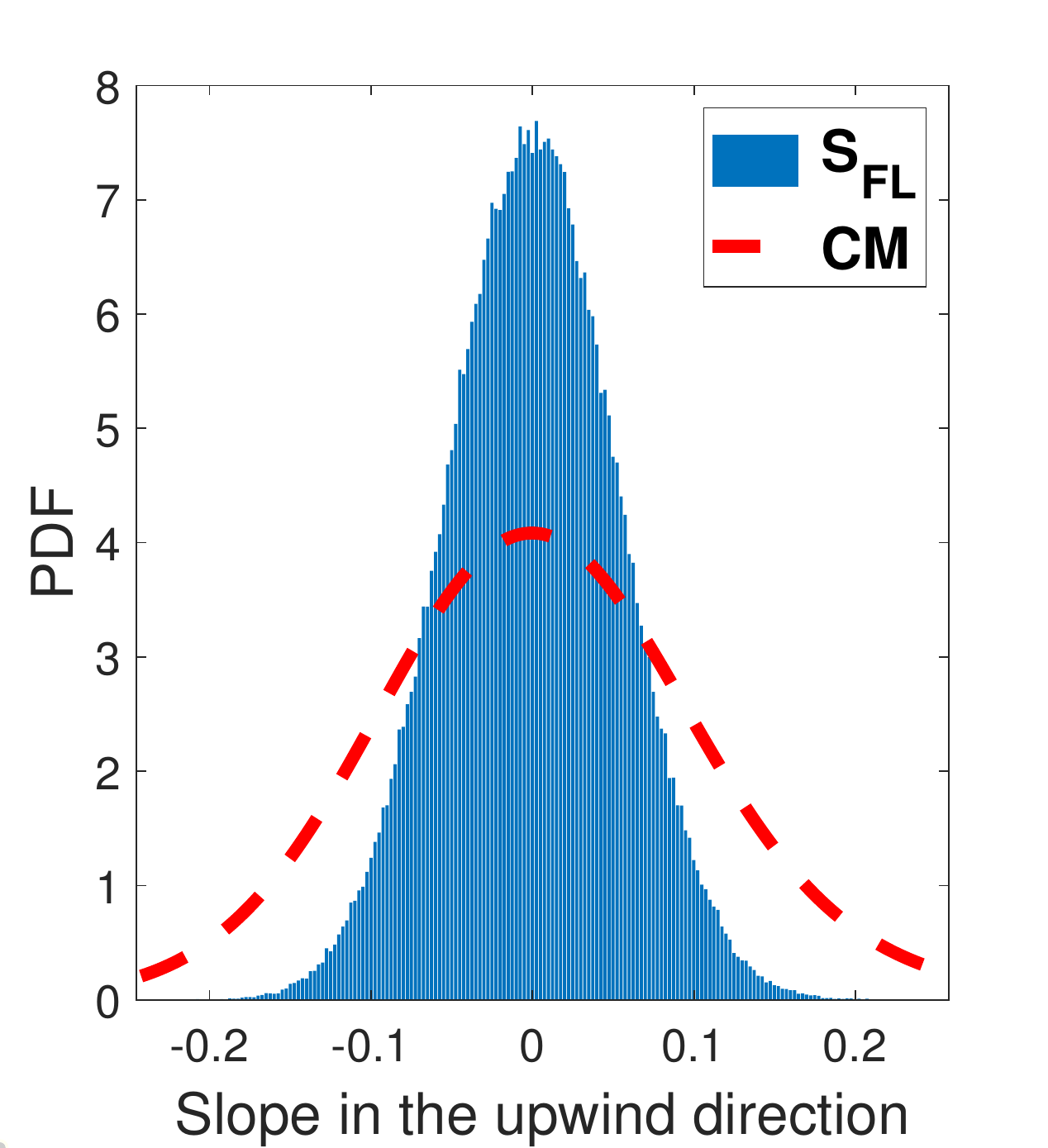}}
\subfigure[]{\includegraphics[width=.19\linewidth]{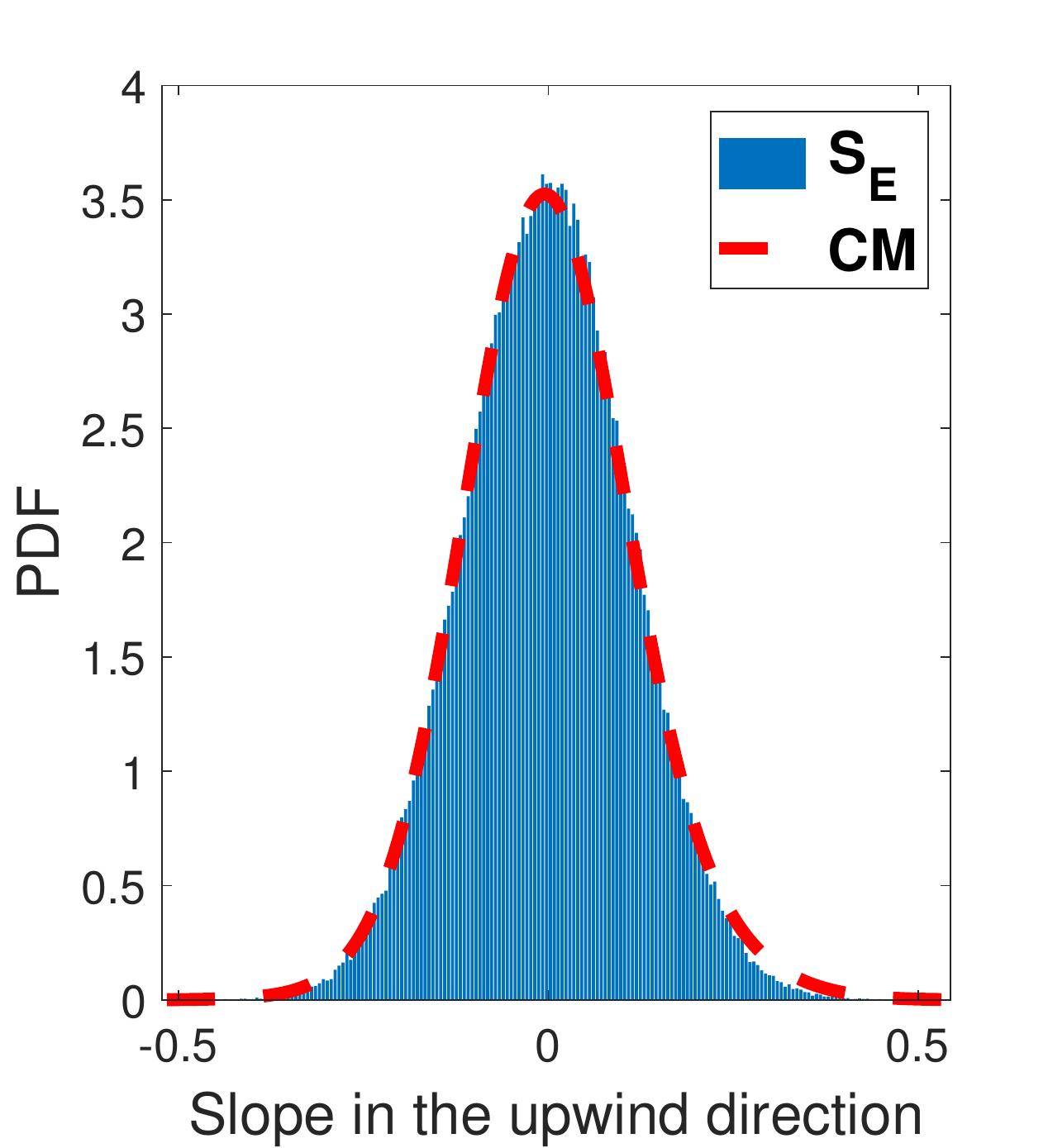}}
\subfigure[]{\includegraphics[width=.19\linewidth]{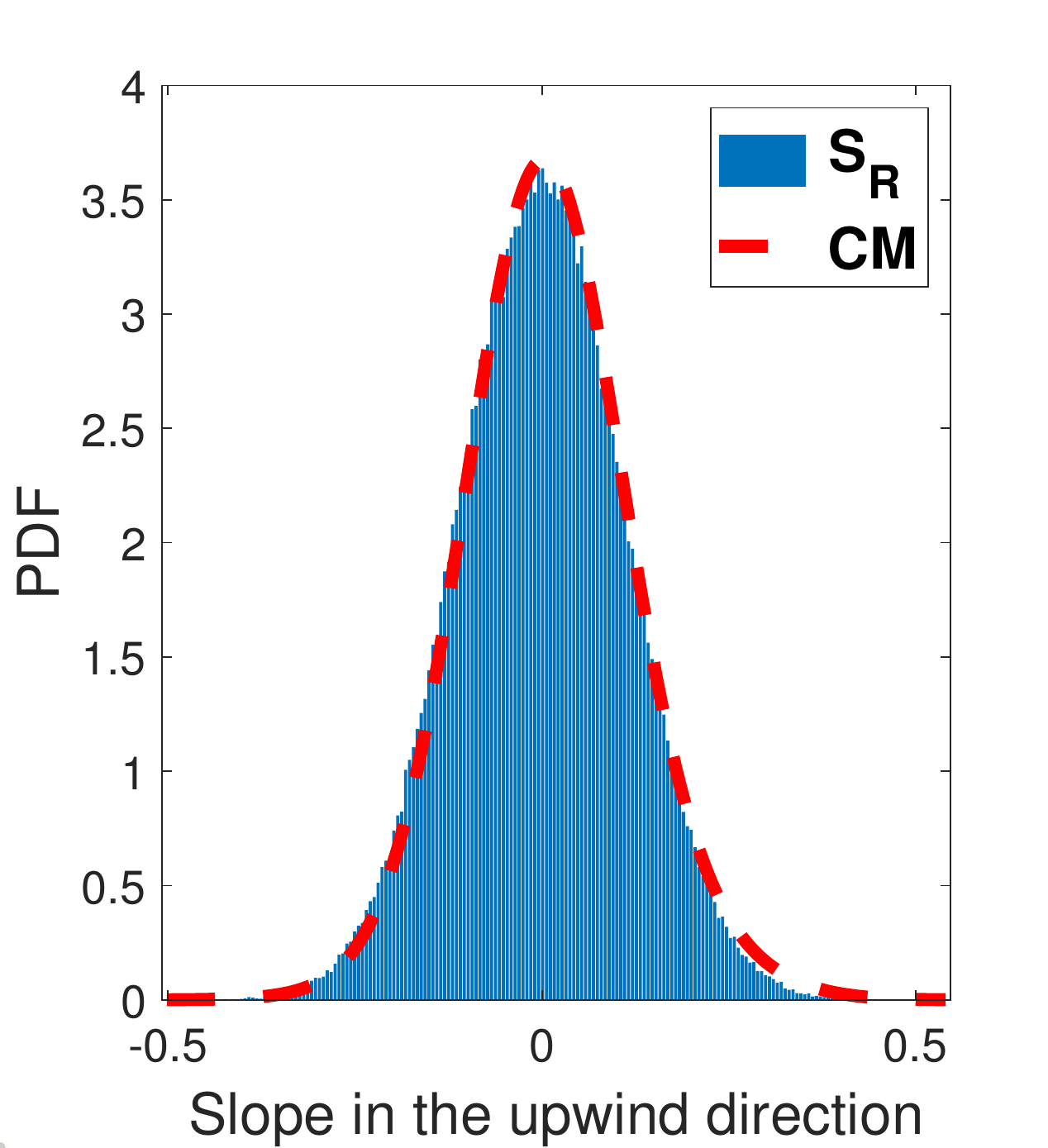}}
\caption{Density comparison for different spectra (a) $S_{PM}$, (b) $S_{J}$ with fetch = 80 km, (c) $S_{FL}$, (d) $S_{E}$ with $\Omega$ = 0.84, and (e) $S_{R}$. Dashed-lines in each sub-figures refer to the Cox and Munk's (CM) model. The wind velocity ${{V}_{w10}}$ = 8 m/s and the size of the sea surface model is 0.25$\times$0.25 km.}
\label{fig:fig12}
\end{figure*}

It is clear that with decreasing facet size the coherency of the slope model also deteriorates; this effect is explained in \cite{mobley2015polarized}. However, if the facet size is increased, for example, to 0.5 m as in \cite{zhang2017electromagnetic}, the distribution of slopes will generally correspond to the PDF model, which can be used for verification. When the facet size is increased to a higher order, for example 3 mm size (capillary waves) then the slopes are consistent with the PDF with sufficient accuracy \cite{kay2011light}. On the other hand, since we use the two-scale SAR model, and the high-frequency part (short waves) is modeled statistically, there is no need for a high-resolution sea surface model calculation. Therefore, here we apply the facet size of 0.5 m following \cite{zhang2017electromagnetic} to test the overall consistency of all spectra of the generated slope models to Cox and Munk's PDF. The wind speed is set as 8 m/s and the size of the surfaces is 0.25$\times$0.25 km (see Section \ref{sec:surfaceModelling} for $L_{min}$ size selection). The result of this comparison is demonstrated in Fig. \ref{fig:fig12}. As is shown, not all spectra give a reasonable match to the Cox and Munk PDF. Only the wave spectra of $S_{E}$ and $S_{R}$ are matched well enough. However, for the other spectra, this agreement was not observed where $S_{PM}$ and $S_{J}$ showed better agreement than $S_{FL}$. The results, in general, are consistent with results shown in previous studies \cite{zhang2017electromagnetic, linghu2018gpu, zheng2018sea}. For example, in \cite{zheng2018sea} the slope variance of different spectra has also been compared with the slope variances according to the Cox and Munk's model. Also, the agreement with Cox and Munk's PDF likely may depend on the spreading functions, as stated in \cite{massel2011geometry, zheng2018sea}. It should be recalled that for the $S_{E}$, $S_{FL}$ and $S_{R}$ spectra, their original spreading functions were applied, while for the $S_{PM}$ cosine-squared function, and for $S_{J}$ the Longuet-Higgins et al. function was used (Section \ref{sec:Directionalfunction}).

\subsection{Spectra comparison: SAR imaging of ship wakes}
In this subsection, we applied quantitative analysis of visualization of ship wakes in SAR images for different sea wave spectra models. The idea behind the comparison consists in the fact that all the spectra are created by empirical or semi-empirical approaches for particular geographical places and conditions. It means that even under the same modeling parameters (first of all, the velocity of wind), the amplitude and as a consequence significant wave height $H_{s}$ may vary. As is mentioned in Section \ref{subsec:windstate} and in various studies in the literature, the ambient amplitude of sea waves can dramatically reduce, or conversely, increase the visibility of vessel signatures in SAR images. On the other hand, different spectra models have been applied for SAR image simulation of ship wake \cite{tunaley1991simulation, oumansour1996multifrequency, arnold2007bistatic, chen2012facet, sun2013scattering, zilman2014detectability, li2015study, zhao2015bistatic, jiang2016ship, wang2016application, wang2017sar}. It is important to note that the results presented here reveal a relative difference in spectra, in terms of a superposition of sea waves with ship wake and are not compared in terms of absolute SAR intensity values. The comparison is presented in two stages: (i) Determination of the borderline condition; (ii) Assessment in terms of standard statistical measures of imagery.

\begin{figure*}[htbp]
\centering
\subfigure[]{\includegraphics[width=.27\linewidth]{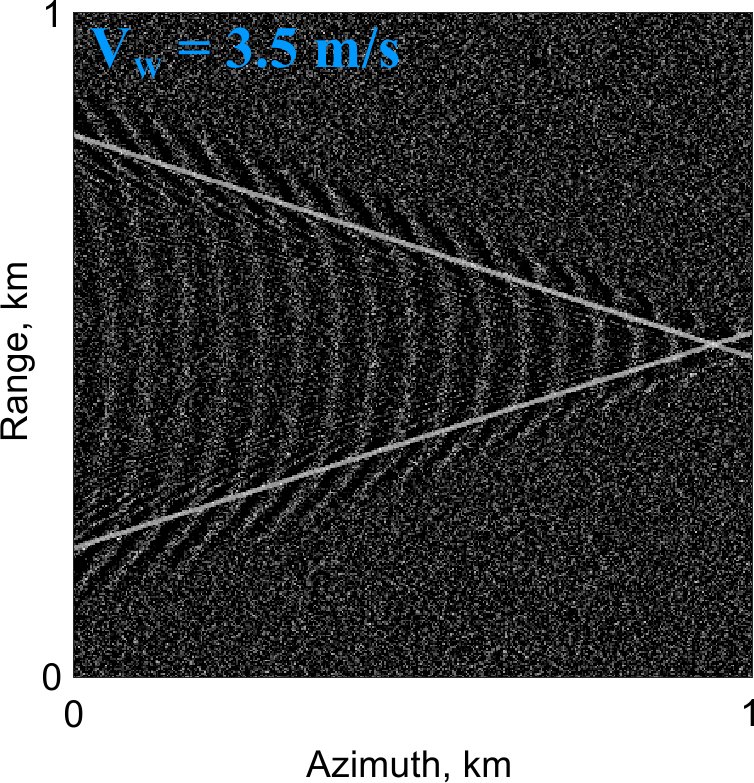}}
\subfigure[]{\includegraphics[width=.27\linewidth]{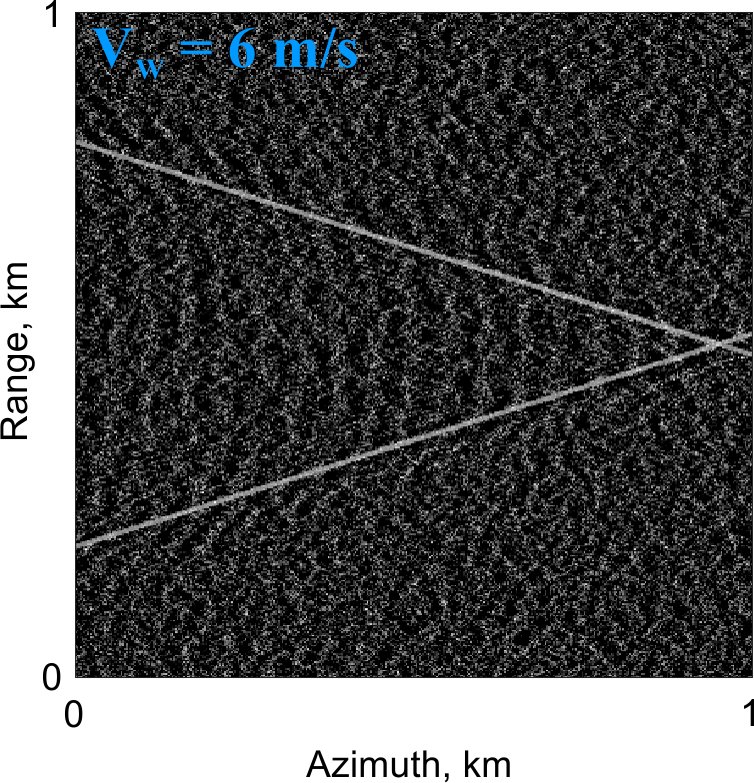}}
\subfigure[]{\includegraphics[width=.27\linewidth]{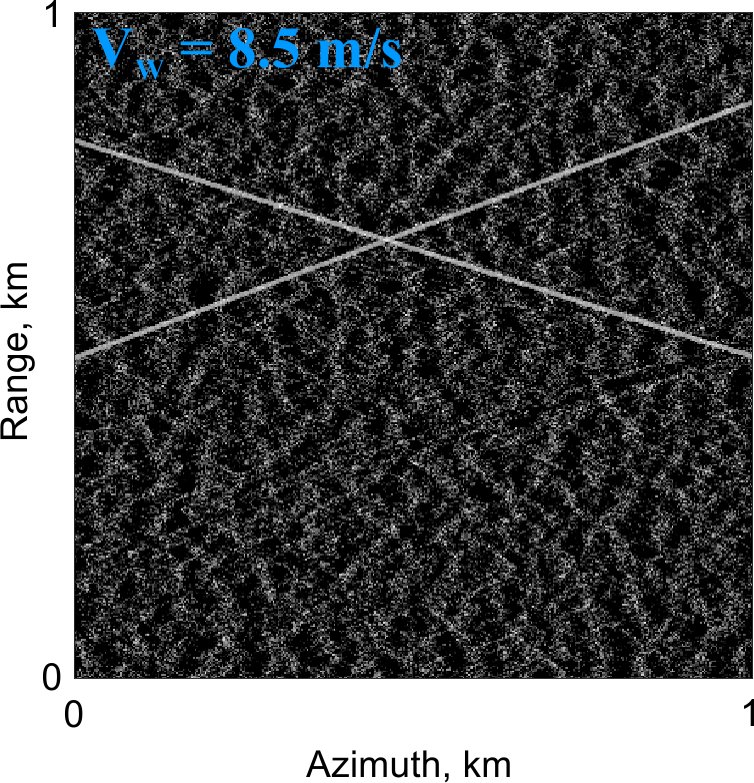}}
\caption{The wake detection results for simulated SAR images (X-band, $\theta_r = 35^{\circ}$, VV polarization) using method based on sparse regularization \cite{karakus2020tgrs} under different velocities of wind: (a) ${{V}_{w10}}$ = 3.5 m/s. (b) ${{V}_{w10}}$ = 6 m/s. (c) ${{V}_{w10}}$ = 8.5 m/s. \textbf{Ship I} with Fr = 0.5. The bright lines represent detected Kelvin envelope lines.}
\label{fig:fig13}
\end{figure*}

In order to provide an evaluation of the contribution of different sea spectra to the SAR imaging of ship wake, we first determined the boundary condition where the wake signatures due to ambient sea waves can disappear or be less noticeable in the SAR image. A recently proposed ship wake detection method based on sparse regularization, and successfully tested on both type of real and simulated SAR images, \cite{karakus2020tgrs} was applied. Noting that the Pierson-Moskowitz spectrum has been the basis of many presented spectra, it was selected as a reference to generate the sea wave models within a range of $V_w = 3.5-11$ m/s. As was mentioned in Section \ref{sec:Spectra}, the conformity of $V_w$ for different spectra was performed by using the Fung and Lee method \cite{fung1982semi}. For the ship model, we selected Ship I with $Fr=0.5$, and calculations are done for the airborne AI platform as the best available SAR configuration with parameters: X-band, incidence angle $\theta_r = 35^{\circ}$ and VV polarization. The speckle model is fixed to an exponential distribution with unit mean for all scenarios. The wind and ship moving directions are tuned to $0^{\circ}$ (azimuth direction). We considered the boundary condition as valid if at least one of the Kelvin arms has not been detected using the method \cite{karakus2020tgrs}. Although, strictly speaking, the boundary condition may change for other SAR platforms and/or ship parameters, this does not reduce the significance of the experiment, since we investigated this approach as a starting point. The results of the detection of ship wake are given in Fig. \ref{fig:fig13} where it is shown that when the sea amplitude reaches $H_s = 1.7$ m ($V_{w10} = 8.5$ m/s), the visualization of the ship's signature in the SAR image becomes difficult, which also affects the detection efficiency (one of the bright lines does not follow the Kelvin envelope line in Fig. \ref{fig:fig13}-(c)).

\begin{figure*}[htbp]
\centering
\subfigure[]{\includegraphics[width=.20\linewidth]{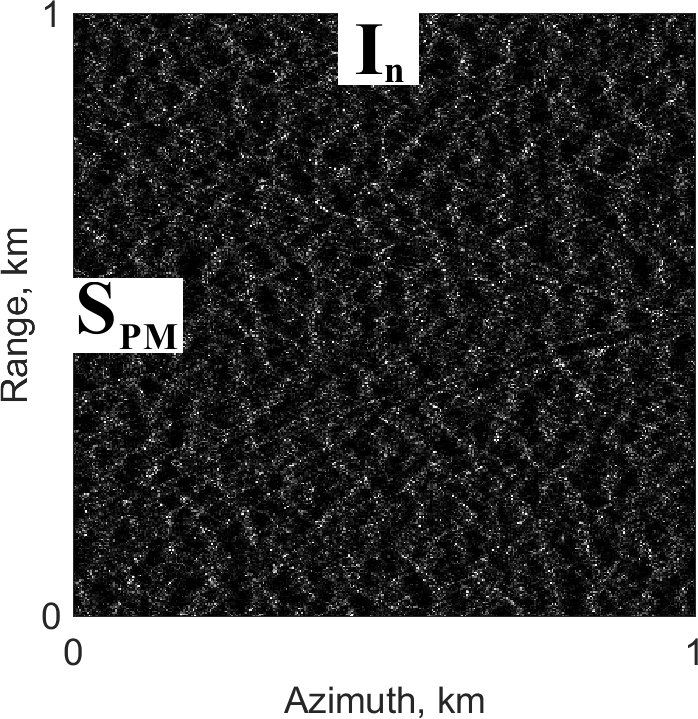}}
\subfigure[]{\includegraphics[width=.20\linewidth]{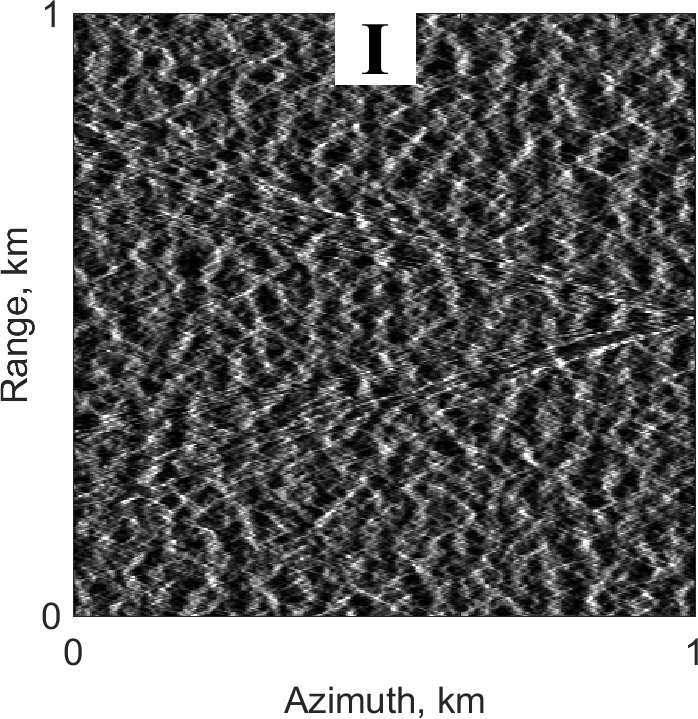}}
\subfigure[]{\includegraphics[width=.23\linewidth]{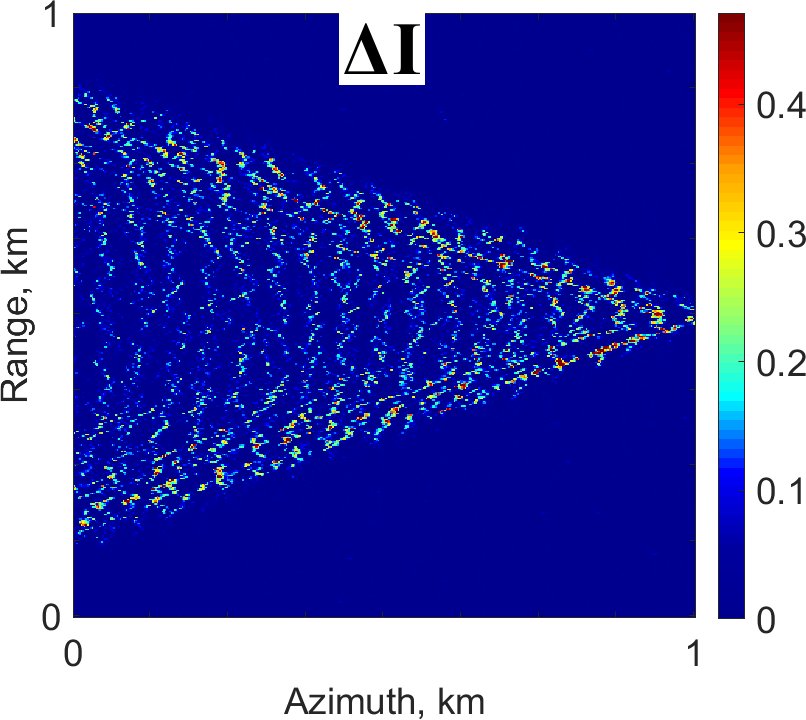}}
\subfigure[]{\raisebox{0.5mm}{\includegraphics[width=.23\linewidth]{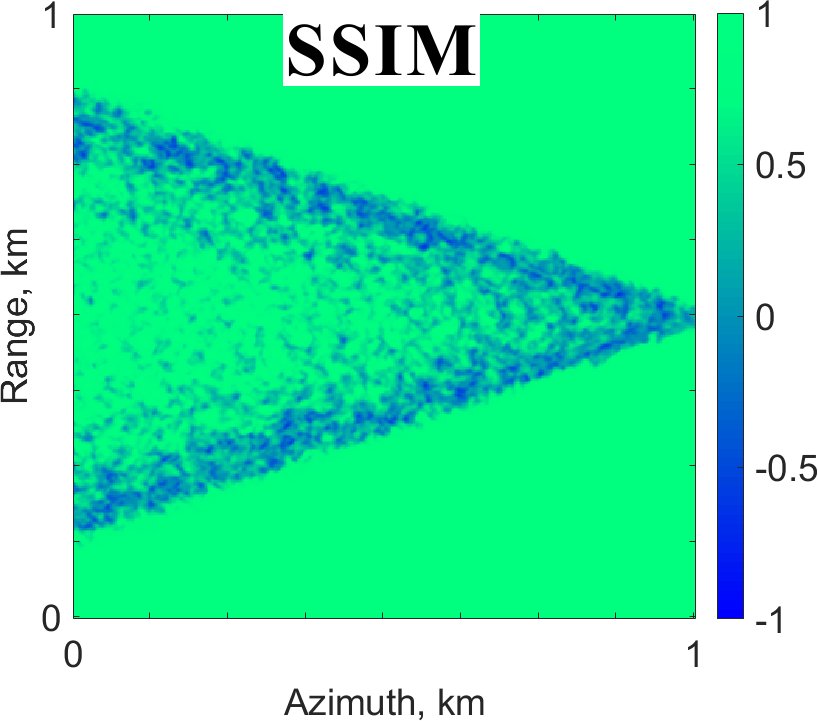}}}
\subfigure[]{\includegraphics[width=.20\linewidth]{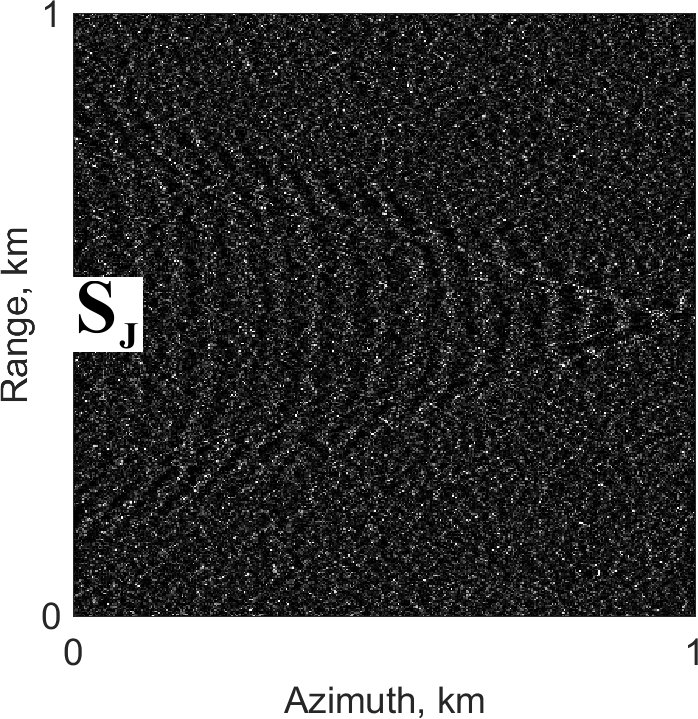}}
\subfigure[]{\includegraphics[width=.20\linewidth]{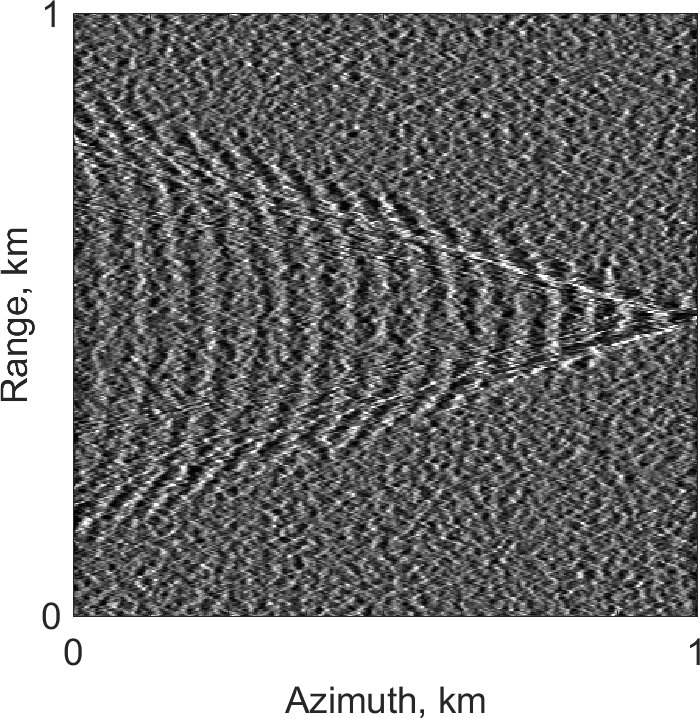}}
\subfigure[]{\includegraphics[width=.23\linewidth]{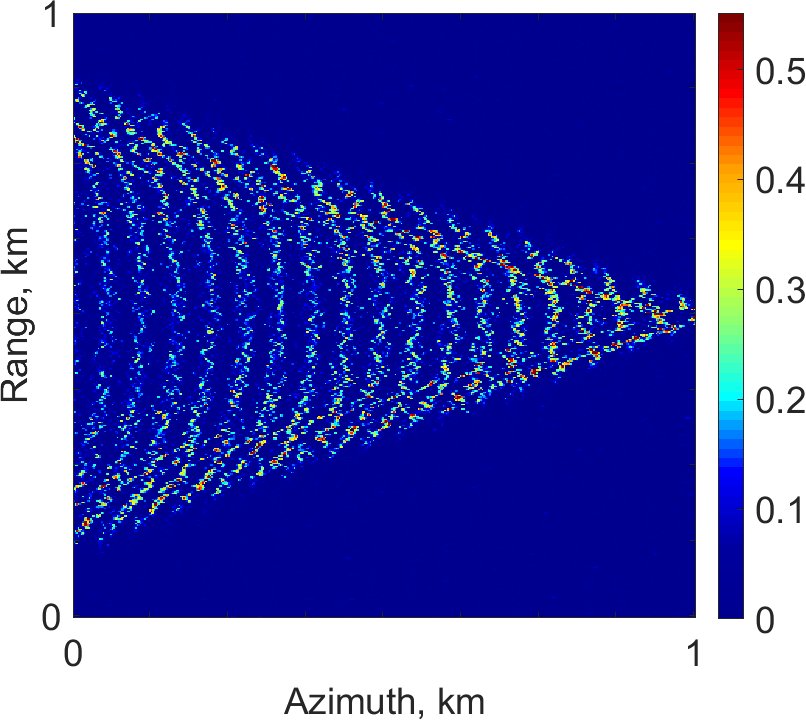}}
\subfigure[]{\raisebox{0.5mm}{\includegraphics[width=.23\linewidth]{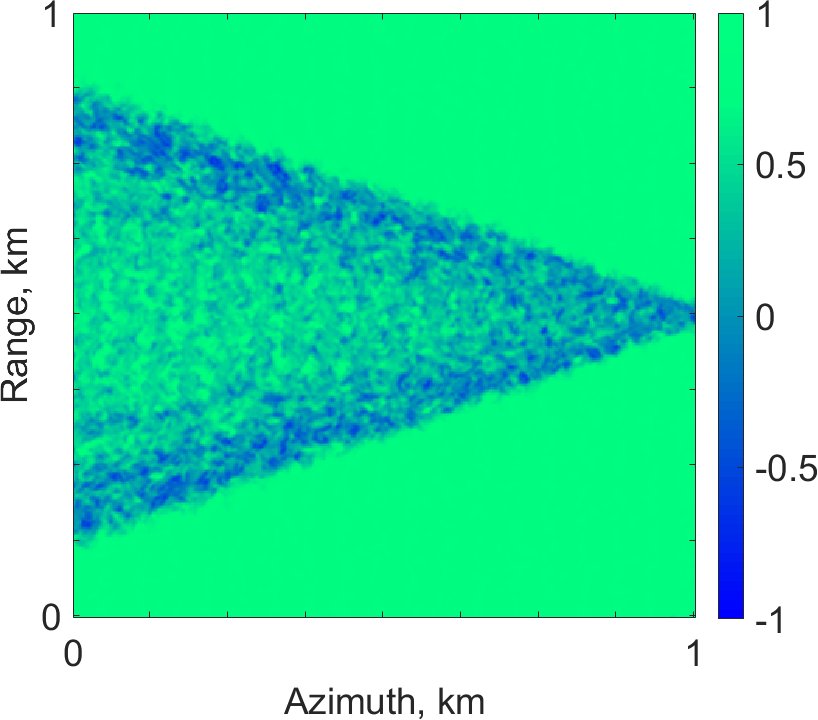}}}
\subfigure[]{\includegraphics[width=.20\linewidth]{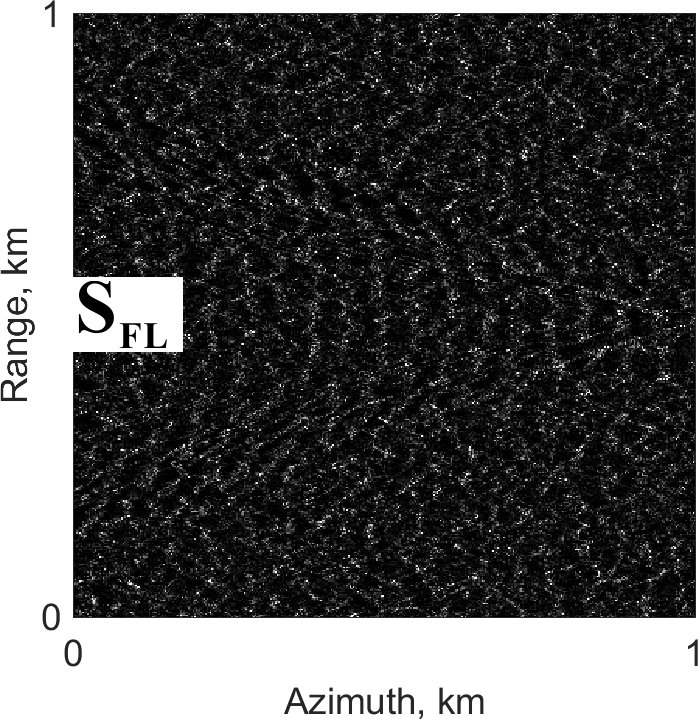}}
\subfigure[]{\includegraphics[width=.20\linewidth]{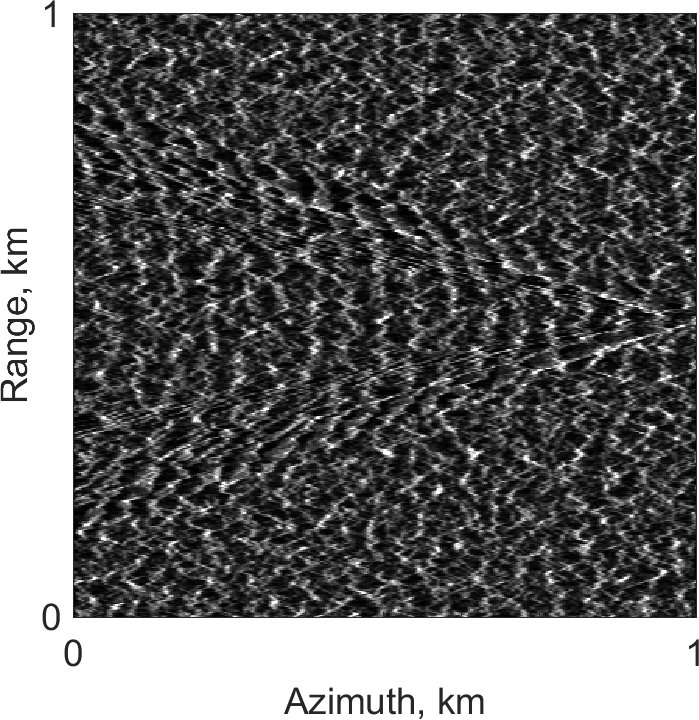}}
\subfigure[]{\includegraphics[width=.23\linewidth]{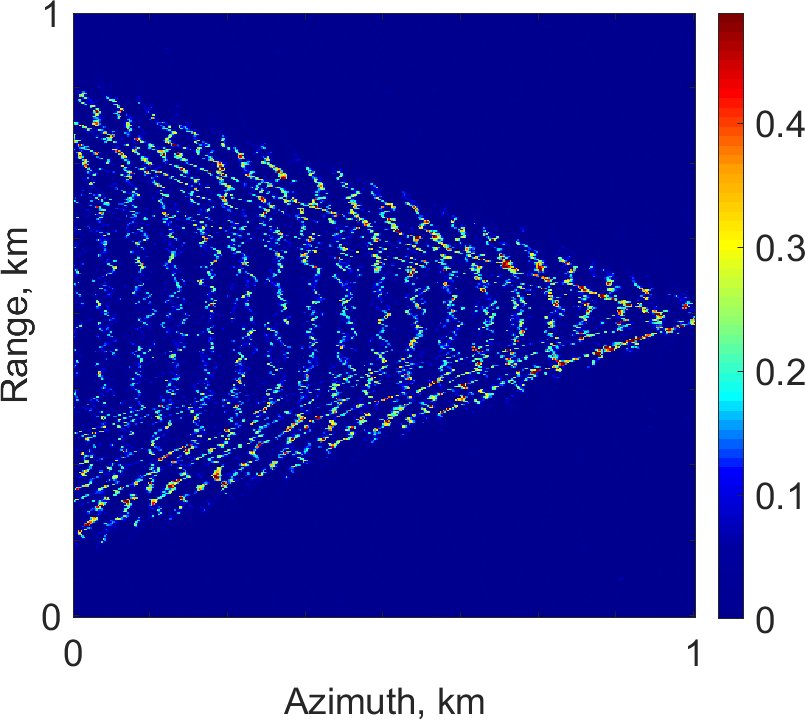}}
\subfigure[]{\raisebox{0.5mm}{\includegraphics[width=.23\linewidth]{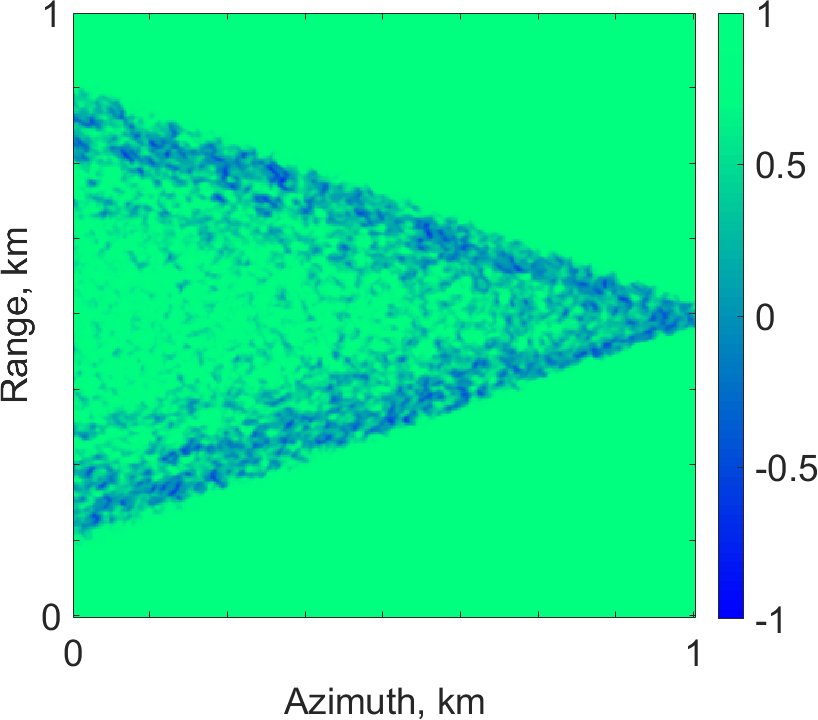}}}
\subfigure[]{\includegraphics[width=.20\linewidth]{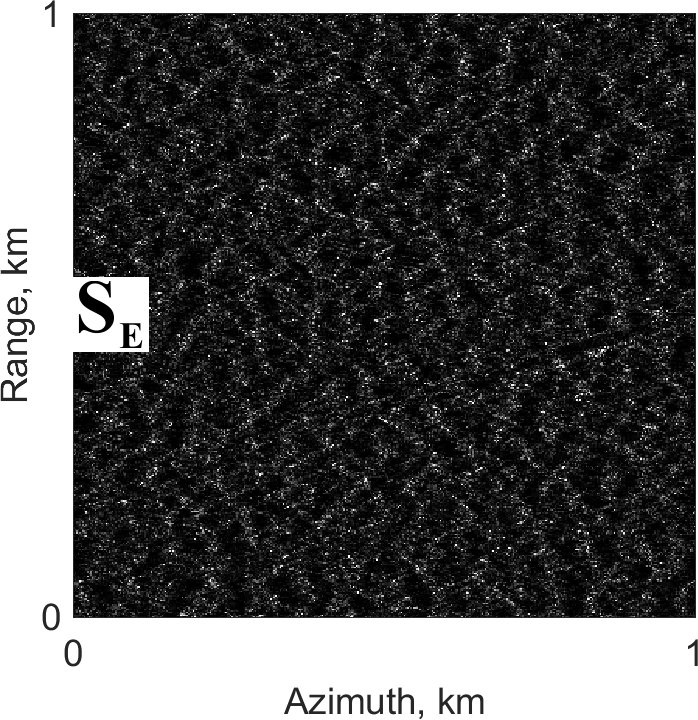}}
\subfigure[]{\includegraphics[width=.20\linewidth]{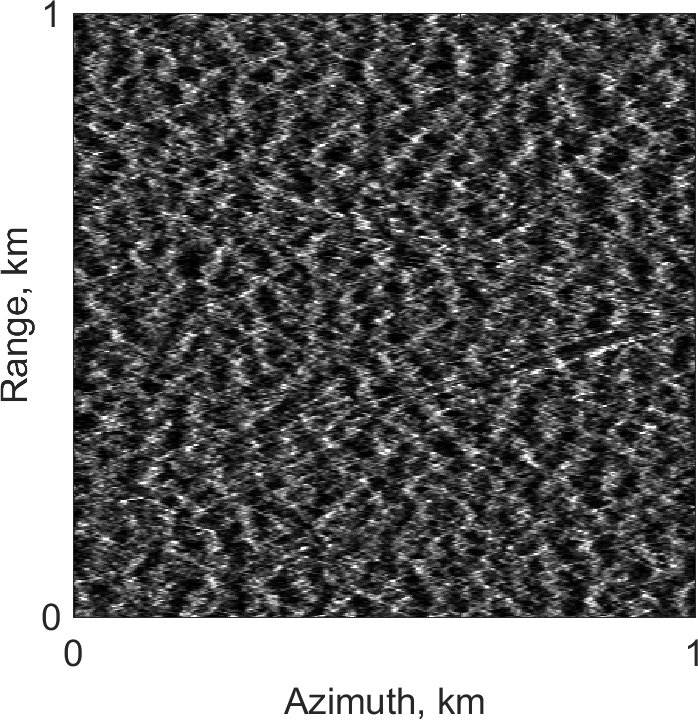}}
\subfigure[]{\includegraphics[width=.235\linewidth]{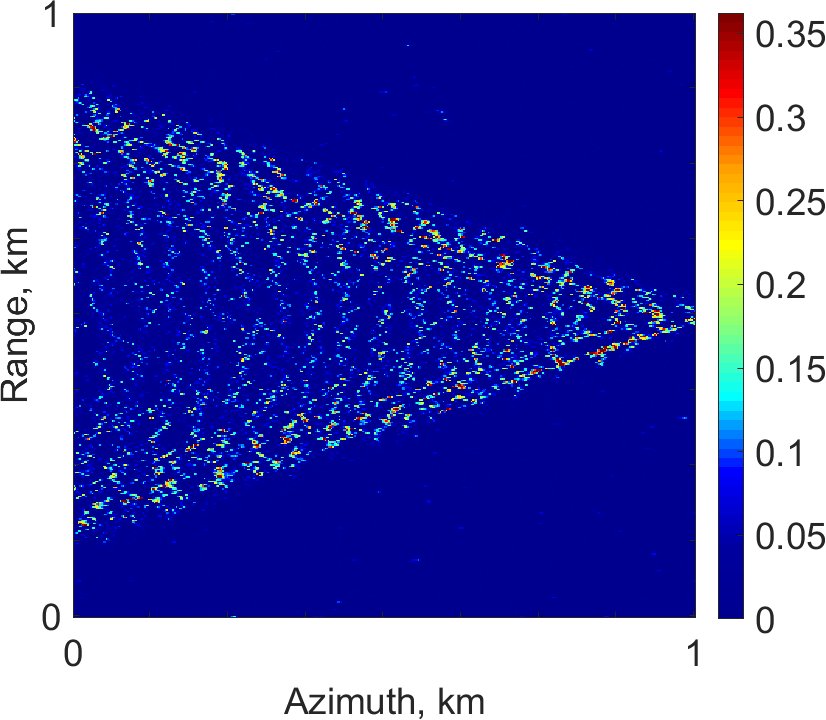}}
\subfigure[]{\raisebox{0.5mm}{\includegraphics[width=.23\linewidth]{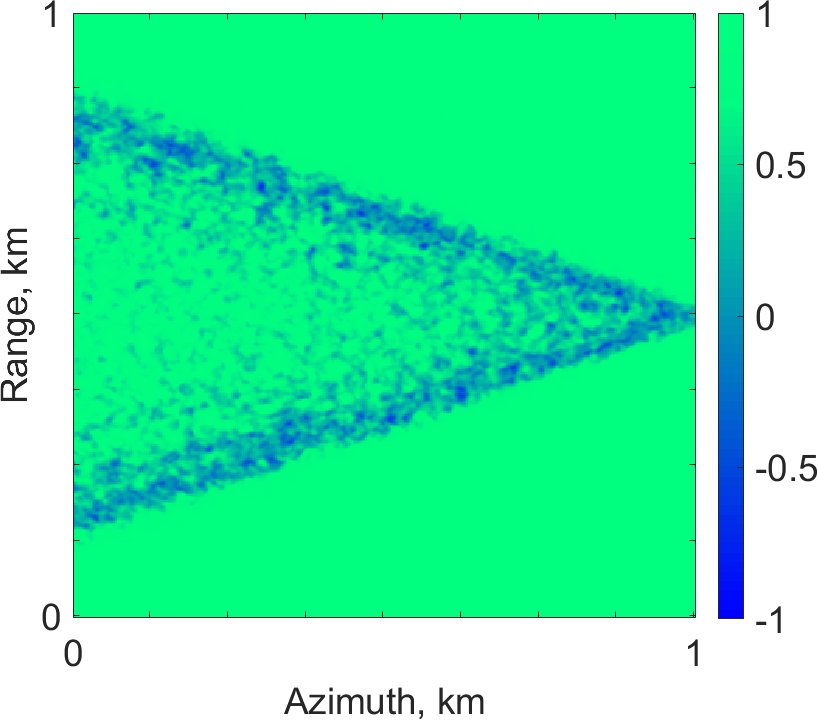}}}
\subfigure[]{\includegraphics[width=.20\linewidth]{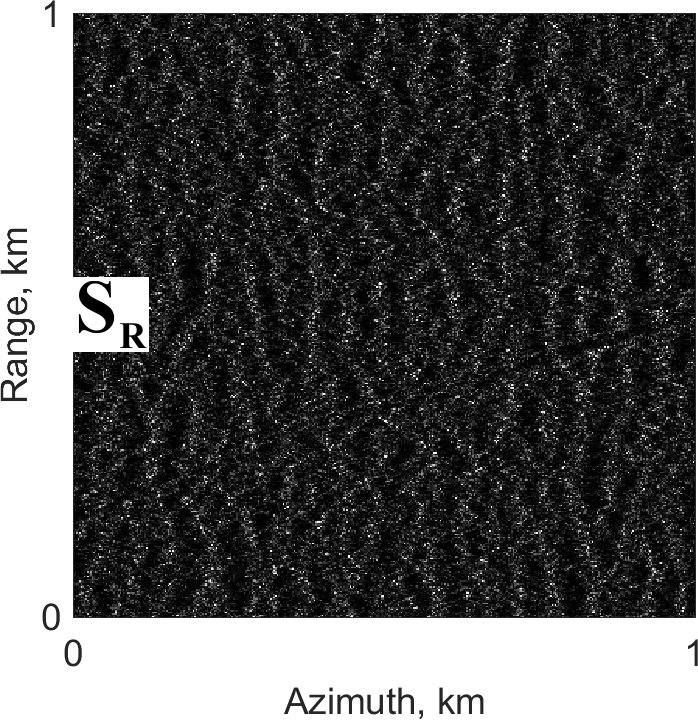}}
\subfigure[]{\includegraphics[width=.20\linewidth]{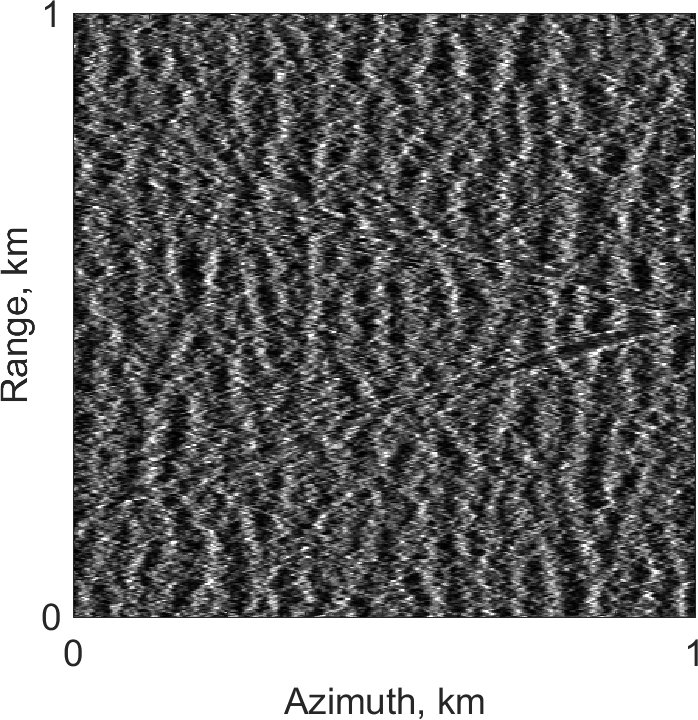}}
\subfigure[]{\includegraphics[width=.23\linewidth]{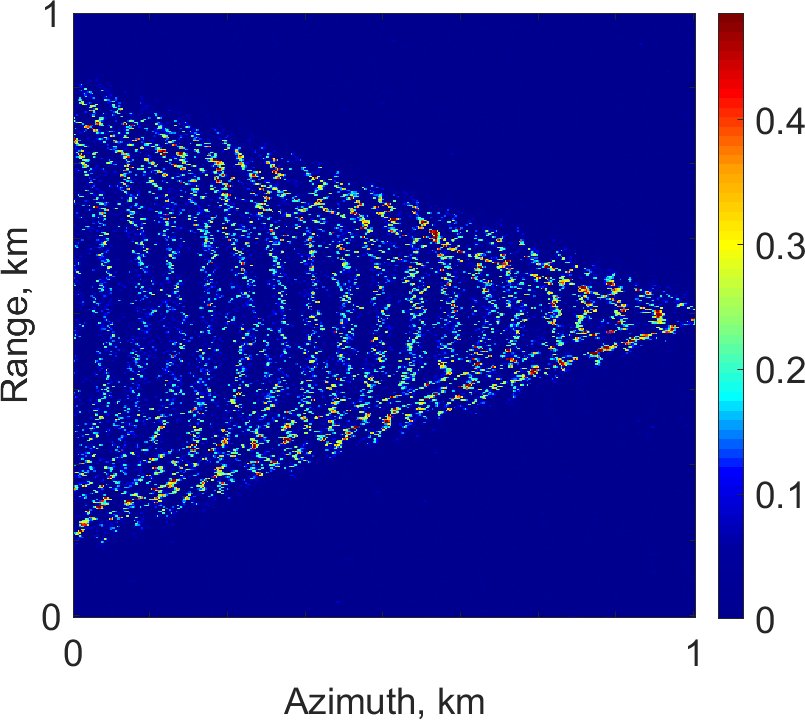}}
\subfigure[]{\raisebox{0.5mm}{\includegraphics[width=.23\linewidth]{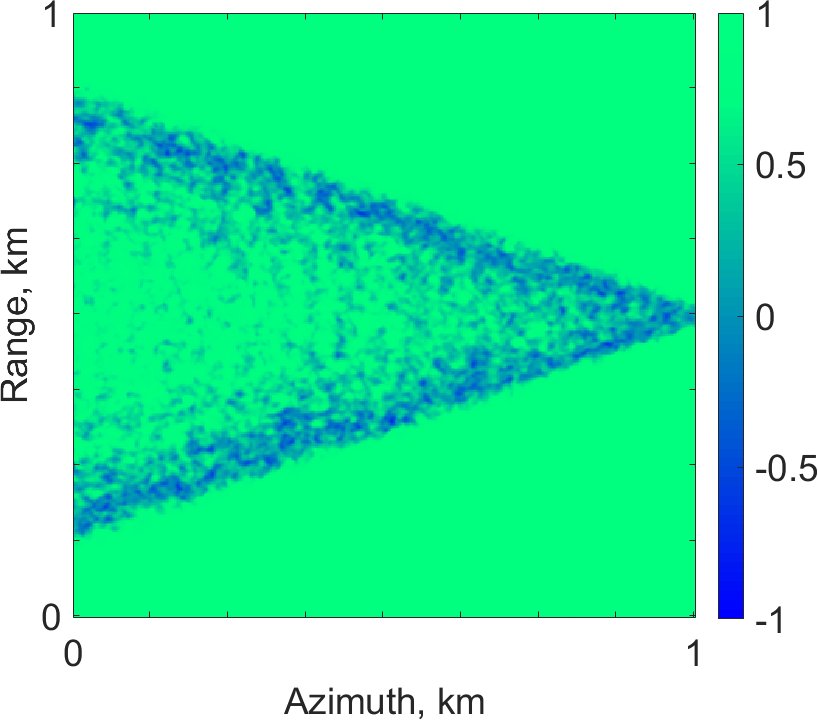}}}
\caption{Simulated SAR images (X-band, $\theta_r = 35^{\circ}$, VV polarization) for different spectra models: (a)-(d) ${{S}_{PM}}$. (e)-(h) ${{S}_{J}}$. (i)-(l) ${{S}_{FL}}$. (m)-(p) ${{S}_{E}}$. (q)-(t) ${{S}_{R}}$. For $S_{J}$ the fetch is 25 km, for ${S_E}$ the inverse wave age $\Omega$ is 0.84. For all spectra ${{V}_{w10}}$ = 8.5 m/s and \textbf{Ship I} with Fr = 0.5.}
\label{fig:fig14}
\end{figure*}

After the boundary condition ($V_{w10} = 8.5$ m/s) was determined (as the point where the visualization of wakes in SAR images becomes unstable, and therefore they may not be detected, as discussed above), the same ship model under the same velocity wind speed was applied for all spectra models. The benefit of the simulation is that it is possible to generate SAR images and reference SAR images to evaluate the comparison. The reference image is considered to be the SAR image with the same sea surface model but without a ship wake model. The analysis is then carried out by comparison, for each spectrum of SAR image (intensity values of superimposed sea-ship waves  $I_s$),  with the reference image (intensity values of ambient sea waves  $I_w$)  in terms of peak signal-to-noise (PSNR), signal-to-noise (SNR), mean-squared error (MSE), standard deviation (STD) and the Structural Similarity (SSIM) index \cite{wang2004image}. Since for different spectra models the absolute NRCS values change (as confirmed in many studies \cite{zheng2018sea, ryabkova2019review, xie2019effects, zhang2020modeling}) and because we are interested in relative changes of intensity values, all images and reference images were normalized before calculating the PSNR, SNR, MSE, STD, and SSIM measures. The STD measure was calculated for the difference image, which was determined as $\Delta I = I_s - I_w$ (in Fig. \ref{fig:fig14}, for better visualization only the positive part of $\Delta I$ values is shown, while in Table \ref{tab:table2} the whole STD values are shown). Also, all statistics are provided for speckle-free images. It is useful to look at the effect of noise in the formation of a SAR image on pairs of speckle-free ($I_n$) and speckle ($I$) intensity images, which are presented in Fig. \ref{fig:fig14}-(a)-(b), (e)-(f), (i)-(j), (m)-(n), (q)-(r). It can be seen that the noise cancels out the details of both sea and ship waves. In contrast to the standard image analysis interpretation, where for example higher PSNR value indicates better denoising performance, here the intensities from ship wakes are considered as a 'positive' noise. In practice, it means that when the noise power is higher (bigger MSE and STD), and hence a lower PSNR value (and SNR), the wakes visualization is improved. The same argument applies to the SSIM, where the lower global values in Table \ref{tab:table2} and more negative local values in Fig. \ref{fig:fig14}-(d), (h), (l), (p), (t), correspond to better identification of ship wake signatures.

According to the results in Table \ref{tab:table2}, for all spectra except the $S_{FL}$ spectrum, the better visualization of wakes on most measures generally follows a decrease of the significant wave height $H_s$, which is consistent with previous simulation studies \cite{zilman2014detectability, wang2016application, zhang2020novel} or real images analysis \cite{hennings1999radar, panico2017sar, tings2018comparison}. As expected, the best performance is shown by $S_{J}$ model (the fetch size is 25 km), where the $H_s$ is around half compared to most of the presented spectra, so the amplitude of the wake is not washed out by the surrounding waves. This spectrum also creates a shorter wavelength of waves on the sea surface thereby allowing better preservation of the ship wake pattern. It is also important to note that there is some variation in the calculated values (Table \ref{tab:table2}) which indicates that the visibility of wake is also associated with the ambient sea wave pattern, which is in turn dependent on the properties of the spectra. For example, we can observe that the pair $S_{PM}$-$S_{R}$ gives similar results. Also, the significant wave height is much lower for the $S_{FL}$ where better wake imaging was expected. One possibility is that this may be attributable to the contribution of the $D_{FL}$ spreading function. Indeed, the $D_{FL}$ spreading function has an isotropic shape for large-scale waves (Fig. \ref{fig:fig4}) and at this scale it does not realistically represent the sea wave propagation \cite{elfouhaily1997unified}. This may be the cause of reduced wake visibility in the results, but this effect requires a separate study. This means that at the boundary condition the contribution of $H_s$ is not of primary importance for modeled SAR images and other modeling parameters can also play a role. Obviously, the obtained results of the comparison of the spectra indicate that for different marine areas and at the same wind speed, the visualization of wakes will differ. However, these results are valid for the boundary condition when the ambient sea waves are comparable in size (wavelength) and amplitude with ship wake (Fig. \ref{fig:fig13}-(c)).

In Fig. \ref{fig:fig14} we show the local maps of the difference image $\Delta I$ and SSIM index. For better visualization, only the positive values of $\Delta I$ are shown, which correspond to the high NRCS backscattering signal. It can be seen that the wake patterns for $S_{PM}$, $S_{E}$, and $S_{R}$ spectra are more smeared compared to $S_{J}$ and $S_{FL}$. In particular, individual divergent and transverse waves are difficult to distinguish because they are mixed with ambient waves. It is interesting to point out that although the statistical results for $S_{FL}$ spectrum are not the best (from Table \ref{tab:table2}), the wake pattern in the difference image $\Delta I$ is better compared to all but the $S_{J}$ spectra. Indeed, the details of transverse and divergent waves are better visualized. However, the relative contrast between backscattering from sea waves and ship wake is bigger for the $S_{J}$ spectrum (Fig. \ref{fig:fig14}-(g)), which is also well supported by the objective results (Table \ref{tab:table2}, MSE, and STD). The same is true for the SSIM index results, where greater difference (or lower global value in Table \ref{tab:table2}) is related to the $S_{J}$ spectrum which means better visualization and detectability of wakes.

\begin{table}[htbp]
  \centering
  \caption{Comparison of different spectra models in simulated speckle-free SAR imagery by various measures.$^{*}$}
    \resizebox{0.75\linewidth}{!}{\begin{tabular}{lrrrrrr}
    \toprule
    Spectrum & \multicolumn{1}{l}{$H_{s}$} & \multicolumn{1}{l}{PSNR} & \multicolumn{1}{l}{SNR} & \multicolumn{1}{l}{MSE} & \multicolumn{1}{l}{STD} & \multicolumn{1}{l}{SSIM} \\
    \toprule
    $S_{PM}$ & 1.732 & 22.289 & 9.104 & 0.006 & 0.046 & 0.762 \\
    \midrule
    $S_{J}$ & 0.795 & \textbf{16.147} & \textbf{6.371} & \textbf{0.0243} & \textbf{0.061} & \textbf{0.567} \\
    \midrule
    $S_{FL}$ & 1.027 & 23.639 & 7.606 & 0.004 & 0.045 & 0.778 \\
    \midrule
    $S_{E}$ & 1.890  & 23.117 & 8.480  & 0.005 & 0.033 & 0.758 \\
    \midrule
    $S_{R}$ & 1.559 & 22.549 & 8.498 & 0.006 & 0.050  & 0.773 \\
    \bottomrule 
    \end{tabular}}\\
    $^{*}$ \footnotesize{Since ship wake considered as a positive noise, the statistics must be interpreted inverted, explanation in the text above.}%
  \label{tab:table2}%
\end{table}%

\subsection{Comparing real vs simulated SAR imagery}

In the literature, comparisons between simulated and real SAR images of ship wakes are very rare and include, e.g., \cite{zilman2014detectability, wang2017sar, liu2017study}. The main reason consists in the limited availability of the necessary parameters for simulations, such as automatic identification system (AIS) data corresponding to a given image, or parameters characterizing the sea state an acquired image. We demonstrate, in the absence of such parameters, how to synthetically generate SAR images based on estimated parameters of ship and measurements of the sea state from real data.
Our experiment indicates that, given the availability of high-resolution SAR data, the key ship parameters can be approximated (as was also shown earlier in \cite{tings2016dynamically}) and a lack of AIS data will not always be a limitation.
We illustrate the reproduction of a real TerraSAR-X image containing examples of two fast moving vessels. The image used is of the Strait of Gibraltar \cite{terra2021} and is a Geocoded Ellipsoid Corrected (GEC) product acquired in StripMap mode with SAR resolution of 3.3 m (pixel spacing is 1.25 m). The image is in HH polarization and with incidence angle $\theta_r = 33.2^{\circ}$. First, a calibration was performed using SNAP software to convert the dataset to an NRCS image. Then we selected two sub-scenes approximately $3 \times 3$ km in size. According to the metadata of the real SAR image (descending pass direction, center heading angle is $\theta_r = 190.31^{\circ}$) we oriented the sub-scenes so that they correspond to the coordinates of the simulator, in which counting goes from the azimuth direction counterclockwise. The ship parameters $L$ and $B$ were carefully measured from the real image as follows: Ship model $\alpha$, $L$ = 100 m, $B$ = 17 m; Ship model $\beta$, $L$ = 60 m, $B$ = 16.5 m. For the draft parameter we selected  $D_{t}$ = 2.7 m for Ship model $\alpha$ and $D_{t}$ = 2.2 for Ship model $\beta$. This was based on analyzing typical high speed crafts using information at www.marinetraffic.com, so that the parameters correspond to real vessels. The heading angles were measured as $\theta_{s}$ = $336^{\circ}$ for Ship model $\alpha$ and $\theta_{s}$ = $151^{\circ}$ for Ship model $\beta$. To estimate the velocities of the vessels we applied two methods based on measurements of the wavelength of the transverse wave \cite{panico2017sar} and length of the waves along the cusp lines \cite{zilman2014detectability}. Thus we estimated the velocities as $V_s \approx 17$ m/s and $V_s \approx 14.5$ m/s for Ship model $\alpha$ and Ship model $\beta$ respectively.

\begin{figure*}[htbp]
\centering
\subfigure[]{\includegraphics[width=.43\linewidth]{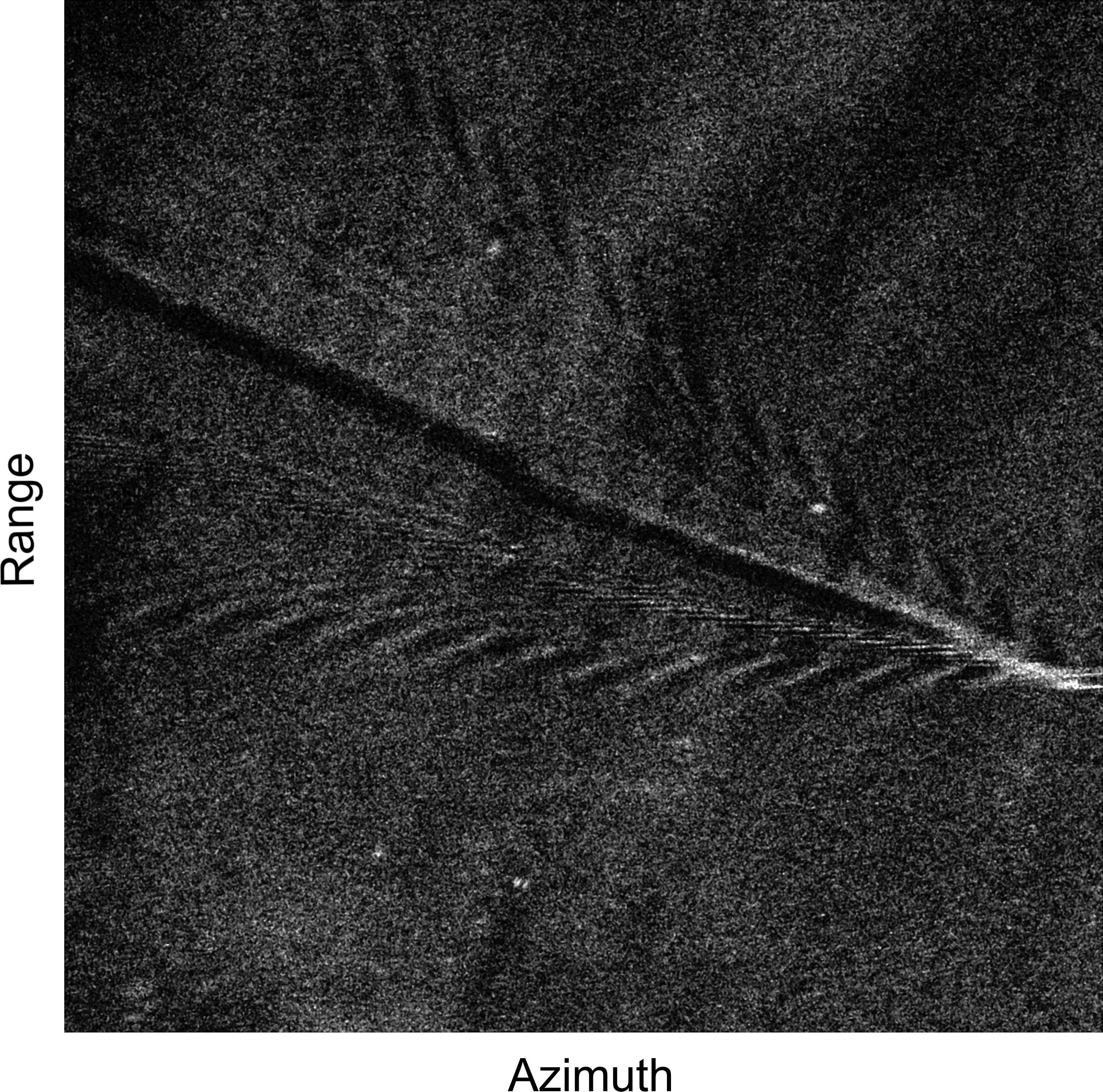}}
\subfigure[]{\includegraphics[width=.43\linewidth]{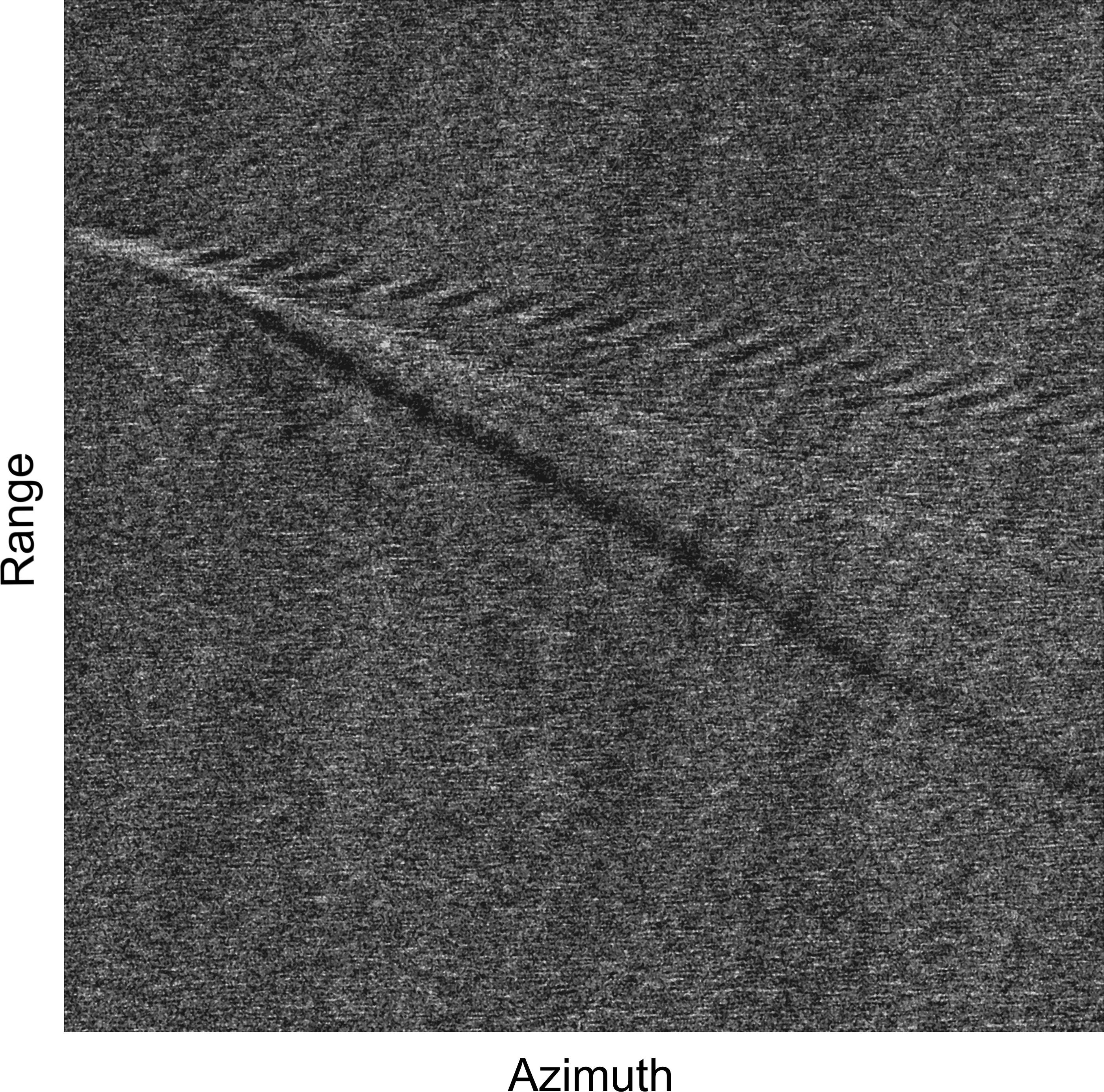}}
\subfigure[]{\includegraphics[width=.43\linewidth]{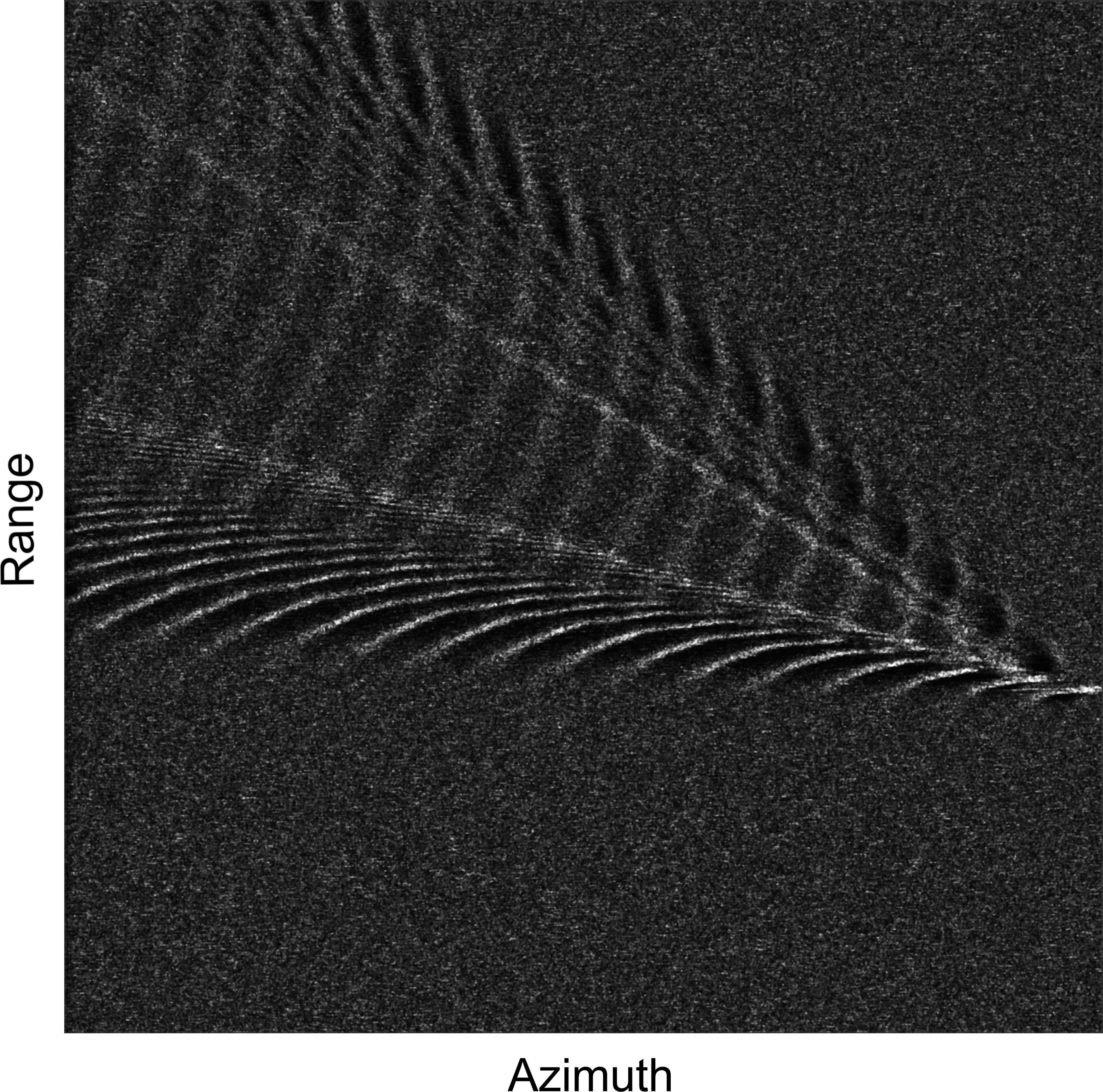}}
\subfigure[]{\includegraphics[width=.43\linewidth]{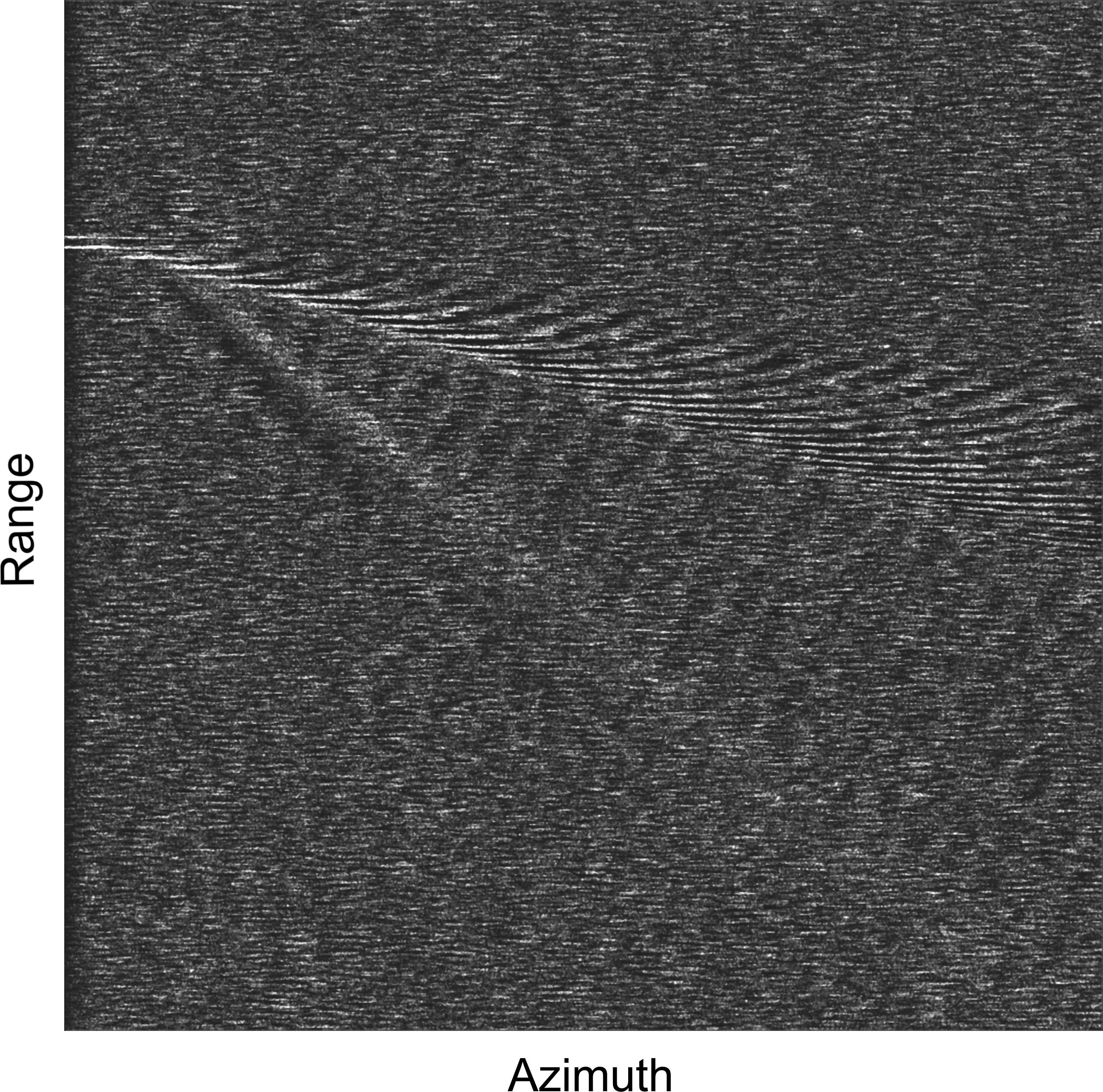}}
\caption{The comparison of TerraSAR-X sub-images (a), (b) with simulated images (c), (d). Parameters: SAR (X-band, $\theta_r = 33.2^{\circ}$, HH polarization); Ship model $\alpha$ ($L$ = 100 m, $B$ = 17 m, $D_{t}$ = 2.7 m); Ship model $\beta$ ($L$ = 60 m, $B$ = 16.5 m, $D_{t}$ = 2.2); Sea state (${{V}_{w10}}$ = 8.9 m/s with fetch = 6 km (c) and ${{V}_{w10}}$ = 14.6 m/s with fetch = 17 km (d)).}
\label{fig:fig15}
\end{figure*}

Finally, the sea state parameters were estimated as follows: Wind fields were derived from a Cross-Calibrated Multi-Platform (CCMP) 6-hourly ocean vector wind analysis model with model resolution 0.25 degree \cite{atlas2011cross}. The dataset corresponds to the four standard times (00:00, 06:00, 12:00, 18:00 UTC), and the closest time to the time the image was acquired was selected. The wind velocity is ${{V}_{CCMPw10}}$ = 5.5 m/s and wind direction $D_{CCMP} = 265.7^{\circ}$.
It should be noted that, due to the coarse resolution of the CCMP model, we preferred to estimate local winds around wakes. We used the XMOD2 algorithm for sea surface wind retrieval from TS-X imagery \cite{li2013algorithm, li2014correction}, in which the value of $D_{CCMP}$ provides the wind direction. We initially converted the NRCS from HH to VV polarization by using the T-PR model \cite{li2013algorithm}. We estimated values of ${{V}_{XMOD2w10}}$ = 8.9 m/s for ship model $\alpha$ and ${{V}_{XMOD2w10}}$ = 14.6 m/s for ship model $\beta$. Because of the type of area modeled (coastal) we employed $S_{J}$ and $D_{J}$ (with S = 8) for the sea waves. We also varied the fetch size (F = 6 km for Ship model $\alpha$ and F = 17 km for Ship model $\beta$) in $S_{J}$, which we measured as the distance between the wake location and shoreline and taking into account $D_{CCMP}$. The effect of fetch size is explained in details in \cite{rizaev2020icip}. The use of these measurements provides realistic sea surface modeling parameters as an input for the simulator. The resulting SAR imagery is presented in Fig. \ref{fig:fig15}. It can be seen that the common Kelvin wake patterns for both ships are similar. The main difference relates to the region of the transverse wave. This is due to the known fact that turbulent wakes generally mask the transverse waves in real images \cite{panico2017sar}. In addition, factors such as deviations in the course of the ship (Fig. \ref{fig:fig15} -(a)) or a change in its speed make modeling difficult. This demonstrates the success of our simulator in modeling real images based on approximated and measured or estimated parameters.

\section{Conclusion}\label{sec:conc}
In this paper, we reviewed the state-of-the-art and presented a comprehensive SAR imagery simulation framework for complex sea-ship waves scenario. This study was performed on the back of decades of research and knowledge that spans a very diverse range of fields including the linear theory \cite{holthuijsen2010waves} and stochastic modeling \cite{ochi2005ocean} of the sea surface, Kelvin ship wake modeling \cite{thomson1887ship, oumansour1996multifrequency, arnold2007bistatic, shemer1996simulation, zilman2014detectability} and the theory of SAR imaging oceans \cite{wright1968new, valenzuela1978theories, hasselmann1985theory, lyzenga1985sar, alpers1981detectability, romeiser1997improved}. In contrast to existing simulation studies for both wind- and ship-induced waves, we have extended our framework to include the analysis of hydrodynamical effects (waves elevation models) as well as SAR imaging effects. For the first time, on the basis of numerical simulations, the most common SAR imaging phenomena related to the sea surface (sea waves and ship wakes) have been integrated and presented. In particular, we investigated different SAR parameters, including frequencies (X, C, and L-band), incidence angles, signal polarizations (VV, HH), image resolutions (2.5, 5 and 10 m), four different platform configurations (two for airborne and two for spaceborne), and velocity bunching effects (azimuthal cut-off, shifting and smearing). Calculations were performed for five sea wave spectra ($S_{PM}$, $S_{J}$, $S_{FL}$, $S_{E}$, and $S_{R}$) and for four different ship models. The simulation results demonstrated a fine agreement with the theory. The proposed simulator package is implemented in MATLAB and publicly available via the University of Bristol Research Data Repository \cite{rizaev2021simulator}.

Our analysis of ship wake visibility for evaluating the various spectra supports the accepted rule that wake observability decreases with the increase of significant wave height. However, there is some variation, for example the lower visibility for the $S_{FL}$ spectrum at lower $H_s$, which means that other modeling parameters could contribute to wake visualization. The important conclusion is not however the perceived differences among spectra, but the very existence of a difference that was not investigated earlier in the SAR simulation of composite sea-ship scenes.

The presented versatile SAR imaging methodology may be more convenient as it allows the selection of different spectra for different tasks, for example, when considering different maritime regions (the fetch limited $S_{J}$ or not $S_{PM}$ spectra), or to obtain greater consistency between spectrum model and real sea surface roughness. This study can also be employed for a better understanding of the visibility and detectability of ship wakes in real SAR images. Work along these lines has been initiated, in particular for ship wake detection \cite{karakus2020tgrs}, despeckling of simulated SAR images \cite{karakus2020igarss}, and for studying the effect of sea state \cite{rizaev2020icip}.

Additionally, we would like to outline some interesting directions for future research. Although it is believed that wind direction has no effect on radar wake imaging, see for example \cite{hennings1999radar}, this conclusion is based on satellite observations and is mainly true for cusp waves only. In this respect, not only the direction but the angular spreading function $D(k, \theta)$ could possibly influence wave-wave interaction, which is directly related to the imaging of a vessel's signature. Thus, a more detailed investigation in this respect is needed. Another interesting direction is a comparison of simulated SAR images of ship wakes with real SAR images, for example using spectral decomposition approaches \cite{sun2018ship}. There have been good results where the simulated SAR spectra of ocean waves have been compared by spectral analysis with the real SAR measurements \cite{zurk1996comparison, bruning1994estimation, alpers1986relative}. However, the methodological basis for direct comparisons of simulated ship wake images with real images has still not been fully explored \cite{tunaley1991simulation, zilman2014detectability} and new approaches can be developed in this direction.

\section*{Acknowledgment}
This work was supported by the Engineering and Physical Sciences Research Council (EPSRC) under grant EP/R009260/1 (AssenSAR). The authors would like to thank Professor G. Zilman for many enlightening scientific discussions.

\bibliographystyle{IEEEtran}
\bibliography{igorReferences.bib}

\end{document}